%% file: EXOT-2016-24-PAPER.tex
\pdfoutput=1
\newcommand*{\ATLASLATEXPATH}{}
\documentclass[cernpreprint, texlive=2016, USenglish]{\ATLASLATEXPATH atlasdoc}

\usepackage[backend=bibtex]{\ATLASLATEXPATH atlaspackage}
\usepackage{\ATLASLATEXPATH atlasbiblatex}

\usepackage[jetetmiss]{\ATLASLATEXPATH atlasphysics}

\usepackage{EXOT-2016-24-PAPER-defs}
\usepackage{bm}
\usepackage{multirow}

\AtlasTitle{Search for heavy particles decaying into a top-quark pair in the fully hadronic final state in \pp collisions at \sqs =13~\TeV\ with the ATLAS detector}

\AtlasAbstract{%
A search for new particles decaying into a pair of top quarks is performed using proton--proton collision data recorded with the ATLAS detector at the Large Hadron Collider at a center-of-mass energy of $\sqrt{s} = 13$~\TeV\ corresponding to an integrated luminosity of 36.1~fb$^{-1}$. Events consistent with top-quark pair production and the fully hadronic decay mode of the top quarks are selected by requiring multiple high transverse momentum jets including those containing $b$-hadrons. Two analysis techniques, exploiting dedicated top-quark pair reconstruction in different kinematic regimes, are used to optimize the search sensitivity to new hypothetical particles over a wide mass range. The invariant mass distribution of the two reconstructed top-quark candidates is examined for resonant production of new particles with various spins and decay widths. No significant deviation from the Standard Model prediction is observed and limits are set on the production cross-section times branching fraction for new hypothetical $Z'$ bosons, dark-matter mediators, Kaluza--Klein gravitons and Kaluza--Klein gluons. By comparing with the predicted production cross sections, the $Z'$ boson in the topcolor-assisted-technicolor model is excluded for masses up to 3.1--3.6~\TeV, the dark-matter mediators in a simplified framework are excluded in the mass ranges from 0.8 to 0.9~\TeV\ and from 2.0 to 2.2~\TeV, and the Kaluza--Klein gluon is excluded for masses up to 3.4~\TeV, depending on the decay widths of the particles.}

\author{The ATLAS Collaboration}

\AtlasRefCode{EXOT-2016-24}

\PreprintIdNumber{CERN-EP-2018-350}

\AtlasJournal{Phys.\ Rev.\ D}
\AtlasJournalRef{Phys.\ Rev.\ D\ 99\ (2019)\ 092004}
\AtlasDOI{10.1103/PhysRevD.99.092004}

\hypersetup{pdftitle={ATLAS document},pdfauthor={The ATLAS Collaboration}}

\begin{document}
\captionsetup[figure]{name={Figure}}

\maketitle

\tableofcontents

\section{Introduction}
\label{sec:intro}
The Large Hadron Collider (LHC), 
currently operating at a center-of-mass energy 
of \sqs = 13~\TeV, has the potential to discover phenomena beyond the Standard Model (SM) 
at the \TeV\ scale. 
The heaviest elementary particle known in the SM, the top quark, is produced abundantly at the LHC\@. 
It is often predicted to be a probe for new physics  phenomena at the \TeV\ scale, 
in models such as the two-Higgs-doublet model (2HDM)~\cite{Branco:2011iw}, topcolor-assisted-technicolor~\cite{Hill:1993hs,Hill:1994hp,topcolor2} and Randall--Sundrum (RS) models of warped extra dimensions~\cite{Randall:1999ee, Lillie:2007yh}. 
Resonant production of a pair of top and antitop quarks (\ttbar) is particularly interesting as it 
provides a clear signature indicating
the existence of new heavy particles decaying into \ttbar. Such new particles 
could manifest themselves as a localized deviation from the SM prediction in the high 
invariant mass distribution of the \ttbar\ system (\mtt). In this paper,  a search for new particles in 
events containing  \ttbar\ pairs, where both the top and antitop quarks 
decay hadronically ($\ttbar \to W^+bW^-\bar{b}$ with $W \to q\bar{q'}$), is presented. 
The analysis is based on \lumi of proton--proton collision data at a center-of-mass energy of \sqs =13~\TeV\ recorded with the ATLAS detector at the LHC in 2015 and 2016.

The fully hadronic final state is characterized by the presence of multiple hadronic jets, two of which contain $b$-hadrons, and the absence of reconstructed leptons.
This {\em all-jets} topology benefits from the largest top-quark decay branching fraction (45.7\% of \ttbar decays), but suffers from large backgrounds due to QCD multijet production.
Dedicated top-quark reconstruction and identification techniques are used to enhance selection of \ttbar\ over
multijet events to maximize the sensitivity to the benchmark signals considered.
Two different search strategies are employed, each targeting a different mass range of the hypothetical resonance.
In the mass range below approximately 1.2~\TeV, 
where the decay products of the top quarks can be resolved as separate small-radius jets,
the ``buckets of tops'' algorithm~\cite{buckets} is used to optimize the reconstruction of top-quark-pair candidates.
At higher masses, top-quark decay products often merge into a single large-radius jet due to the high transverse momentum (\pt) of the top quarks,
hence a second strategy with a jet-substructure-based top-quark identification technique~\cite{Thaler:2010tr, Thaler:2011gf} is exploited. 
In the intermediate mass range of about 1.1 to 1.6~\TeV, signals are searched for using both strategies separately. 
The two results are compared at each mass point and the one with the better expected sensitivity is selected.

The ATLAS and CMS collaborations performed searches for heavy particles decaying into \ttbar\ using \pp\ collision data 
recorded at \sqs = 7~\TeV~\cite{TOPQ-2011-23,TOPQ-2012-14,CMS-EXO-11-006,CMS-TOP-11-010,CMS-TOP-12-017}, 
8~\TeV~\cite{CMS-B2G-13-001,TOPQ-2012-18,CMS-B2G-13-008,EXOT-2016-04} and 
13~\TeV~\cite{EXOT-2015-04, CMS-B2G-16-015, CMS-B2G-17-017} and set lower limits on the masses for several benchmark signal models.  The ATLAS search
at 13~\TeV~\cite{EXOT-2015-04}, using data equivalent to 36.1~\ifb, exploits the {\em lepton-plus-jets} topology, where a high-\pt electron or muon and large missing transverse momentum are required, and excludes masses below 3.0 (3.8)~\TeV\ for the new \zprime boson with an intrinsic decay width\footnote{In the rest of this paper, the decay width of a resonance divided by the resonance mass is referred to  as the width.} 
of $\Gamma=1\%$ (3\%) in the topcolor-assisted-technicolor model~\cite{Hill:1993hs,Hill:1994hp} (described in Section~\ref{sec:signal}).
The CMS search with the {\em lepton-plus-jets}, {\em all-jets}, and {\em dilepton} topologies 
at 13~\TeV~\cite{CMS-B2G-17-017} excludes the \zprime boson with $\Gamma=1\%$ up to 3.8~\TeV\ using 35.9~\ifb.
The Kaluza--Klein (KK) excitation of the graviton \kkG\ predicted in the specific ``bulk'' RS model~\cite{Agashe:2007zd, Fitzpatrick:2007qr} decaying into \ttbar (see details in Section~\ref{sec:signal})
 was also searched for by the ATLAS Collaboration and the mass range from 0.45 to 0.65~\TeV\ is excluded assuming $k/\overline{M}_{\mathrm{Pl}} = 1$, where $k$ is the curvature of the warped extra dimension and 
$\overline{M}_{\mathrm{Pl}} = M_{\mathrm{Pl}}/\sqrt{8 \pi}$ is the reduced Planck mass. 
The KK excitation of the gluon, \kkg, predicted in an RS model with a single warped extra
dimension~\cite{Lillie:2007yh} with $\Gamma=15\%$ (30\%) 
is excluded by the ATLAS search up to 3.8 (3.7)~\TeV. The CMS search~\cite{CMS-B2G-17-017} considered a slightly different 
model~\cite{Agashe:2006hk}, including a KK gluon with $\Gamma=20\%$ and larger production cross section, 
and set a lower limit of 4.55~\TeV\ on the mass.

The paper is organized as follows. The signal models considered are discussed in Section~\ref{sec:signal}. After a brief description of the ATLAS detector in Section~\ref{sec:detector}, 
the data and simulation samples are summarized in Section~\ref{sec:data_sim}. 
The analysis strategy including event selection, reconstruction and categorization is presented 
in Section~\ref{sec:strategy}.
The background estimation is described in Section~\ref{sec:background} and
the systematic uncertainties in the background and signal predictions in Section~\ref{sec:systematics}. 
After describing the signal search and 
the statistical procedure in Section~\ref{sec:stat_analysis}, the results are presented in Section~\ref{sec:result} with
the conclusions given in Section~\ref{sec:conclusion}.


\section{Signal models}
\label{sec:signal}
Several benchmark signal models are considered in this analysis, in which new spin-1 or spin-2 color-singlet and color-octet
bosons with masses ranging from 0.5 to 5~\TeV\ are introduced. The width of these bosons can vary from $\Gamma=1\%$ to 30\% to cover 
resonances narrower or wider than the typical detector resolution of about 10\%.

As the first benchmark,
a topcolor-assisted-technicolor (TC2) model~\cite{Hill:1993hs,Hill:1994hp} is considered, which predicts a spin-1 color-singlet boson.
This leptophobic $Z'$ boson (denoted by \zprime), referred to as Model IV in Ref.~\cite{topcolor2},
couples only to first- and third-generation quarks and is mainly produced by $q\bar{q}$ annihilation.
The model parameters are chosen to maximize the branching fraction for the $\zprime\rightarrow\ttbar$ decay, which reaches 33\%, and the width is set to $\Gamma=1\%$ or 3\%.

A framework of simplified models for dark matter (DM) interactions is considered as the second benchmark.
An axial-vector mediator \zprimeaxvec and a vector mediator \zprimevec are used,
following the recommendation of the LHC Dark Matter Working Group in Ref.~\cite{Albert:2017onk}. In the simplified model there are five parameters
relevant for $pp\to\zprimemed \to \ttbar$ processes (\zprimemed is either \zprimeaxvec or \zprimevec): the mediator mass $m_{\text{med}}$, the dark-matter mass $m_{\text{DM}}$, and the mediator
couplings to quarks $g_q$, to leptons $g_\ell$, and to dark matter $g_{\text{DM}}$. This search considers the coupling parameters 
defined in the A1 (V1) scenario of Ref.~\cite{Albert:2017onk} for the axial-vector (vector) mediator.
The branching fraction of the mediators into \ttbar is 8.8\% and the width is approximately constant at $\Gamma=5.6\%$ over the search range considered. The DM mass $m_{\text{DM}}$ is fixed to 10~\GeV.

An RS model with the SM fields propagating in the bulk of a single warped extra dimension~\cite{Lillie:2007yh} 
is used as the third benchmark, which predicts a 
spin-1 color-octet boson, the first KK excitation of the gluon, \kkg.
The \kkg is primarily produced in $q\bar{q}$ annihilation and 
decays predominantly into \ttbar with a branching fraction of approximately 92.5\% as predicted in Ref.~\cite{Lillie:2007yh}. 
In this analysis, the coupling of the KK gluon to quarks is set  to $g_q=-0.2 g_\mathrm{s}$,
where $g_\mathrm{s}$ is the strong coupling constant in the SM\@. The left-handed coupling to the top quark is fixed to $g_\mathrm{s}$ while 
the right-handed coupling is varied to change the intrinsic width.

The ``bulk'' RS model~\cite{Agashe:2007zd, Fitzpatrick:2007qr} with 
the SM fields propagating in the bulk, inherited from the original RS model, is used as the fourth benchmark
to predict a spin-2 color-singlet boson.
The first KK excitation of the graviton, \kkG, 
in this model is mainly produced in gluon--gluon fusion, and the production rate and width are controlled by a dimensionless coupling
constant $k/\overline{M}_{\mathrm{Pl}}$.
In this analysis $k/\overline{M}_{\mathrm{Pl}}$ is chosen to be 1, resulting in the \kkG width varying from $\Gamma=3\%$ to 6\% in the mass 
range between 0.5 and 3~\TeV. The branching fraction of the \kkG into \ttbar increases from 18\% to 50\% 
between 400 and 600~\GeV\ and stays approximately constant at 68\% for masses larger than 1~\TeV. 
In addition, the \kkG can decay into a pair of $W$, $Z$ or Higgs bosons and, with negligible branching fraction, into light fermions 
or photons.

Representative leading-order (LO) Feynman diagrams of the benchmark signals are presented in Figure~\ref{fig:intro_feynman}.

\begin{figure}[t]
\centering
\subfloat[\label{fig:intro_feynman_zprime}]{
        \includegraphics[width=0.32\textwidth]{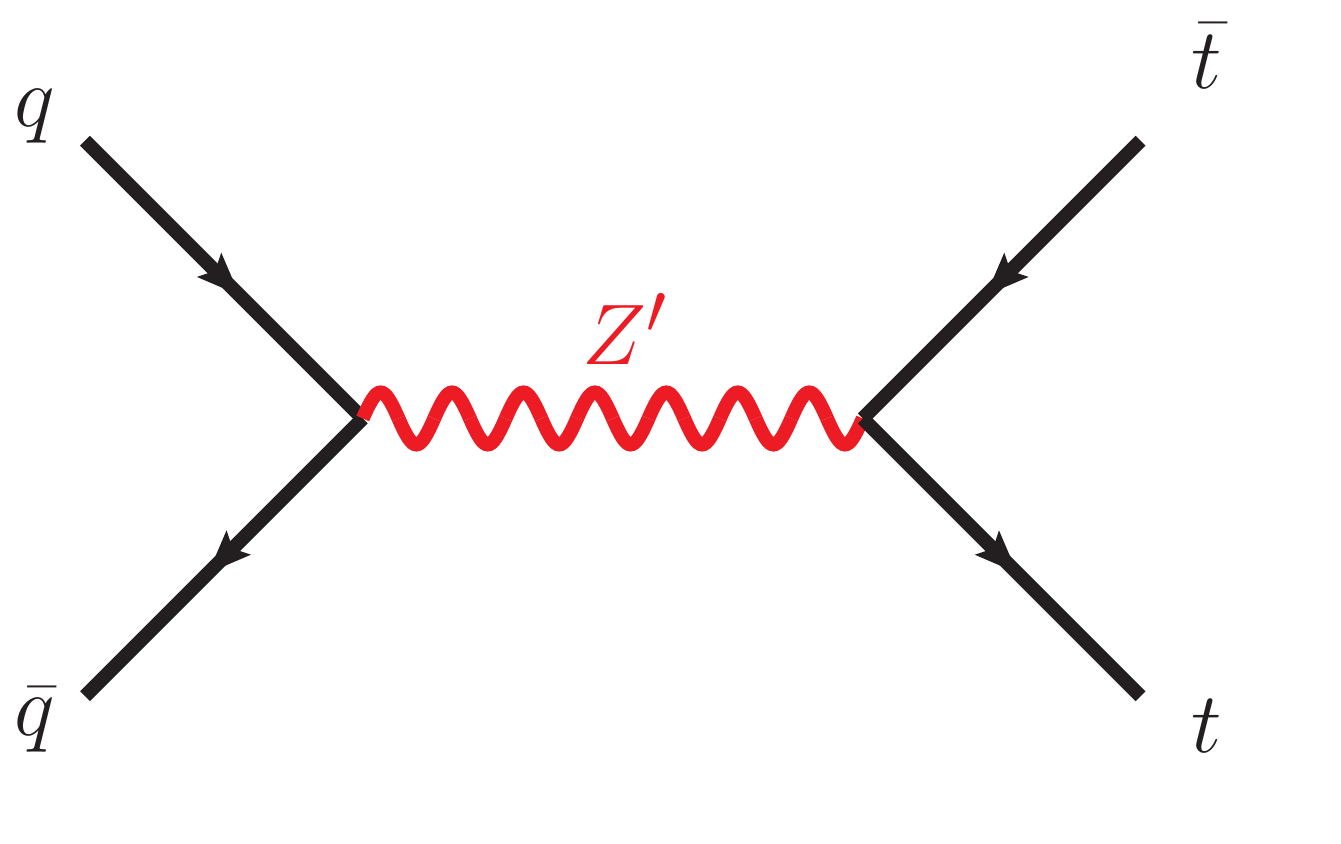}
}
\subfloat[\label{fig:intro_feynman_kkgluon}]{
        \includegraphics[width=0.32\textwidth]{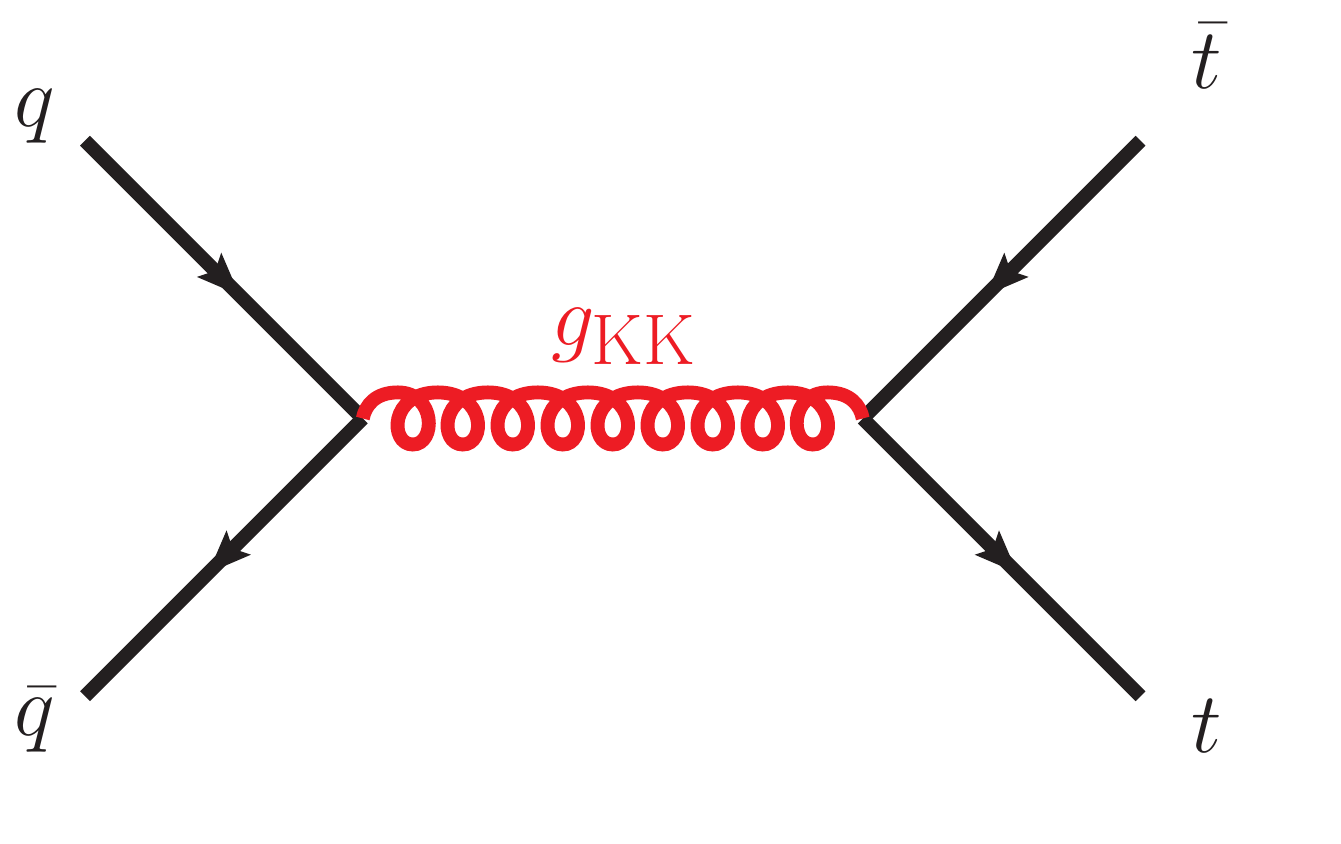}
}
\subfloat[\label{fig:intro_feynman_kkgraviton}]{
        \includegraphics[width=0.32\textwidth]{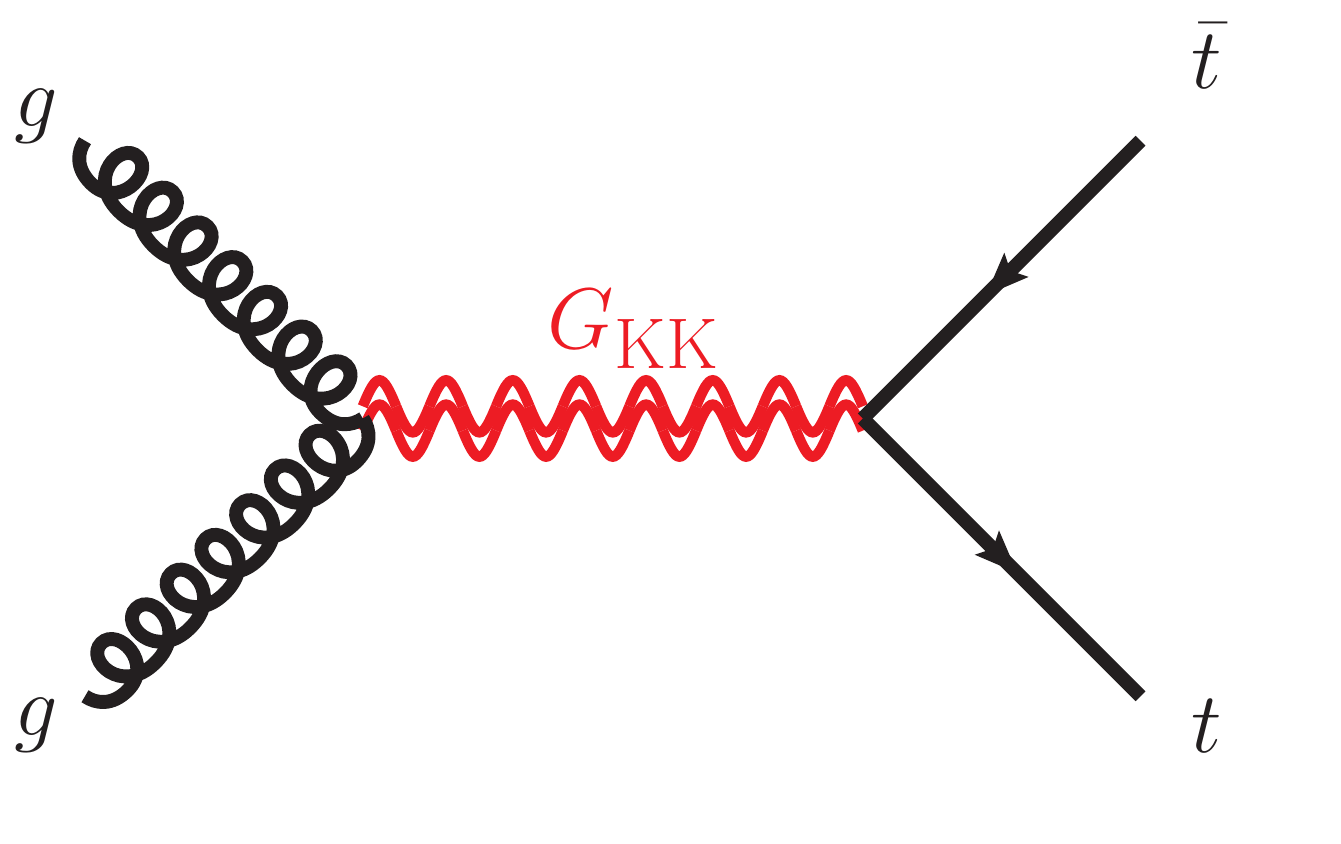}
}
\caption{Representative Feynman diagrams for leading-order production in the selected signal models: (a) $Z^\prime$, (b) \kkg and (c) \kkG. The details of each signal model are described in the text.}
\label{fig:intro_feynman}
\end{figure}


\section{ATLAS detector}
\label{sec:detector}
\newcommand{\AtlasCoordFootnote}{%
ATLAS uses a right-handed coordinate system with its origin at the nominal interaction point (IP)
in the center of the detector and the $z$-axis along the beam pipe.
The $x$-axis points from the IP to the center of the LHC ring,
and the $y$-axis points upwards.
Cylindrical coordinates $(r,\phi)$ are used in the transverse plane, 
$\phi$ being the azimuthal angle around the $z$-axis.
The pseudorapidity is defined in terms of the polar angle $\theta$ as $\eta = -\ln \tan(\theta/2)$.
Angular distance is measured in units of $\Delta R \equiv \sqrt{(\Delta\eta)^{2} + (\Delta\phi)^{2}}$.}

The ATLAS detector at the LHC is a multipurpose, forward--backward symmetric detector~\footnote{\AtlasCoordFootnote} 
with nearly full solid angle coverage, as described in Refs.~\cite{PERF-2007-01,ATLAS-TDR-19,Abbott_2018}. 
It consists of an inner tracking detector (ID) surrounded by a thin superconducting solenoid, a calorimeter system composed of 
electromagnetic (EM) and hadronic calorimeters, and a muon spectrometer.

The ID consists of a silicon pixel detector, 
a silicon microstrip tracker and a transition radiation tracker, all immersed in a \SI{2}{\tesla} axial magnetic field,
and provides charged-particle tracking in the range $|\eta|<2.5$.
The EM calorimeter is a lead/liquid-argon (LAr) sampling calorimeter with accordion geometry.
It is divided into a barrel section covering $|\eta|<1.475$ and two endcap sections covering $1.375<|\eta|<3.2$. 
For $|\eta|<2.5$ it is divided into three layers in depth,  which are finely segmented in $\eta$ and $\phi$. 
In the region $|\eta|<1.8$, an additional
thin LAr presampler layer is used to correct for energy losses in the material upstream 
of the calorimeters.
The hadronic calorimeter is a sampling calorimeter composed of steel/scintillator tiles 
in the central region ($|\eta|<1.7$), while 
copper/LAr modules are used in the endcap ($1.5<|\eta|<3.2$) regions. The forward region ($3.1<|\eta|<4.9$) is instrumented 
with copper/LAr and tungsten/LAr calorimeter modules optimized for electromagnetic and hadronic measurements,
respectively.
Surrounding the calorimeters is a muon spectrometer that includes three air-core superconducting toroidal magnets
and multiple types of tracking chambers,
providing precision tracking for muons with $|\eta|<2.7$ and trigger capability in the range $|\eta|<2.4$. 

A  two-level trigger system is used to select events for offline analysis~\cite{TRIG-2016-01}. 
Events are first selected by the level-1 trigger implemented in custom electronics, which uses a subset of the detector information
to reduce the event rate to \SI{100}{\kilo\hertz}. This is followed by a software-based trigger that 
reduces the accepted event rate to \SI{1}{\kilo\hertz} on average by refining the first-level trigger selection.


\section{Data and simulation}
\label{sec:data_sim}
This analysis is based on \lumi\ of \pp\ collisions recorded by the ATLAS experiment at the LHC at a center-of-mass energy of 13 \TeV\ in 2015 and 2016. 
A number of quality criteria were imposed to ensure that the data were collected during stable beam conditions 
with the relevant detectors operational.
Simulated signal and background events are used to optimize the event selection,
to estimate the background contribution and to perform the hypothesis test of the benchmark signal models considered.

The main backgrounds after applying criteria to enhance potential signals originate from SM \ttbar and multijet production. The \ttbar contribution and the related modeling uncertainties are
evaluated using Monte Carlo (MC) simulated events, 
while the multijet contribution is estimated directly from data. 
However, simulated events of multijet processes are used 
to optimize selection criteria 
and derive residual corrections to the
multijet distributions.

For the generation of SM $t\bar{t}$ events, 
the next-to-leading-order (NLO) generator \POWHEGBOX\ v2~\cite{Nason:2004rx,Frixione:2007vw,Alioli:2010xd} was used
with the CT10~\cite{Lai:2010vv,Gao:2013xoa} parton distribution function (PDF) set 
in the matrix element calculations. 
The \ttbar production cross section in $pp$ collisions at $\sqrt{s} = 13$~\TeV\ is
$\sigma_{t\bar{t}}= 832^{+46}_{-52}$~pb for a top-quark mass of $172.5$~\GeV.
It was calculated at next-to-next-to leading order (NNLO) in QCD including resummation of
next-to-next-to-leading logarithmic soft gluon terms with
Top++2.0~\cite{Cacciari:2011hy,Beneke:2011mq,Baernreuther:2012ws,Czakon:2012zr,Czakon:2012pz,Czakon:2013goa,Czakon:2011xx}.
Parton showering, hadronization and the underlying event were 
simulated using \PYTHIAV{v6.428}~\cite{Sjostrand:2006za} with the CTEQ6L1~\cite{Pumplin:2002vw} PDF set and the corresponding 
Perugia 2012 set of tuned parameters~\cite{Skands:2010ak}. 
The $h_{\mathrm{damp}}$ parameter, which controls the transverse momentum of the first additional parton emission 
beyond the Born configuration, was set equal to the top-quark mass. 
The top-quark kinematics in $t\bar{t}$ events were corrected to account for electroweak higher-order 
effects~\cite{Kuhn:2013zoa}.
The generated events were weighted by this correction factor as a function of the flavor and center-of-mass energy of the initial partons, 
and of the decay angle of the top quarks in the center-of-mass frame of the initial partons. 
The value of the correction factor decreases with increasing \mtt from 0.98 at $\mtt=0.4$~\TeV\ to 0.87 at 
$\mtt=3.5$~\TeV.
Multijet processes were simulated with the \PYTHIAV{v8.186}~\cite{Sjostrand:2007gs} generator
using the LO NNPDF2.3~\cite{Ball:2012cx} PDF set.

Simulated signal samples of spin-1 color-singlet \zprime bosons decaying into \ttbar were generated with \PYTHIAV{v8.165}~\cite{Sjostrand:2007gs}
with the LO NNPDF2.3 PDF set and 
the A14 set~\cite{ATL-PHYS-PUB-2014-021} of tuned parameters. To account for higher-order contributions, the LO calculation of the cross section was 
multiplied by a factor 1.3 obtained at NLO in QCD~\cite{Bonciani:2015hgv} 
using the PDF4LHC2015 PDF set~\cite{Butterworth:2015oua}. 
For the spin-1 mediators \zprimemed in the DM simplified model, the same samples are used after being reweighted
to have the approximate mediator width and cross section as simulated by \MGMCatNLO~\cite{Alwall:2014hca}.  The production cross sections were calculated at LO accuracy using the LO NNPDF2.3 PDF set.
The production of a spin-2 bulk RS graviton \kkG was performed
using \MGMCatNLO\ with the LO NNPDF2.3 PDF set, interfaced to \PYTHIAV{v8.165}
with the A14 set of tuned parameters for parton shower and hadronization. Simulated samples of spin-1 color-octet KK gluons 
\kkg with $\Gamma=30\%$ were generated with \PYTHIAV{v8.165} with the same PDF and 
tuned parameters as those used for the \zprime samples. 
Samples of \kkg with different widths (from 10\% to 40\%) were derived by reweighting the shapes of corresponding samples with $\Gamma=30\%$ and adjusting their normalization according to the appropriate prediction.
The \zprime and \kkg samples were generated for the mass range between 0.5 and 5~\TeV.
Signal masses were sampled at intervals of 100--150~\GeV\ below 1~\TeV, 250~\GeV\ between 1 and 3~\TeV\ and 500~\GeV\ above 3~\TeV\ for the \zprime.
The \kkg samples were produced at fixed intervals of 500~\GeV\ in all mass ranges.
The \kkG samples were generated between 0.5 and 3~\TeV\ in steps of 250~\GeV\ (1~\TeV) below (above) 1~\TeV.
The simulated samples are also used to evaluate the acceptance and selection efficiencies for the signals considered 
in the search.

The {\textsc{EvtGen}}  v1.2.0 program~\cite{Lange:2001uf} was used in all simulated samples to model the properties of heavy-flavor hadron decays.
All simulated samples include the effects of multiple \pp interactions in the same and neighboring 
bunch crossings (pileup) and are processed through the ATLAS detector simulation ~\cite{SOFT-2010-01} 
based on {\textsc{Geant4}}~\cite{Agostinelli:2002hh}. 
Pileup effects were emulated by overlaying simulated minimum-bias events generated with \PYTHIAV{v8.186}, 
using the MSTW2008LO PDF set~\cite{Martin:2009iq} and the A2 set of tuned parameters~\cite{ATL-PHYS-PUB-2012-003}.
The number of overlaid minimum-bias events was adjusted to match the luminosity profile of the recorded data.
Simulated events were processed through the same reconstruction software as the data, and corrections are applied so that the object identification efficiencies, energy scales and energy resolutions match those determined from control samples of data.


\section{Event reconstruction, selection and categorization}
\label{sec:strategy}
The production of a pair of hadronically decaying top quarks is characterized by the presence of multiple hadronic jets.
When the top quarks have moderate transverse momentum, \pt, of less than approximately 500~\GeV, the decay products can be reconstructed as separate jets,
which is referred to as the ``resolved'' event topology. 
At higher transverse momentum, the decay products of each of the two top or antitop quarks are merged into a single large-radius jet, referred to as the ``boosted'' event topology.
For both topologies the identification and reconstruction of the jets originating from the top quarks is crucial for reconstructing
the top-quark pair, resulting in a better separation of signal from background. 
The resolved and boosted event analyses are employed in parallel in the analysis. 

\subsection{Object reconstruction and event preselection}
\label{subsec:selection}
Events are required to have at least one $pp$ interaction vertex associated with two or more tracks with $\pt>400$~\MeV. 
If more than one vertex is found in an event, the one with the largest $\sum \pt^2$ of associated tracks is chosen
as the primary interaction vertex.
Depending on the kinematic regime of the top quarks, resolved or boosted, different jet reconstruction techniques are applied.
Events containing leptons (electrons or muons) are  included in the complementary search targeting 
the {\em lepton-plus-jets} topology~\cite{EXOT-2015-04} but are rejected in the analyses presented here.

{\textbf{Small-$R$ jets}} are built from three-dimensional topological clusters of energy deposits in the calorimeter~\cite{PERF-2014-07}, 
calibrated at the electromagnetic energy scale, using the anti-$k_t$ algorithm~\cite{Cacciari:2008gp} with a radius parameter $R=0.4$.  
These jets are calibrated to the hadronic energy scale by applying \pt- and $\eta$-dependent corrections 
derived from MC simulations and in situ measurements obtained from $Z/\gamma$+jets and 
multijet events at $\sqs = 13$~\TeV~\cite{PERF-2016-04}. Jets from pileup interactions are suppressed by 
applying the jet vertex tagger~\cite{PERF-2014-03}, 
which uses information from tracks associated with the hard-scatter and pileup vertices,
to jets with $\pt < 60$~\GeV\ and $|\eta| < 2.4$. Events containing jets from calorimeter noise or non-collision backgrounds are 
removed by discarding events containing at least one jet failing to satisfy the loose quality criteria defined in Ref.~\cite{ATLAS-CONF-2015-029}.
Jets that satisfy all the selection requirements and have $\pt > 25$~\GeV\ and $|\eta| < 2.5$
are considered in the resolved analysis. 
Small-$R$ jets containing $b$-hadrons are identified using an
algorithm~\cite{PERF-2016-05} based on multivariate techniques to combine information from the impact parameters of displaced tracks as well as topological properties of secondary and tertiary decay vertices reconstructed within the jet.
Two working points with 70\% (tight) and 85\% (loose) efficiencies for $b$-quark-induced jets are chosen, where the efficiencies are averaged values derived from simulated SM \ttbar events.
The corresponding misidentification rates of the tight (loose) working point are 0.26\% (3\%) and 8\% (32\%) for jets containing hadrons composed of light-flavor quarks and $c$-quarks, respectively.
Efficiencies to tag jets from $b$- and $c$-quarks in the simulation are corrected to match the efficiencies in data using
\pt-dependent factors, whereas the light-jet efficiency is scaled by \pt- and $\eta$-dependent factors~\cite{PERF-2016-05}.

{\textbf{Large-$R$ jets}} are built from three-dimensional topological clusters of energy deposits in the calorimeter calibrated with the local cluster weighting (LCW) procedure~\cite{PERF-2014-07} using the anti-$k_t$ algorithm with a radius parameter $R=1.0$. 
The non-compensating response of the calorimeter and the energy loss in dead material and due to out-of-cluster leakage 
from charged and neutral particles are corrected in the LCW procedure before jet reconstruction. 
The reconstructed jets are ``trimmed''~\cite{Krohn:2009th} to mitigate  contributions from pileup and soft radiation. 
In the trimming procedure, the jet constituents are reclustered into subjets using the $k_t$ 
algorithm~\cite{Catani:1991hj,Ellis:1993tq,Catani:1993hr} with a radius parameter $R=0.2$ and subjets with \pt less 
than 5\% of the \pT of the parent jet are removed~\cite{ATLAS-CONF-2015-035}. 
Finally, the large-$R$ jets are formed from the momentum vectors of the remaining subjets and selected by requiring   
$\pt > 200$~\GeV\ and $|\eta| < 2.0$ in the boosted analysis. 
For highly boosted top quarks, the mass resolution of a large-$R$ jet containing 
the top-quark decay products deteriorates with increasing top \pt due to the limited angular granularity of the 
calorimeter. To overcome this the mass of the large-$R$ jet, $m_\text{J}$, is calculated by combining the calorimeter
energy measurement with the track information from the ID, as described in Ref.~\cite{ATLAS-CONF-2016-035}. 
The two jets with the highest \pt in the event are required to have 50~\GeV$<m_\text{J}<350$~\GeV.

{\textbf{Track-jets}} are built from charged-particle tracks using the anti-$k_t$ algorithm with 
a radius parameter $R=0.2$. Tracks used in the reconstruction are selected by requiring that 
they are associated with the primary vertex, and have $\pt>400$~\MeV\ and $|\eta|<2.5$. 
Track-jets composed of at least two constituent tracks and having $\pt>10$~\GeV\ and $|\eta|<2.5$ are used to
identify jets containing $b$-hadrons in the boosted analysis.
In the dense environment characteristic of the boosted topology, the $b$-tagging is more efficient if performed on track-jets than 
on calorimeter jets~\cite{ATL-PHYS-PUB-2014-013}.
The same $b$-tagging algorithm as used for small-$R$ jets with 77\% (tight) and 85\% (loose) efficiency working points from $b$-quark-induced jets is employed.
The training of the multivariate algorithm and the evaluation of systematic uncertainties 
associated with the track-jet $b$-tagging efficiency are performed separately from those for the small-$R$ calorimeter jets.
The corresponding misidentification rates at the tight (loose) working point are 1.7\% (5.3\%) and 23.8\% (40.5\%) for 
light-flavor quarks and $c$-quarks, respectively.

{\textbf{Electrons}} are reconstructed from clusters of EM calorimeter energy deposits matched to 
an ID track with $|\eta|<2.47$, excluding the barrel and endcap transition region of
$1.37<|\eta|<1.52$. 
The electron candidates are required to have $\et>25$~\GeV\ and to
satisfy the ``tight'' identification criteria defined in Ref.~\cite{ATLAS-CONF-2016-024}. 
To suppress contamination from misidentified hadrons, the electron candidates 
are further required to be isolated from other hadronic activity in the event. This is achieved by requiring the 
scalar sum of track \pt within a cone around the electron direction, excluding the track associated with the electron, 
to be less than 6\% of the electron transverse momentum $\pt^e$. 
The cone size is given by the minimum of $\Delta R=10\,\text{GeV}/\pt^e$ and $\Delta R=0.2$.

{\textbf{Muons}} are reconstructed by combining tracks separately reconstructed in the ID and the muon spectrometer.
The muon candidates are required to have $\pt>25$~\GeV\ and $|\eta|<2.5$, and 
satisfy the ``medium'' quality requirements defined in Ref.~\cite{PERF-2015-10}. The muons are also required to be isolated
by using the same track-based isolation conditions as for electrons, 
except that the value of $\Delta R=0.2$ is replaced with $\Delta R=0.3$.

Electron and muon candidate tracks are required to be associated with the primary vertex using criteria based
on the longitudinal and transverse impact parameters.
To avoid the misidentification of jets  as electrons and electrons from heavy-flavor decays, the closest small-$R$ jet within $\Delta R_y = \sqrt{(\Delta y)^{2} + (\Delta\phi)^{2}} = 0.2$ around
a reconstructed electron is removed.\footnote{The rapidity is defined as $y= \frac{1}{2} \ln\frac{E+p_z}{E-p_z}$ where $E$ is the energy and $p_z$ is the longitudinal component of the momentum along the beam direction.} If an electron is then found within $\Delta R_y = 0.4$ of a jet, the electron
is removed. If a muon is found within $\Delta R_y = 0.04+10\,\text{\GeV}/\pt^{\mu}$ of a jet (where
$\pt^{\mu}$ is the muon transverse momentum), the muon is removed if the jet contains at least three
tracks, otherwise the jet is removed.

In the resolved analysis, the event selection is based on multijet triggers requiring the presence of at least five small-$R$ jets 
with $\pt>60$--65~\GeV\ depending on the data-taking periods. Events are further required to have at least six jets
with $\pt > 25$~\GeV\ and $|\eta| < 2.5$, out of which the five highest-\pt jets must have $\pt > 75$~\GeV\ and $|\eta| < 2.4$.
Among those six jets at least two of them are required to be $b$-tagged with $|\eta|<1.6$ 
using the loose efficiency working point. 
The trigger efficiency for the events satisfying the offline selection criteria is estimated using a lower-threshold multijet trigger.
The trigger efficiency is above 99\% and consistent between data and the simulated events.

In the boosted analysis, events are selected using triggers that require at least one large-$R$ jet with 
$\pt>360$--420~\GeV\ depending on the data-taking periods. Events are required to have at least two large-$R$ jets with $\pt>400\,\GeV$ to ensure that the jets can fully contain the top-quark decay products. 
The large-$R$ jets with the highest and the second-highest \pt in the event are referred to as the leading and sub-leading jets, respectively.
The leading jet has to satisfy $\pt>500$~\GeV\ to ensure a nearly full trigger efficiency.
The trigger efficiency is measured using a control sample in data and found to be approximately 100\% in this \pt~range.
The invariant mass $m_\text{JJ}$ of the two leading large-$R$ jets is required to be $m_\text{JJ}>1$~\TeV\ to avoid a kinematic 
bias caused by the jet \pt~requirements.
The two leading jets are required to have an azimuthal angle difference larger than 1.6. 
In addition, each jet is required to have at least one track-jet within $\Delta R = 1.0$ satisfying the loose $b$-tagging efficiency working point.
The fraction of events with more than two $b$-tagged jets is negligibly small, and those events are rejected to simplify the data-driven multijet background estimation.

\subsection{Top-quark pair reconstruction}
\label{subsec:reconstruction}
In the resolved analysis, the top-quark pair reconstruction is achieved by exploiting 
the ``buckets of tops'' algorithm~\cite{buckets} using small-$R$ jets. 
In this algorithm, all jets in the event are assumed to originate from \ttbar events, 
including those from  initial- or final-state radiation, and are assigned to one of three groups, referred to as ``buckets''. 
The first two buckets correspond to reconstructed candidates of the two top quarks in \ttbar events and the third bucket 
contains all jets from extra radiation.
The assignment of small-$R$ jets to buckets is performed by taking all jet combinations and minimizing a metric based on the 
difference between the invariant mass of jets falling into one of the first two buckets and the top-quark mass. In this analysis 
the metric $\Delta^2$ is defined as 

\begin{equation*}
  \Delta^2 = \omega\Delta_{B_1}^2 + \Delta_{B_2}^2,\, \Delta_{B_{1(2)}} = |m_{B_{1(2)}} - m_{\text{top}}|,\, \omega = 100,
\end{equation*}

where $m_{B_{1(2)}}$ is the invariant mass of the jets falling into bucket 1(2), denoted by $B_{1(2)}$, and $\mtop=173.5$~\GeV\ 
is the top-quark mass. The difference from \mtop used in the simulation (172.5\,\GeV) does not affect the performance of the \ttbar reconstruction.
The $\omega$ factor is introduced to ensure that $B_{1}$ has a mass closer to \mtop than $B_{2}$, i.e.\ 
$\Delta_{B_1}<\Delta_{B_2}$, as described in Ref.\ \cite{buckets}.
No restriction is imposed on the multiplicity of jets falling into the buckets except that 
$B_1$ and $B_2$ are required to contain exactly one $b$-tagged jet each.
Furthermore, the mass window requirements of 

\begin{equation*}
155\,  \text{\GeV} < m_{B_{1,2}} < 200\, \text{\GeV}
\label{eq:m_Bi}
\end{equation*}

are applied to increase the fraction of \ttbar events.
The preferred two ``top buckets'' $B_{1,2}$ are further classified according to the hadronic $W$-boson decay. 
If the following condition is satisfied for at least one combination of two non-$b$-tagged jets ($k$,$l$), the  bucket is considered to contain a $W$-boson candidate and labeled \tw, 
otherwise it is labeled \tm:

\begin{equation*}
  \left| \frac{m_{kl}}{m_{B_i}} - \frac{m_W}{m_{\text{top}}} \right| < 0.15,
\end{equation*}

where $m_{kl}$ is the invariant mass of the  ($k$, $l$) jet combination inside $B_i$, and
$\mw=80.4$~\GeV\ is the $W$-boson mass. To retain \ttbar events where one of the jets originating from the top-quark decay, 
presumably the softer quark from $W \to q\bar{q'}$, falls outside the top buckets, two-jet top buckets are formed. 
The metric used to form the bucket is adjusted to be  

\begin{equation*}
  \Delta_B^{bj} = |m_B - 145\, \text{\GeV}|
\end{equation*}

if the bucket mass $m_B$ is smaller than 155~\GeV, otherwise $\Delta_B^{bj} $ is set to an arbitrary large number. 
The mass criteria are  
based on the top-decay kinematics in which only the $b$-quark and the harder quark from $W \to q\bar{q'}$ 
fall inside the bucket. 
When the two top buckets are classified as (\tw, \tw)  the event is kept. 
If the buckets are classified as (\tw, \tm) or (\tm, \tw) with the notation that
the first bucket in the parentheses is always chosen to be $B_1$, the $t_-$ bucket 
is recalculated using the new metric $\Delta_B^{bj}$ from all jets excluding those belonging to any \tw bucket 
in the event.
Hereafter these two categories are collectively referred to as (\tw, \tm).
If the two top buckets are (\tm, \tm), the new buckets are formed from all jets in the event by minimizing the sum of a
new metric $\Delta_{B_1}^{bj}+\Delta_{B_2}^{bj}$.
The new two-jet bucket is finally required to satisfy the mass window requirement of 

\begin{equation*}
75\, \text{\GeV} < m_{B_i}^{bj} < 155\, \text{\GeV}.
\label{eq:m_bj}
\end{equation*}

If an event has no buckets satisfying the mass window requirements, 
the event is classified as (\tz, \tz). 
Finally, the top-quark candidate, reconstructed as the sum of the momentum vectors of the jets in the \tw, \tm or \tz bucket,
is required to have $\pt>200$~\GeV\ to suppress
 multijet backgrounds. The performance of the resolved \ttbar reconstruction is summarized in Table \ref{tab:bucketsPerformance}. 
The resolution of the reconstructed \ttbar mass for the resolved analysis is typically 6\%.
 
\begin{table}[t]
\centering
\caption{
Performance of the resolved \ttbar\ reconstruction with the ``buckets of tops'' algorithm estimated using simulated SM \ttbar and \zprime (850~\GeV) events in the fully hadronic final state. The fraction of events in each of the five possible top bucket categories is shown for all events satisfying the selection criteria described in Section~\ref{subsec:selection}. For each event category the relative fraction of events that have correctly matched top-quark pairs is presented. The measure of accuracy is based on a geometrical matching in the $\eta$--$\phi$ plane. Specifically the matched top buckets are required to be within $\Delta R = 0.3$ of a simulated top quark. The momenta of the simulated top quarks are evaluated immediately before the decay. The errors indicate the statistical uncertainty only. }
\label{tab:bucketsPerformance}
\begin{tabular}{lrr p{10pt} rr}
\toprule
Top buckets category & \multicolumn{2}{c}{Fraction of events [$\%$]} & & \multicolumn{2}{c}{Matched top-quark pairs [$\%$]}  \\
                     & \multicolumn{1}{r}{SM \ttbar} & \multicolumn{1}{r}{\zprime (850~\GeV)}             & & \multicolumn{1}{r}{SM \ttbar} & \multicolumn{1}{r}{\zprime (850~\GeV)}                    \\
\midrule
$(\tz, \tz)$  &  $16.5 \pm 0.3$  &  $12.6 \pm 0.7$  & &  $57.1 \pm 1.0$ &  $63.6 \pm 2.7$ \\
$(\tm, \tm)$  &  $17.5 \pm 0.3$  &  $15.0 \pm 0.9$  & &  $66.7 \pm 0.9$ &  $74.2 \pm 2.6$ \\
$(\tm, \tw)$  &  $ 7.8 \pm 0.2$  &  $ 7.9 \pm 0.8$  & &  $72.2 \pm 1.3$ &  $80.0 \pm 3.9$ \\
$(\tw, \tm)$  &  $30.2 \pm 0.4$  &  $30.9 \pm 1.2$  & &  $78.9 \pm 0.6$ &  $82.6 \pm 1.5$ \\
$(\tw, \tw)$  &  $28.0 \pm 0.4$  &  $33.6 \pm 1.3$  & &  $88.7 \pm 0.5$ &  $90.7 \pm 1.1$ \\
\bottomrule
\end{tabular}
\end{table} 

For the boosted analysis, a top-quark pair is reconstructed using the top-quark tagging requirements based on the 
jet mass and a jet substructure variable called $n$-subjettiness, $\tau_n$~\cite{Thaler:2010tr, Thaler:2011gf}. 
For each large-$R$ jet, $\tau_n$ is calculated 
by reconstructing exactly $n$ subjets with the ``winner-take-all'' recombination scheme~\cite{Larkoski:2014uqa} from 
the large-$R$ jet constituents using the $k_t$ algorithm~\cite{Catani:1991hj,Ellis:1993tq,Catani:1993hr} with a radius parameter of $R=0.2$:

\begin{equation*}
\tau_n = \frac{1}{d_0}\sum_i \pt^i \times \min(\Delta R_{1,i}, \Delta R_{2,i}, \cdots, \Delta R_{n,i}),
\end{equation*}

where $\pt^i$ is the transverse momentum of the $i$-th large-$R$ jet constituent and $\Delta R_{j,i}$ is the 
$y$--$\phi$ distance between the subjet $j$ and the $i$-th constituent. The $\tau_n$ variable is scaled by
$d_0^{-1} = (\sum_i \pt^i \times R)^{-1}$ with $R=1.0$, the radius parameter of the large-$R$ jet.
To distinguish fully contained top quarks with a three-prong structure
from other backgrounds dominated by a single-prong or two-prong structure, the \ttt variable defined as
$\tau_{32}=\tau_3/\tau_2$ is used as a discriminant. Since there are two top quarks in signal events, the
\ttt variables from the two leading large-$R$ jets are used to construct a single likelihood ratio $L_{\ttt}$, which is then used to suppress the multijet background.
The likelihood ratio is computed as $L_{\ttt} = P_{\text{s}}/(P_{\text{s}}+P_{\text{b}})$ where 
$P_{\text{s}}$ and $P_{\text{b}}$ are the probability 
density functions for the signal and background, respectively, 
obtained from MC simulations (see Section~\ref{sec:data_sim}).
The performance of the \ttbar reconstruction in the boosted analysis is summarized in Table \ref{tab:boostedPerformance}, where  signal regions as defined in Section~\ref{subsec:categorization} are used for illustration.
The resolution of the reconstructed \ttbar mass for the boosted analysis is typically 10\%.
 
\begin{table}[t]
\centering
\caption{
Performance of the boosted \ttbar reconstruction in the boosted analysis estimated using simulated SM \ttbar and \zprime (3~\TeV) events in the fully hadronic final state. The fraction of events in each of the eight possible boosted signal regions is shown for all events satisfying the selection criteria described in Section~\ref{subsec:selection}, together with the relative fraction of events that have correctly matched top-quark pairs. The measure of accuracy is based on a geometrical matching in the $\eta$--$\phi$ plane. Specifically the matched large-$R$ jets are required to be within $\Delta R = 0.4$ of a simulated top quark. The notation used to define each signal region is described in Section~\ref{subsec:categorization}. The momenta of the simulated top quarks are evaluated immediately before the decay. The errors indicate the statistical uncertainties only. }
\label{tab:boostedPerformance}

\begin{tabular}{lrr p{10pt} rr}
\toprule
Signal region category & \multicolumn{2}{c}{Fraction of events [$\%$]} & & \multicolumn{2}{c}{Matched top-quark pairs [$\%$]}  \\
                     & \multicolumn{1}{r}{SM \ttbar} & \multicolumn{1}{r}{\zprime (3~\TeV)}             & & \multicolumn{1}{r}{SM \ttbar} & \multicolumn{1}{r}{\zprime (3~\TeV)}                    \\
\midrule
Medium R1 1b  & $1.80 \pm 0.07$ & $2.41 \pm 0.08$  & & $89.8 \pm 4.4$ & $86.7 \pm 4.1$\\
Medium R1 2b  & $5.24 \pm 0.11$ & $4.39 \pm 0.10 $ & & $94.0 \pm 2.7$ & $84.3 \pm 2.8$\\
Tight R1 1b   & $2.55 \pm 0.08$ & $2.07 \pm 0.10 $ & & $93.8 \pm 4.0$ & $83.5 \pm 4.2$\\
Tight R1 2b   & $7.75 \pm 0.14$ & $4.18 \pm 0.10 $ & & $97.2 \pm 2.3$ & $83.5 \pm 2.8$\\
Medium R2 1b  & $1.20 \pm 0.06$ & $1.99 \pm 0.07 $ & & $83.8 \pm 5.3$ & $86.4 \pm 4.4$\\
Medium R2 2b  & $3.13 \pm 0.09$ & $3.08 \pm 0.08 $ & & $91.4 \pm 3.3$ & $86.3 \pm 3.3$\\
Tight R2 1b   & $0.89 \pm 0.05$ & $1.54 \pm 0.06 $ & & $90.0 \pm 6.6$ & $89.8 \pm 5.2$\\
Tight R2 2b   & $2.25 \pm 0.07$ & $2.59 \pm 0.07 $ & & $93.9 \pm 4.1$ & $86.5 \pm 3.6$\\
\bottomrule
\end{tabular}
\end{table}

\subsection{Event categorization}
\label{subsec:categorization}
For both the resolved and boosted analyses, the reconstructed events are categorized into several subsamples 
used for the signal search and background estimation.

In the resolved analysis, events satisfying the preselection criteria in Section~\ref{subsec:selection} are classified according to the reconstructed top buckets and number of $b$-tagged jets in the events. The combination of four possible pairs of top buckets, 
(\tw, \tw), (\tw, \tm), (\tm, \tm) and (\tz, \tz), and the two $b$-tagging criteria, i.e., (1) satisfying the tight or (2) satisfying 
the loose but failing to satisfy the tight efficiency working points for both $b$-tagged jets, are used to classify events into eight different regions A--D,  \cAz, \cAf, \cCz\ and \cCf
defined in Table~\ref{tab:bucketsABCDPurity}.
By construction those regions have no overlapping events. 
Region D, which contains events with (\tw, \tw) buckets and tight $b$-tagged jets, is the
most sensitive to the benchmark signals and hence chosen to be the main signal region (SR) for the resolved analysis. 
Regions A--C are used in a joint likelihood fit with the SR to extract the multijet background in the SR as detailed in Section~\ref{sec:background}.
The regions with the (\tm, \tm) and (\tz, \tz) buckets (\cAz, \cAf, \cCz
and \cCf) are used to estimate systematic uncertainties associated with the multijet background modeling (see Section~\ref{sec:systematics}).

\begin{table}[t]
\centering
\caption{Event categorization in the resolved analysis. The multijet-enriched regions A--C and the main signal region D, as well as the additional validation regions \cAz, \cAf, \cCz, \cCf selected with looser requirements on the top-quark pair candidates are shown. 
The events are also classified according to the two $b$-tagging criteria, i.e, satisfying the tight or satisfying the loose but failing to satisfy the tight efficiency working points for both $b$-tagged jets.
The expected fraction of
\ttbar events to the total background events in each region, as estimated from the simulation, is given in parentheses. 
The error indicates the statistical uncertainty only.}
\label{tab:bucketsABCDPurity}
\begin{tabular}{lcccc}
\toprule
Top buckets category &$(t_{0}, t_{0})$ &$(t_{-}, t_{-})$& $(t_W, t_{-})$ & $(t_W, t_W)$\\
\midrule
Loose $b$-tag & \cAz $(2.1 \pm 0.0)\%$   & \cAf \hspace{0.5em}$(4.2 \pm 0.1)\%$  & A $(12.3 \pm 0.2)\%$  & B $(38.9 \pm 0.9)\%$\\
Tight $b$-tag & \cCz $(8.0 \pm 0.1)\%$   & \cCf $(16.9 \pm 0.2)\%$  & C $(44.9 \pm 0.5)\%$  & D $(79.6 \pm 1.3)\%$\\
\bottomrule
\end{tabular}
\end{table}

In the boosted analysis, preselected events are first categorized by the number of tight $b$-tagged track-jets ($n_b$) and the \ttt-likelihood ratio (\lrt)  as shown in Figure~\ref{fig:categorization_boosted_a}.
Most signal events have $n_b=1$ or 2, which define the $1b$ and $2b$ regions. The events with $n_b=0$ ($0b$ region) are used to model the multijet background.
\begin{figure}[t]
\centering
\subfloat[][\label{fig:categorization_boosted_a}]{
        \includegraphics[width=0.87\hsize]{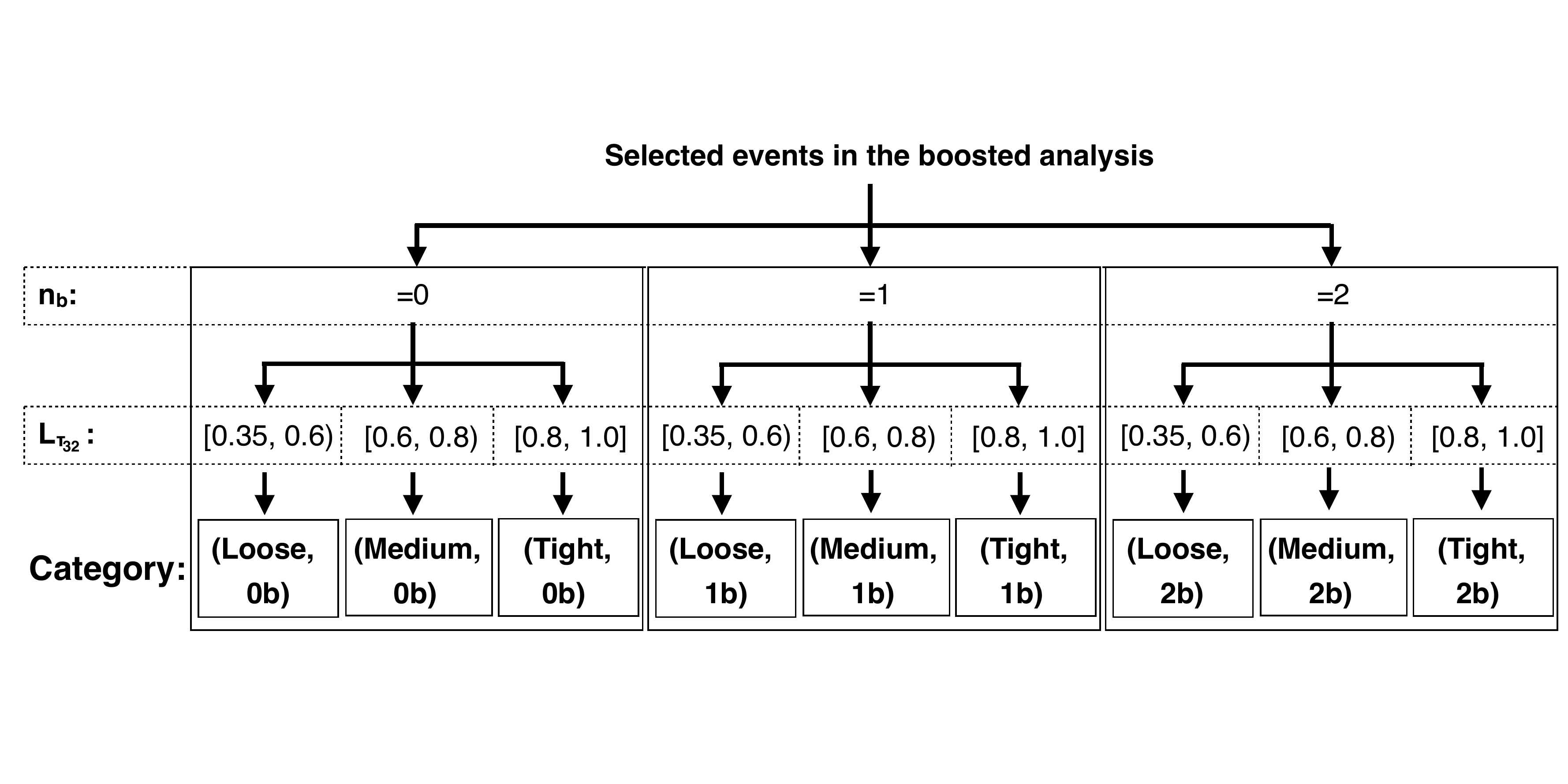}
}\\
\subfloat[][\label{fig:categorization_boosted_b}]{
        \includegraphics[width=0.57\hsize]{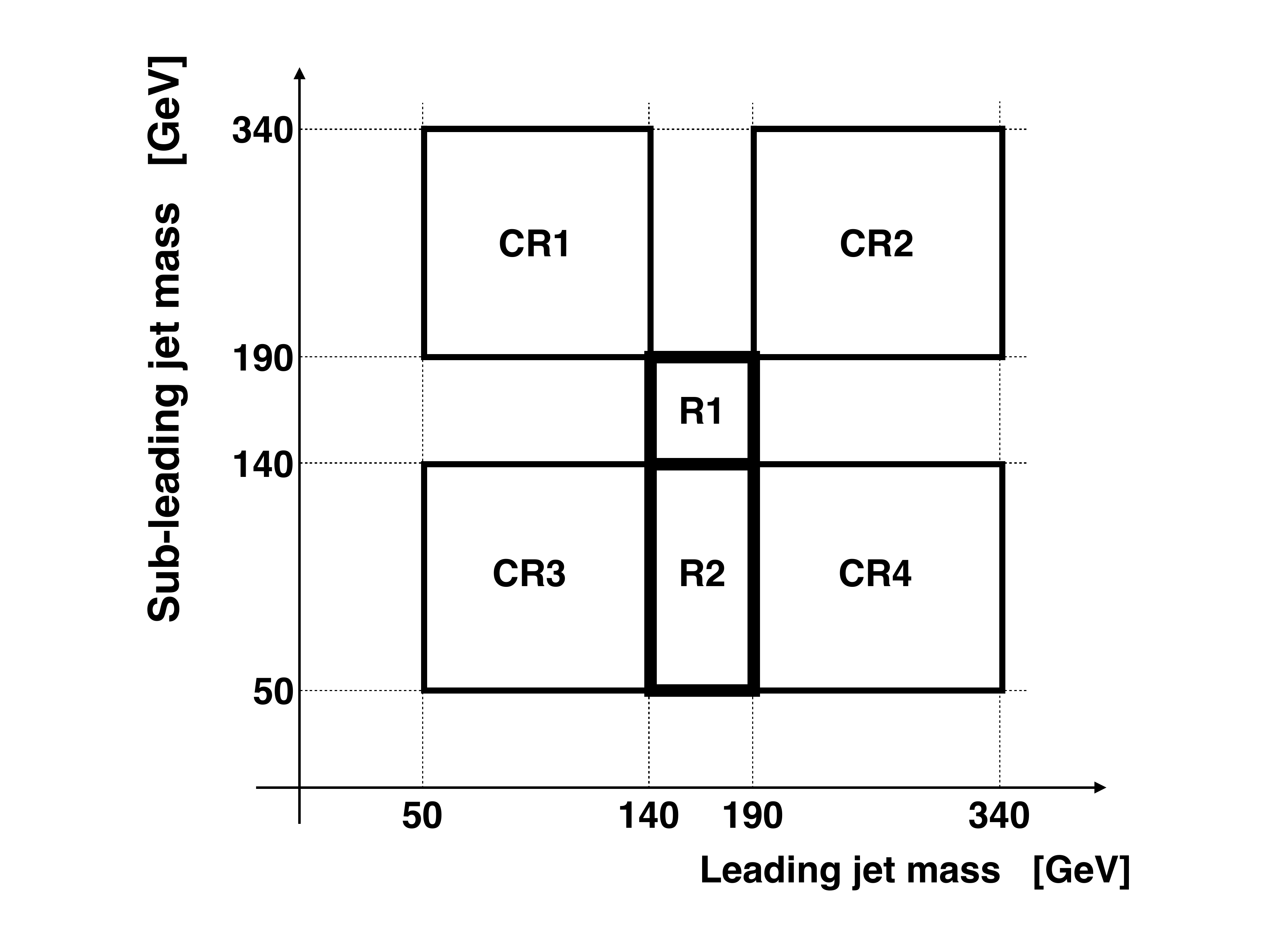}
}
\caption{Schematic diagram of the event categorization in the boosted analysis. (a) Events selected in the boosted analysis are classified into nine categories based on the number of tight $b$-tagged jets ($n_b$) and $L_{\tau_{32}}$, i.e, Loose, Medium and Tight regions for $n_b=0,1$ and 2. At least two loose $b$-tagged jets are already required in the preselection.  The region $0.35\leq\lrt<1.0$ is referred to as Inclusive. (b) In each category, events are further classified into three regions, R1, R2 and CR1--4, according to the leading and sub-leading large-$R$ jet masses.}
\label{fig:categorization_boosted}
\end{figure}

For the \lrt variable, the three criteria $0.35\leq\lrt<0.6$, $0.6\leq\lrt<0.8$ and $0.8\leq\lrt\leq1.0$ define Loose, Medium and Tight regions,
respectively, while $0.35\leq\lrt<1.0$ is referred to as Inclusive. The lower boundaries of the Tight and Medium regions are determined by optimizing the signal sensitivity 
while the lower boundary of the Loose region is used to ensure that events have kinematic properties similar to those in the Tight and Medium regions.
The Loose region is used for validation of the background estimation across the \lrt regions (see Section \ref{sec:background} for details).
The possible contamination from \zprime signal events in the Loose region is a few percent as estimated for a signal
with a cross section that has already been excluded by previous analyses. It is hence negligible for the signals with higher masses, which have lower predicted cross sections, and also for other benchmark signals with
kinematic properties similar to the \zprime.
In each category, events are further classified into
different regions using the masses $m_{\text{J}_{1}}$ and $m_{\text{J}_{2}}$ of large-$R$ jets with the leading and sub-leading \pt as shown in Figure~\ref{fig:categorization_boosted_b}. Representative distributions of the jet masses are shown in Figure~\ref{fig:BoostedKinematicsMass} for events satisfying the $(\text{Tight},1b)$ or $(\text{Tight},2b)$ requirements. The jet mass distributions are shown for the data and background predictions 
obtained after the fit to data (``Post-Fit''), as detailed in Section~\ref{sec:stat_analysis}.
\begin{figure}
\centering
\subfloat[]{\includegraphics[width=.45\textwidth]{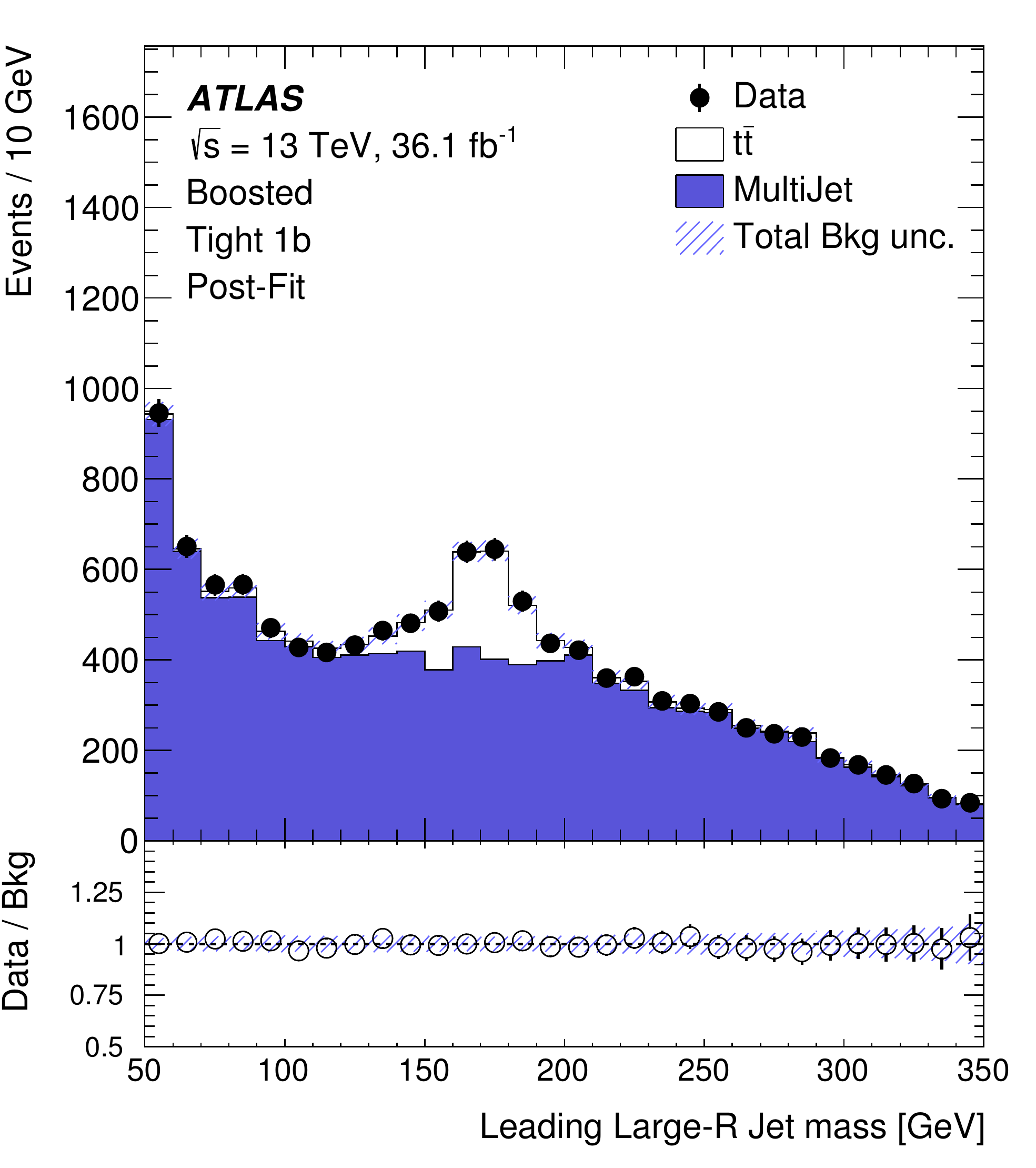}}
\subfloat[]{\includegraphics[width=.45\textwidth]{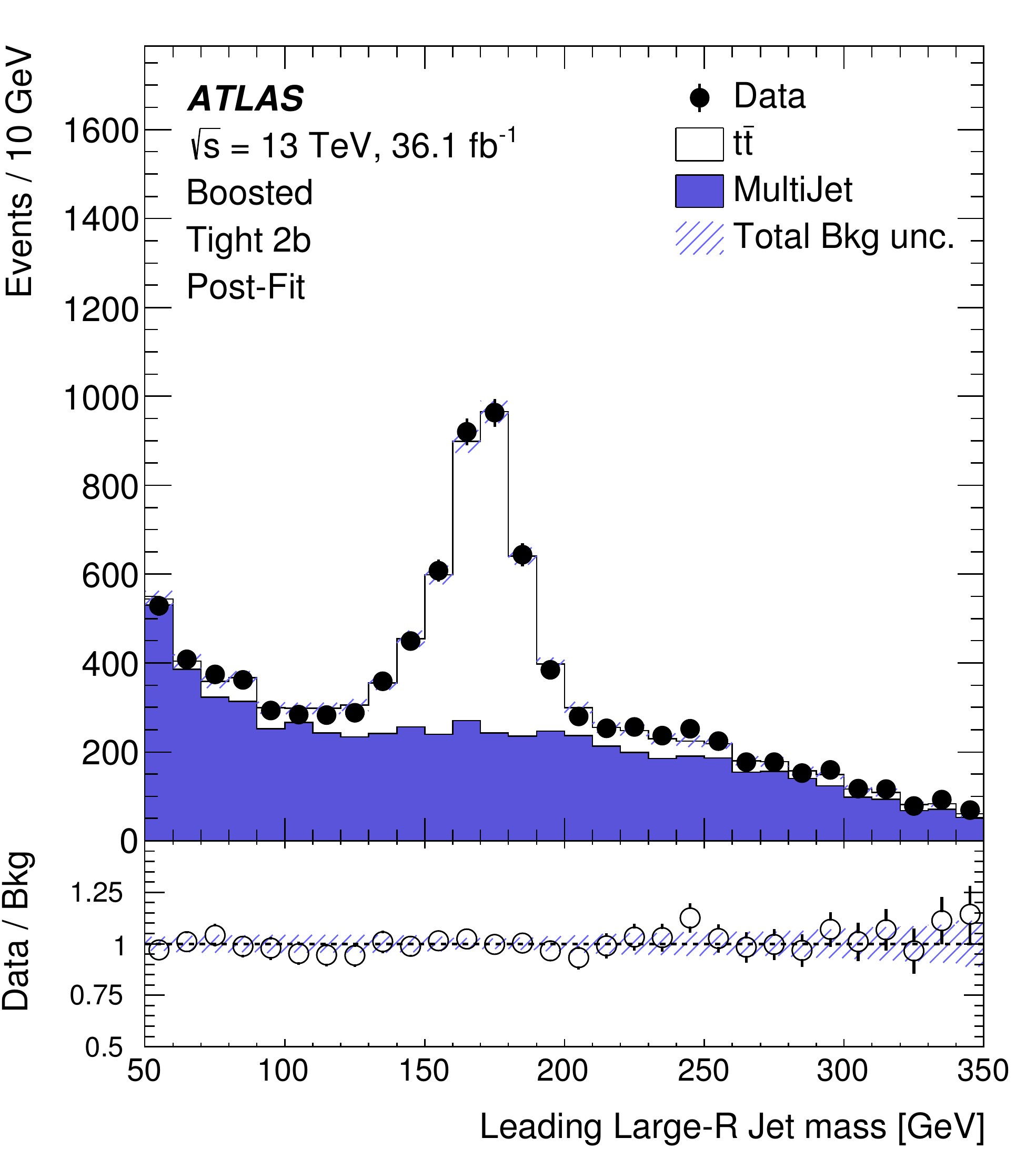}}\\
\subfloat[]{\includegraphics[width=.45\textwidth]{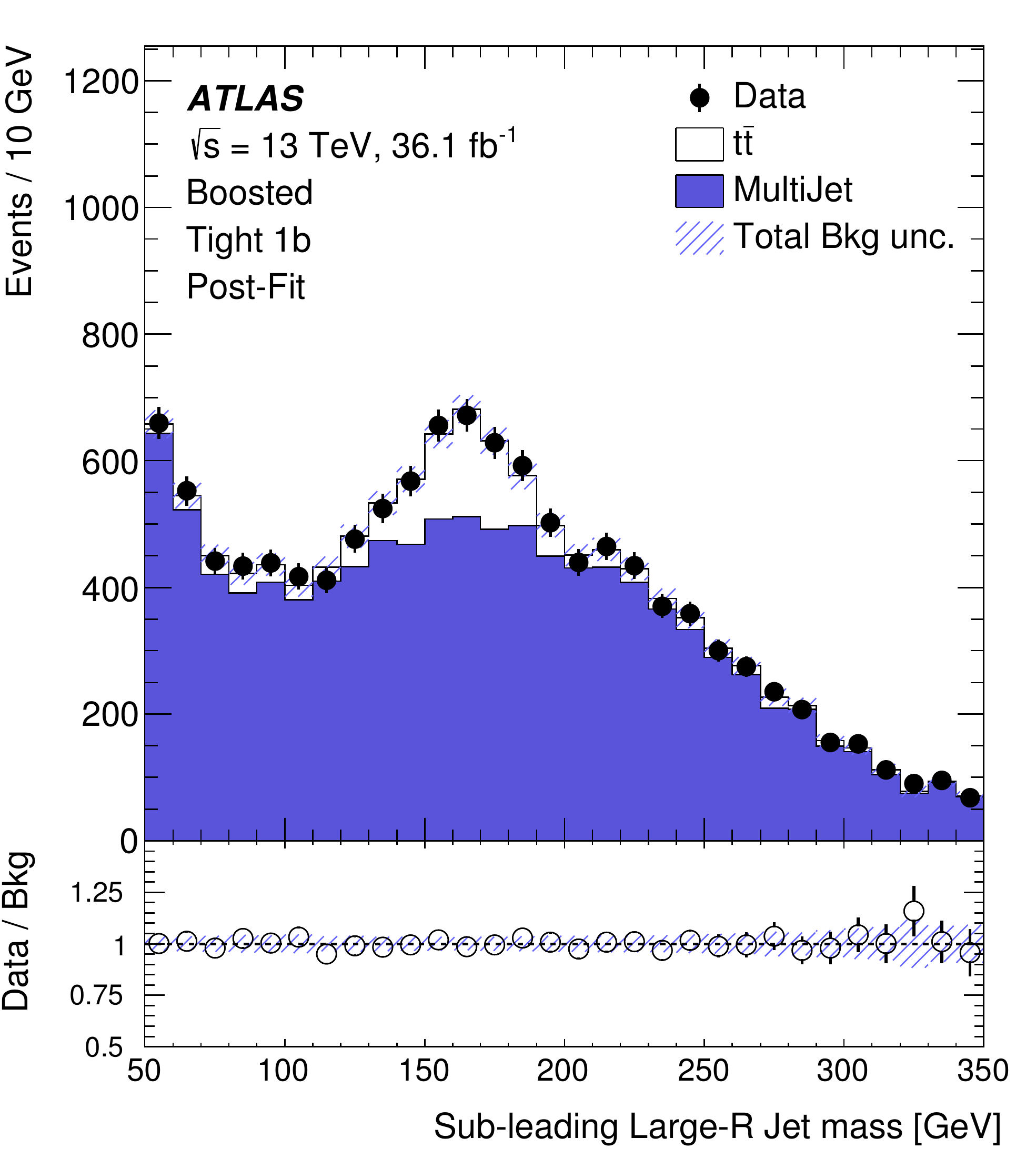}}
\subfloat[]{\includegraphics[width=.45\textwidth]{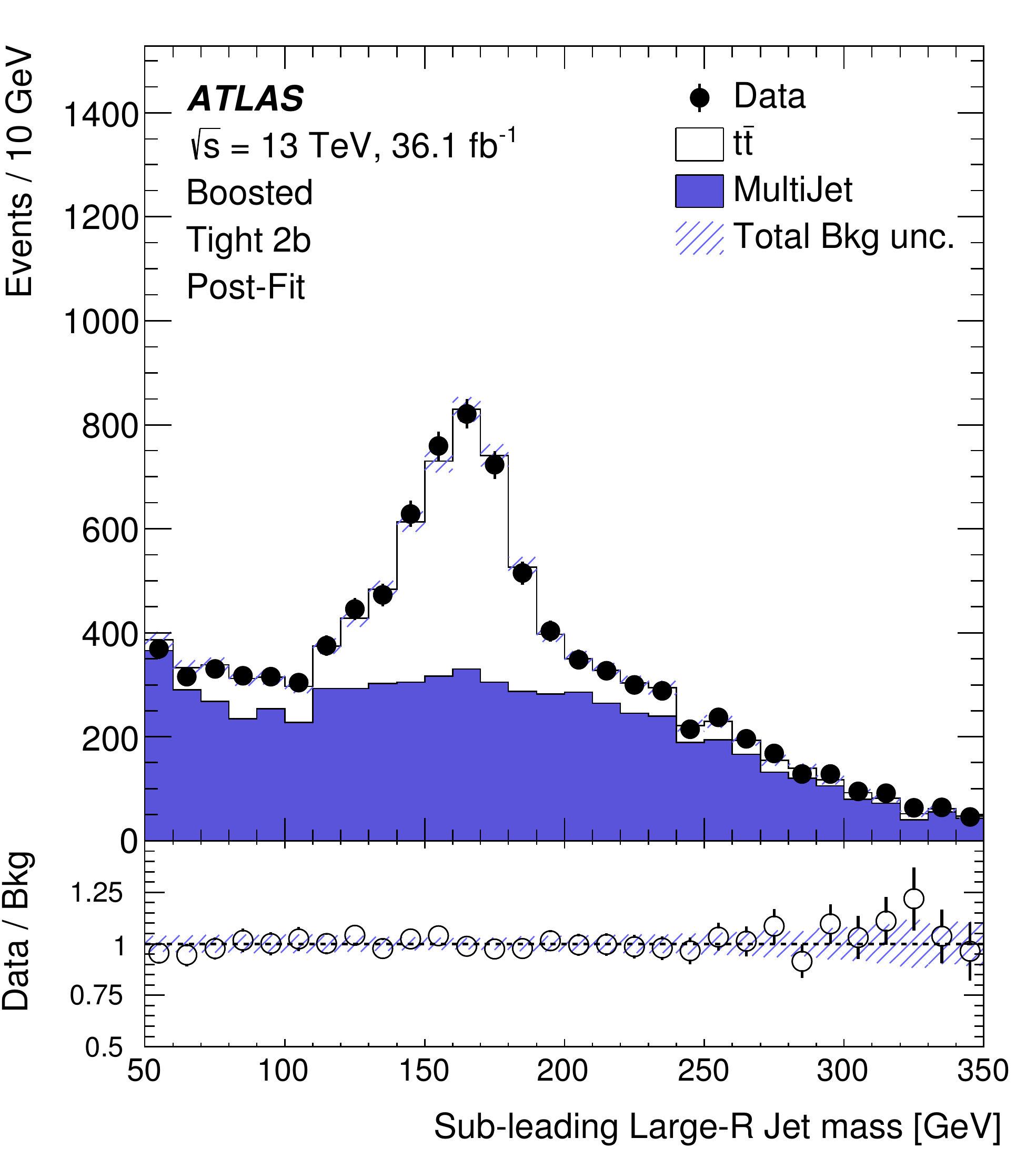}}\\
\caption{Comparison between data and predicted background after the fit (``Post-Fit'') in events satisfying the criteria for the Tight \lrt requirement and one (a, c) or two (b, d) $b$-tagged jets in the boosted analysis. Shown are (a, b) the mass of the leading reconstructed top-quark candidate, and (c, d) the mass of the sub-leading reconstructed top-quark candidate. The background components are shown as stacked histograms and the shaded areas around the histograms indicate the total systematic uncertainties after the fit. The lower panel of the distribution shows the ratio of data to the background prediction. The multijet contribution also contains all other small non-\ttbar backgrounds.
}
\label{fig:BoostedKinematicsMass}
\end{figure}
Signal regions are defined in the ranges
$140<m_{\text{J}_{1,2}}<190$~\GeV\ (denoted by R1) or $140<m_{\text{J}_{1}}<190$~\GeV\ and $50<m_{\text{J}_{2}}<140$~\GeV\ (denoted by R2). 
About 38\% (34\%) of the \zprime signal events with $m_{\zprime}=1.5$~\TeV (3~\TeV) fall into the region R1.
In some cases, not all partons from the top-quark decay ($q\bar{q}'b$) are fully contained within the large-$R$ jet, 
in particular at low \pt.
In the higher-\pt\ region above 1.2~\TeV, the large-$R$ jets contain all the decay products of the top quark more than
90\% of the time, but the mass resolution deteriorates and the number of jets lost due to final state radiation increases as a function of \pt.
Consequently, a significant fraction of signal events 
(28\% and 27\% at $m_{\zprime} = 1.5$~\TeV\ and 3~\TeV, respectively) 
have a lower mass for the sub-leading large-$R$ jets, falling into the region R2 of 
$50<m_{\text{J}_{2}}<140$~\GeV.
Therefore, eight SRs are considered in the boosted analysis, namely the R1 and R2 mass regions for each combination of 
the Tight or Medium \lrt requirement, and one or two tight $b$-tagged jets, as illustrated in Figure~\ref{fig:categorization_boosted} and Table~\ref{tab:boosted_categorization}.
The same categories but with the Loose \lrt requirement are collectively called the validation region (VR).
The regions labelled as control regions CR1--4 in Figure~\ref{fig:categorization_boosted_b} are used to determine the normalization of multijet backgrounds 
separately for the SR and VR\@.
The mass regions R1 and R2 in the $0b$ region are used to extract the shape of the multijet backgrounds in the SR and VR and are collectively called the template region (TR).
The details of the multijet background estimation are discussed in Section~\ref{sec:background}.

\begin{table}[t]
\centering
\caption{List of the event categories considered in the boosted analysis. The index $i$ is the region number defined in Figure~\ref{fig:categorization_boosted_b}. The indices $j$ and $k$ correspond to the \lrt and $n_b$ categories, respectively,
defined in Figure~\ref{fig:categorization_boosted_a}. The TR$i(\text{Inclusive},0b)$ is used to estimate the multijet background shape in the SR$i(j,k)$ and the CR$i(\text{Inclusive},k)$ are used to estimate the shape correction.}
\label{tab:boosted_categorization}
\begin{tabular}{lccccc}
\toprule
 & Category & Mass region & $i$ & $j$ & $k$ \\
\midrule
Signal Region (SR) & $\text{SR}i(j,k)$ & $\text{R}i$ & 1, 2 & Medium, Tight & $1b$, $2b$ \\
\midrule
Validation Region (VR) & $\text{VR}i(j,k)$ & $\text{R}i$ & 1,2 & Loose & $1b$, $2b$ \\
\midrule
Control Region (CR) & $\text{CR}i(j,k)$ & $\text{CR}i$ & 1, ..., 4 & Loose, Medium, & $0b$, $1b$, $2b$ \\
 & & & & Tight, Inclusive &  \\
\midrule
Template Region (TR) & $\text{TR}i(j,k)$ & $\text{R}i$ & 1, 2 & Loose, Medium, & $0b$ \\
 & & & & Tight, Inclusive &  \\
\bottomrule
\end{tabular}
\end{table}

The normalized reconstructed \mtt distributions, \mttreco, in the resolved main SR (region D) and one of the most sensitive boosted SRs ($\text{R}1(\textrm{Tight},2b)$)
are shown in Figure~\ref{fig:categorization_mttReco_buckets} for different masses of the hypothesized particle in each of the benchmark signal scenarios considered.
The acceptance times efficiency as a function of the top-quark pair invariant mass, \mtt, at the generator level for SR selections 
are shown in Figure~\ref{fig:categorization_accpetanceXefficiency}. 
Due to the spin nature of the resonance,  the two top quarks from the spin-2 graviton \kkG (spin-1 \zprime) are likely to be produced in the barrel (endcap) region. 
Hence the acceptance for the \kkG signal is higher than that of the \zprime or \kkg\ signals.

\begin{figure}[t]
\centering
\subfloat[\label{fig:categorization_mttReco_zprime}]{
        \includegraphics[width=0.49\textwidth]{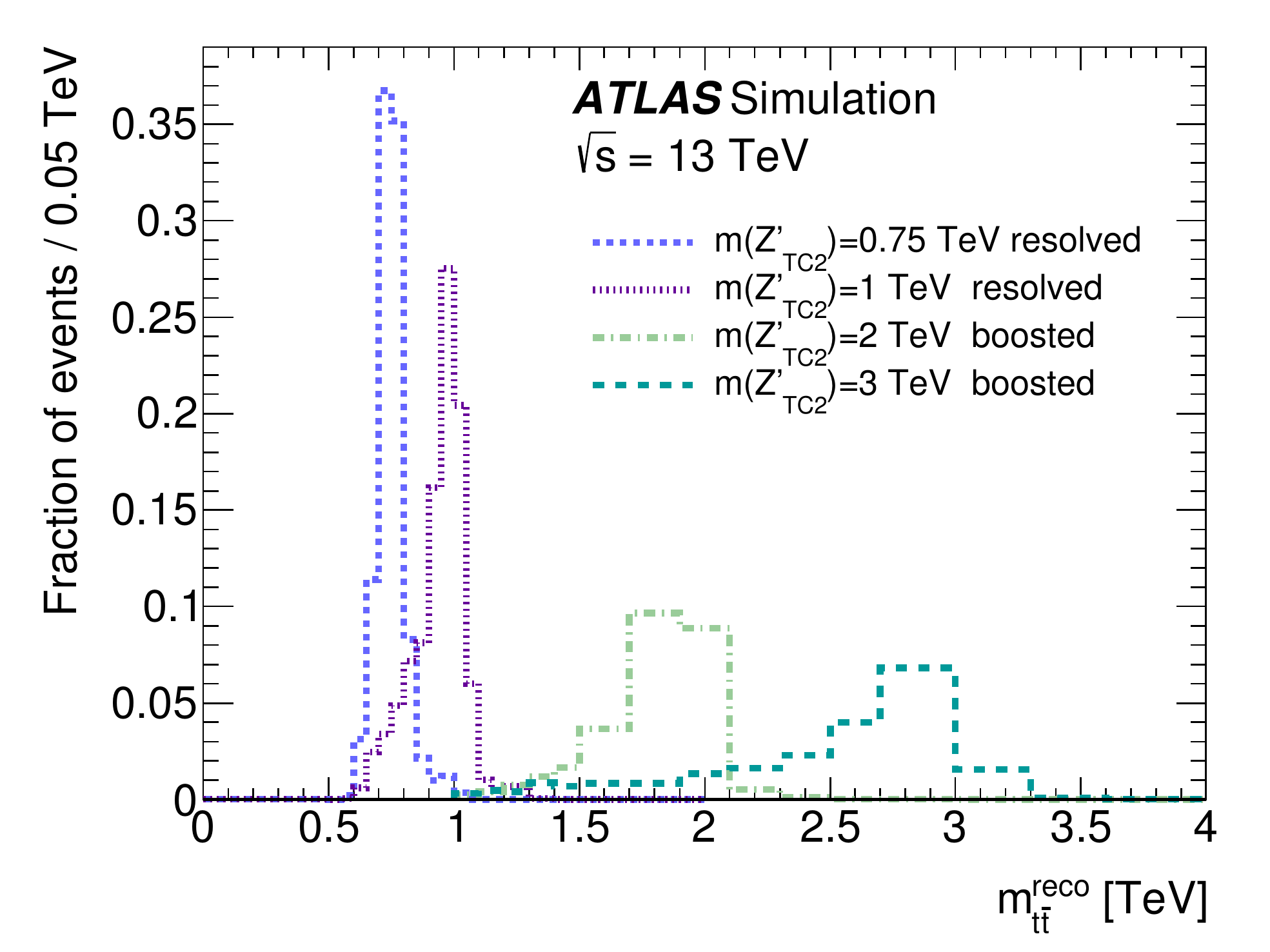}
}
\subfloat[\label{fig:categorization_mttReco_kkgraviton}]{
        \includegraphics[width=0.49\textwidth]{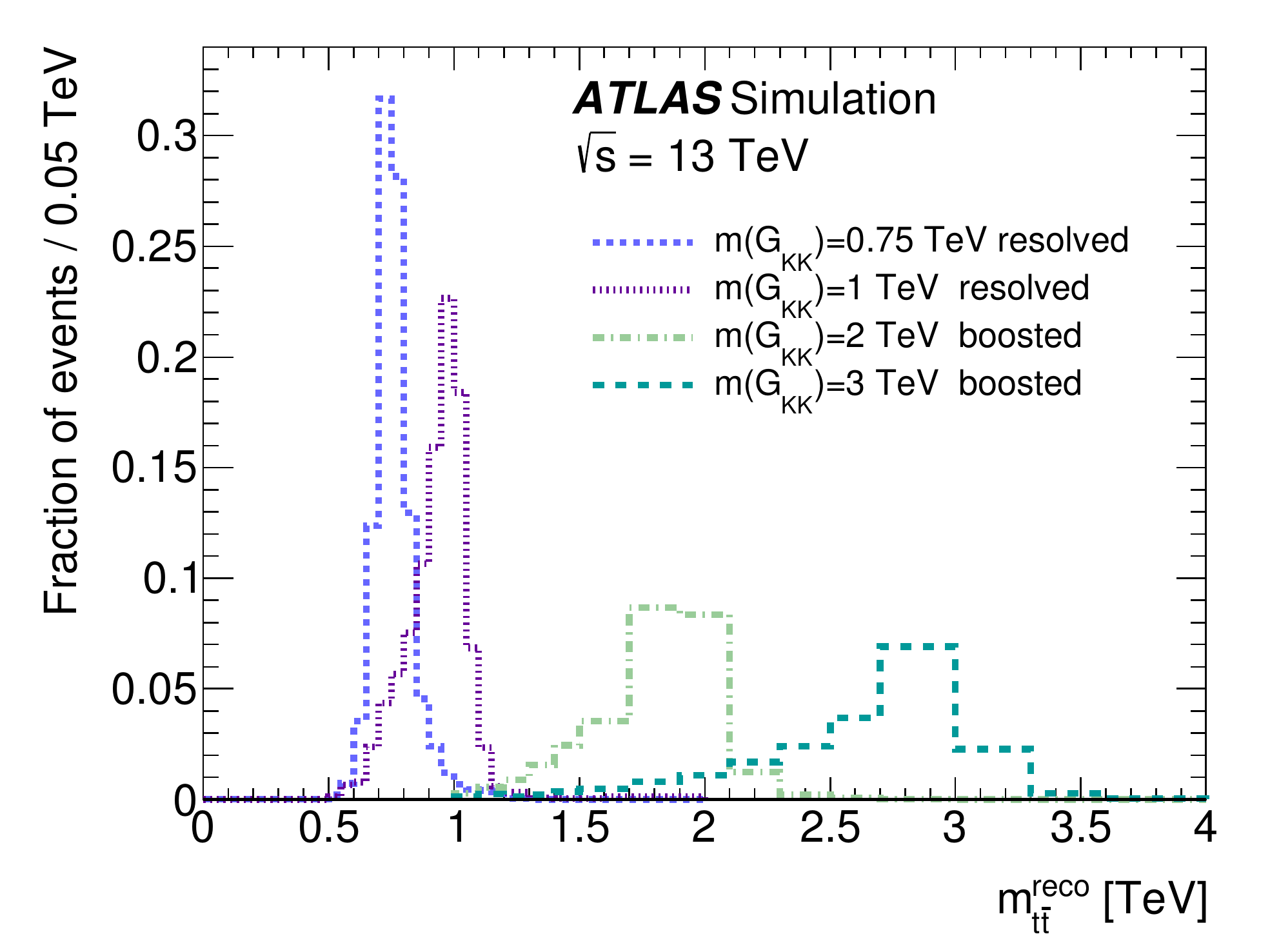}
} \\
\subfloat[\label{fig:categorization_mttReco_kkgluon}]{
        \includegraphics[width=0.49\textwidth]{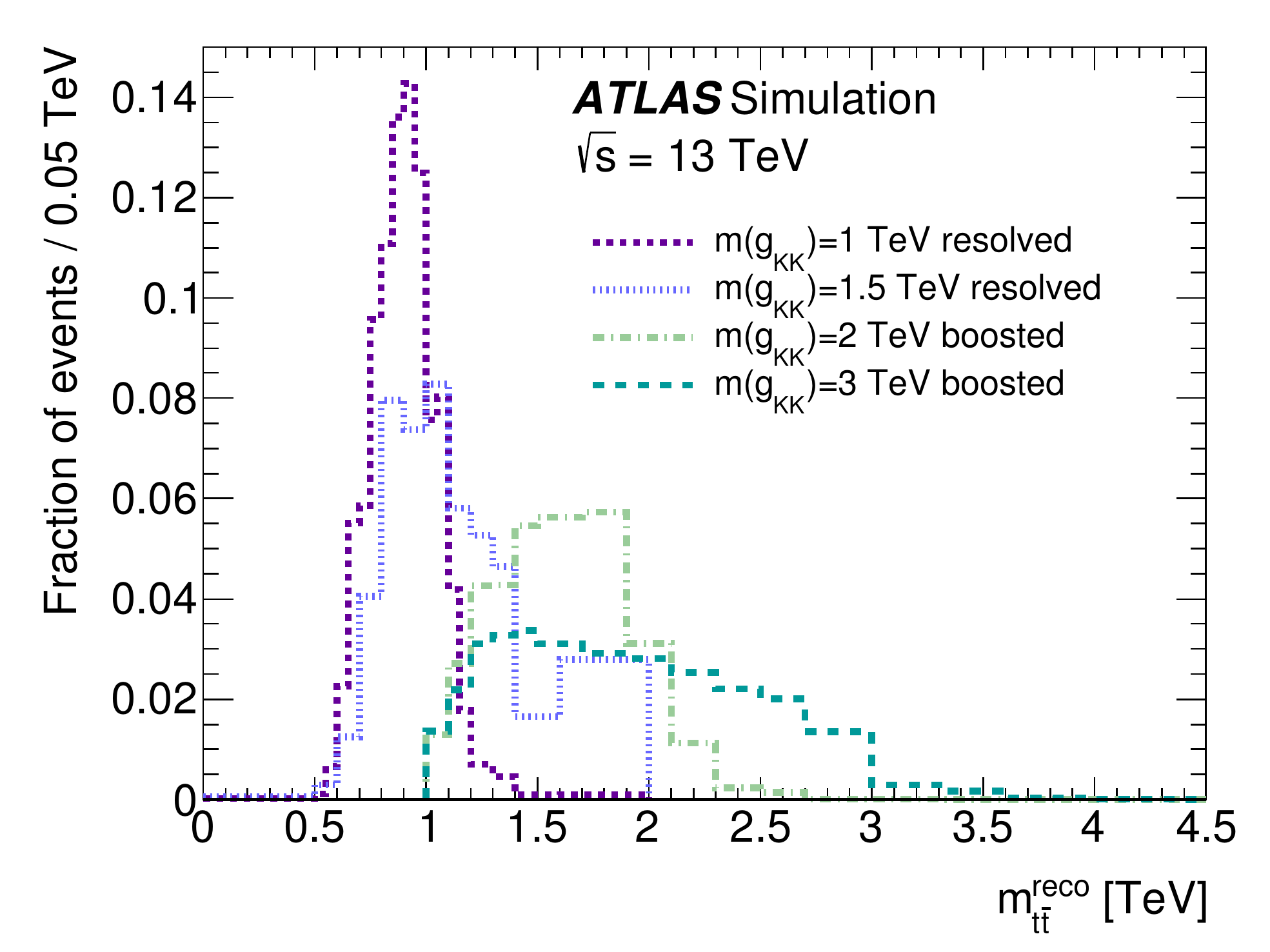}
} 
\caption{Normalized \mttreco distributions for simulated signal samples of (a) $pp \to \zprime \to \ttbar$, (b) $pp \to \kkG \to \ttbar$  and (c) $pp \to \kkg \to \ttbar$. The benchmark signals with masses of 0.75, 1 or 1.5~\TeV\ reconstructed in region D of the resolved analysis, and with masses of 2 and 3~\TeV\ reconstructed in the $\textrm{R}1(\textrm{Tight},2b)$ region of the boosted analysis are shown. The 3~\TeV\ \kkg signal has a broader \mttreco distribution without an apparent peak at the generated mass because the \kkg signal is much wider than other signals and the lower mass region is further enhanced by the parton luminosity effect.}
\label{fig:categorization_mttReco_buckets}
\end{figure}

\begin{figure}[t]
  \centering
   \subfloat[]{
      \includegraphics[width=0.49\textwidth]{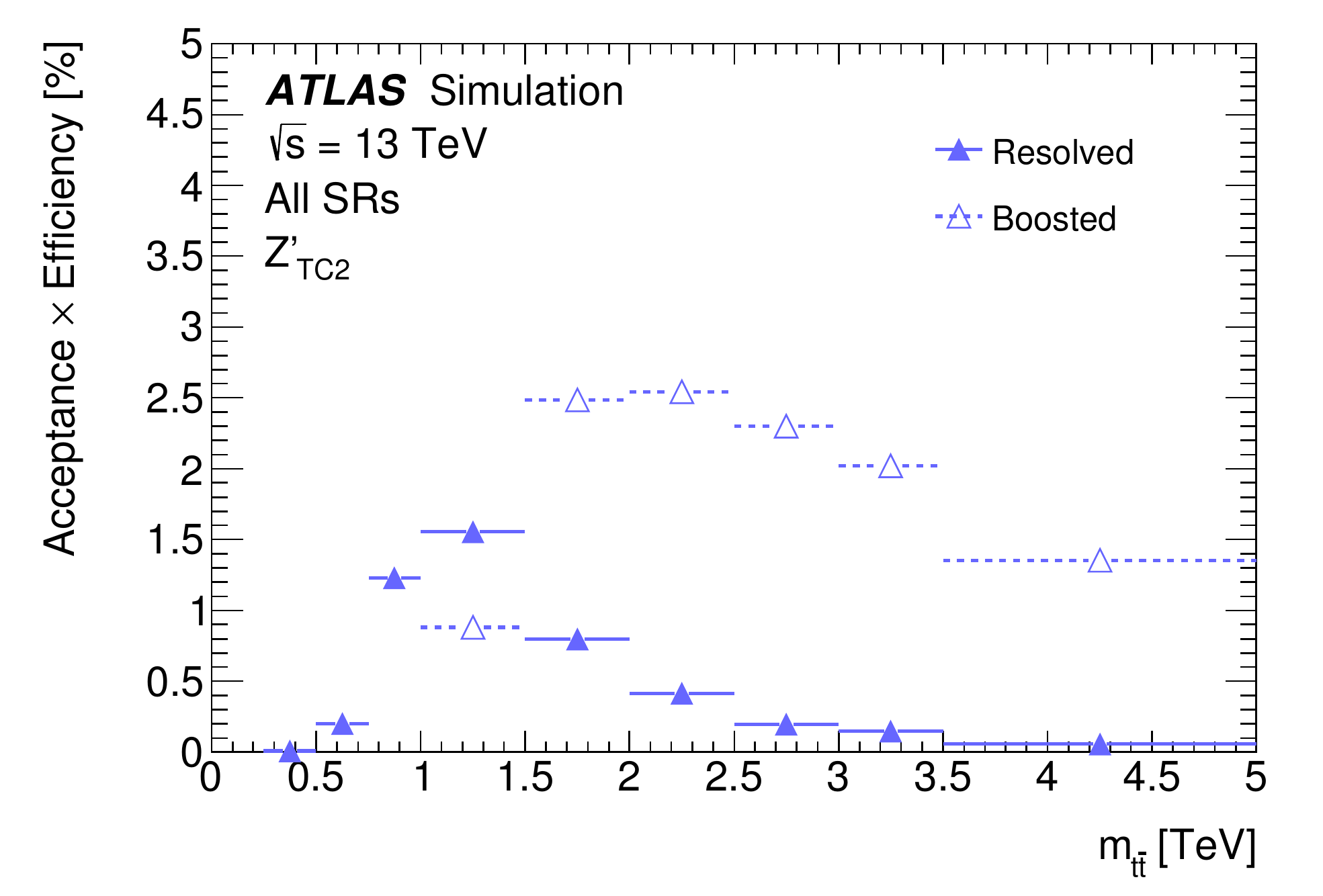}
   }    
   \subfloat[]{
      \includegraphics[width=0.49\textwidth]{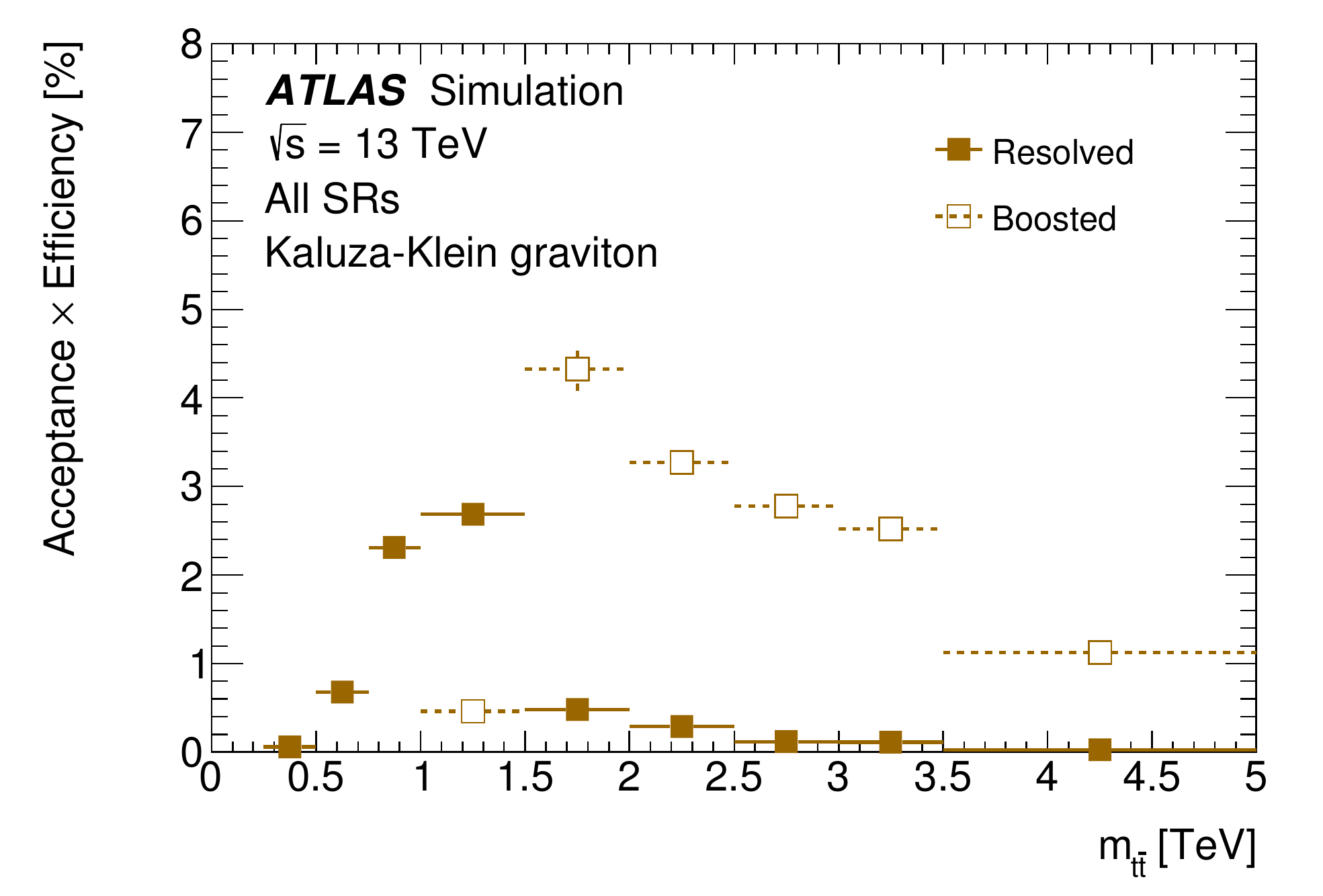}
   }\\
   \subfloat[]{
      \includegraphics[width=0.49\textwidth]{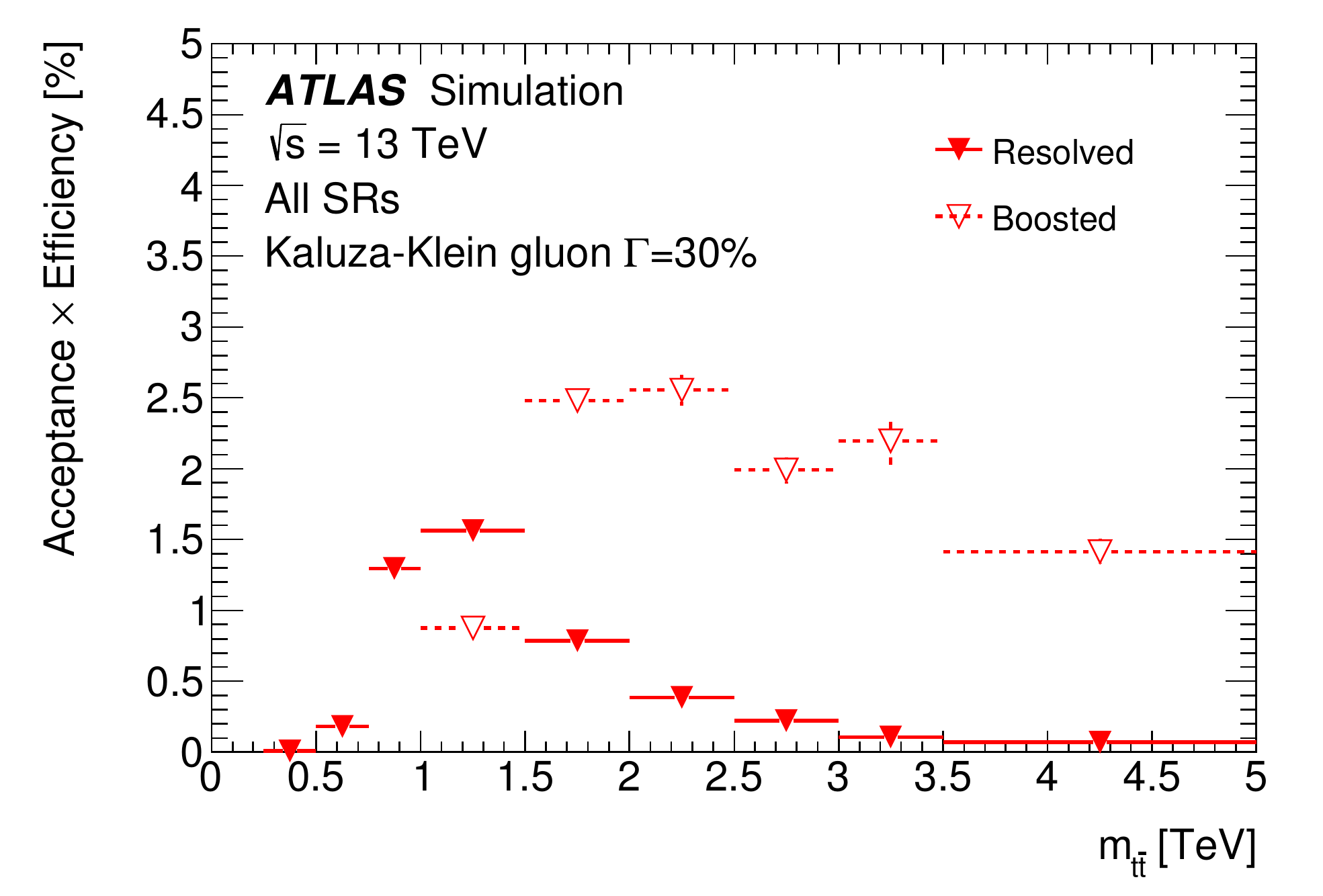}
   }    
   \caption{Acceptance times selection efficiency as a function of \mtt for all regions A--D in the resolved analysis and the combination of all SRs in the boosted analysis. The momenta of top and antitop quarks evaluated at the generator level before final state radiation are used to define \mtt. The efficiency calculation includes the branching fractions of the \ttbar system into all possible final states. (a) is \zprime, (b) is \kkG, and (c) is \kkg.}
\label{fig:categorization_accpetanceXefficiency}
\end{figure}


\section{Background estimation}
\label{sec:background}
The main SM backgrounds in both the resolved and boosted analyses
are from SM production of \ttbar pairs and multijet processes.
The \ttbar events are predicted from simulation as described in Section~\ref{sec:data_sim}.
The multijet backgrounds are estimated using multijet-enriched regions A--C. The data-driven estimation methods are validated in dedicated validation regions.
Contributions from the production of single top quarks, $W$/$Z$ bosons in association with jets, and dibosons ($WW$, $WZ$ and $ZZ$) are
negligibly small and are accounted for in the multijet background estimate.

The resolved analysis exploits a double-sideband likelihood method to estimate the multijet background contribution in each of the regions A--D, defined in Table~\ref{tab:bucketsABCDPurity}. 
The \mttreco templates extracted from the regions A and B, by subtracting the simulated SM \ttbar contribution, are used to model the multijet background shape in the region C and the main signal region D, respectively.
It is confirmed that the simulated SM \ttbar sample can model the data well by comparing the kinematic 
distributions observed in the \ttbar-enriched data and the \ttbar simulation sample.
The multijet yields in the main signal region D are first estimated by multiplying the yield in B by the ratio of the yields in C and A, assuming no contamination from signal in the regions A--C and no correlation between top- and $b$-tagging requirements.
This first estimation is used to get the input values of the unconstrained normalization parameters in the following likelihood fit.
The presence of a possible contamination from signal in the multijet-enriched regions A--C, the correlation between the top- and $b$-tagging variables and the subtraction of the SM \ttbar background in the multijet background estimate
are then taken into account by performing a likelihood fit to the data \mttreco distributions in all the regions A--D.
This simultaneous likelihood fit 
allows the multijet background from the three multijet-enriched regions A--C to be estimated and the probability of compatibility of expected
backgrounds with observed data in the main signal region D to be quantified at the same time, as described in Section~\ref{sec:stat_analysis}. 
Systematic uncertainties associated with the data-driven method discussed in Section~\ref{subsec:background_modeling_uncertainties} are considered in the fit as nuisance parameters.

For the boosted analysis, the
multijet yield in a SR is estimated by multiplying the multijet yield in the corresponding TR by the normalization factor ($F_\text{N}$) obtained by comparing the data yields in the CR between $1b$ or $2b$ and $0b$ regions.
For a SR$i(j,k)$ with the jet mass requirement $i$, \lrt requirement $j$ and $n_b$ requirement $k$ (defined in Table~\ref{tab:boosted_categorization}),
the multijet yield $N^{\text{MJ}}_{\text{SR}i(j,k)}$ is obtained by

\begin{equation*}
N^{\text{MJ}}_{\text{SR}i(j,k)} = F_\text{N}(j,k) \times N^{\text{MJ}}_{\text{TR}i(j,0b)},
\end{equation*}

where the $N^{\text{MJ}}_{\text{TR}i(j,0b)}$ is the event yield in the TR$i(j,0b)$.
The normalization factor for the SR with the selection $(j,k)$, $F_\text{N}(j,k)$, is defined as 

\begin{equation*}
F_\text{N}(j,k) = \frac{ \sum_{i} N^{\text{MJ}}_{\text{CR}i(j,k)} }{ \sum_{i} N^{\text{MJ}}_{\text{CR}i(j,0b)} },
\end{equation*}

where the $N^{\text{MJ}}_{\text{CR}i(j,k)}$ is the multijet yield in the CR$i(j,k)$. The $N^{\text{MJ}}_{\text{CR}i(j,k)}$ is obtained from data by subtracting 
the simulated SM \ttbar background. 
The normalization factors obtained separately from the four CRs (CR$i$; $i=1,...,4$) with the selection $(j,k)$ are found to be comparable 
within the statistical uncertainty; therefore they are averaged into a single $F_\text{N} (j,k)$ value for improved statistical accuracy. 
The obtained $F_\text{N}(j,k)$ is about 2.4 (1.4) with a relative uncertainty of about 2\% for $k=1b$ ($2b$), and is the same  
for both $j=$\,Medium and Tight within the statistical uncertainty.
Contributions from the SM \ttbar\ background in the TR are about 3\% and 1\% for mass regions R1 and R2, respectively.
The contamination from the SM \ttbar in the CR is less than 1\% for the $0b$ region and a few percent for 
the $1b$ and $2b$ regions, and at most 9\% in the CR$(\text{Tight},2b)$ category.

For the multijet background shape, the inclusive \lrt range $[0.35, 1.0]$ is used in the TR ($\text{TR}i(\text{Inclusive},k)$) to improve
the statistical accuracy after checking the compatibility of the \mttreco shapes in the three \lrt regions.
However, the templates are extracted separately for R1 and R2 as they have non-negligible differences.
The estimated multijet shapes are further corrected to account for the \pt-dependence of the $b$-tagging efficiency as observed in the simulation. 
This is performed by using the scalar sum of the \pt of the two leading large-$R$ jets, $\pt^{\text{sum}}$,
and comparing the $\pt^{\text{sum}}$ distributions of the CR events in the $1b$ and $2b$ regions with the ones in the $0b$ region in the simulated multijet events.
The inclusive \lrt range and the sum of the four CRs (CR1--4) are used for this study.
The shape correction is then extracted separately for the $1b$ and $2b$ regions by performing a fit to the ratio of the distributions. 
Finally, in order to reduce the statistical fluctuation of the predicted multijet contribution at high mass, the estimated \mttreco distribution 
in the SR is fit in the range from 1.2~\TeV\ to 4~\TeV\ using an exponential function and  the prediction replaced
with the fit result above 1.5~\TeV. The same procedure is applied to the simulated
SM \ttbar events to improve the statistical accuracy.
The method used to estimate the multijet background is validated in the VR$i(\text{Loose},k)$, 
where  good agreement is seen between the observed data and the prediction from the TR$i(\text{Inclusive},0b)$ for $i=1$ and 2 and $k=1b$ and $2b$.


\section{Systematic uncertainties}
\label{sec:systematics}
There are two categories of systematic uncertainties considered in the analysis: experimental uncertainties associated with the detector response and reconstruction algorithms, and uncertainties in the background modeling.

Each source of systematic uncertainty is considered to be uncorrelated with other sources, while it is treated as being fully correlated across event categories and between processes, whenever appropriate.
In addition, statistical uncertainties in the signal and background predictions due to the limited amount of simulated data are taken into account.

\subsection{Experimental uncertainties in simulated samples}
The SM \ttbar and signal predictions are subject to experimental systematic uncertainties because they are estimated using 
simulated events. Dominant sources of the experimental systematic uncertainty are associated with the small-$R$ and large-$R$ jet energy scales (JES), jet energy resolutions (JER) and $b$-tagging. 

The small-$R$ JES uncertainty is derived using a combination of simulation,
test-beam data, and in situ measurements~\cite{PERF-2016-04}.
Additional contributions from jet flavor composition, punch-through,
single-particle response, calorimeter response to different jet flavors 
and pileup are taken into account, resulting in a total of 21  systematic uncertainty components. 
The total JES uncertainty is typically 4\% at $\pt=25$~\GeV\ and varies from 1\% to 3\% at $\pt>75$~\GeV.
The small-$R$ JER uncertainty (typically 2\%--3\% at $\pt=50$~\GeV) obtained 
from an in situ measurement of jet response using 
dijet events~\cite{PERF-2016-04} is also included.
The uncertainty in the efficiency of the jet vertex tagger (Section~\ref{subsec:selection}) is also considered following Ref.~\cite{PERF-2014-03}.
The impact on the total background yield (for a 850~\GeV\ \zprime signal) in the resolved analysis 
is about 9\% (11\%) for the JES uncertainty and 3\% (11\%) for the JER uncertainty.

The large-$R$ JES uncertainties 
are 
estimated with the $R_{\mathrm{trk}}$ method using dijet data control samples~\cite{PERF-2015-04,ATLAS-CONF-2016-035}.
The method assumes that the track-related uncertainties are uncorrelated with the calorimeter cluster-related uncertainties.
The procedure works by measuring the ratio $r_{\mathrm{ trk}}$ of an observable (which can be the \pt, $m_\text{J}$ or $\tau_{32}$ variables) using 
calorimeter jets to that using track-jets reconstructed within the same detector region.
The deviation of the average data-to-simulation ratio 
$\left<R_{\mathrm{ trk}}\right> = \left<r_{\mathrm{trk}}^{\textrm{data}}\right>/\left<r_{\mathrm {trk}}^{\textrm{MC}}\right>$ 
from unity is taken as the uncertainty, together with the uncertainties associated with the track measurement, 
charged particle multiplicity modeling in simulation and the statistical uncertainty of the dijet sample.
The impact on the total background yield (for a 3~\TeV \zprime signal) in the boosted analysis 
is about 3\% (4\%) for the large-$R$ JES uncertainty and 3\% (2\%) for the large-$R$ JER uncertainty.

Correction factors to the simulated event samples are applied, separately for small-$R$ jets and track-jets,  to compensate for
differences observed between data and simulation
in the $b$-tagging efficiency of $b$-, $c$- and light-quark and gluon-induced jets~\cite{PERF-2016-05}.
The correction factor for $b$-jets is derived from \ttbar events with
final states containing two leptons, and is consistent
with unity within uncertainties at the level of a few percent over most of the jet \pt range.
Uncertainties in the correction factors for the $b$-tagging identification efficiency result 
in a variation of the total background yield of about 5\% (4\%) for the resolved (boosted) analysis.
Uncertainties due to possible correlations between the correction factors in the signal and control regions are checked to have a negligible impact on the final results.
An additional term is included to extrapolate the measured uncertainties to the high-\pt region of interest. This term 
is calculated from simulated events by considering variations of the quantities affecting the $b$-tagging performance such as the impact parameter resolution, percentage of poorly measured tracks, description of the detector material, and track multiplicity per jet. 
The impact on the 3~\TeV\ \zprime signal yield due to such high-\pt extrapolation uncertainty is about 3\%.

In addition, smaller uncertainties associated with the luminosity measurement and the trigger efficiency are  considered.
The uncertainties associated with electron and muon reconstruction and identification are found to be negligible.

The uncertainty in the combined 2015+2016 integrated luminosity is 2.1\%. It is derived, following a methodology similar to
that detailed in Ref.~\cite{DAPR-2013-01}, and using the LUCID-2 detector for the baseline luminosity measurements \cite{LUCID2},
from calibration of the luminosity scale using $x$--$y$ beam-separation scans.
The pileup modeling uncertainty is considered by varying the average number of 
$pp$ collisions in simulated events.

In the resolved analysis the trigger efficiency is corrected around the jet \pt threshold at the trigger level.
The uncertainty in the correction factor, estimated to be below 1\%, is dominated by the statistical uncertainty of the lower-threshold trigger data.
In the boosted analysis the uncertainty in the trigger efficiency is found to be negligible.

\subsection{Background modeling uncertainties}
\label{subsec:background_modeling_uncertainties}
In this section, uncertainties associated with the data-driven estimates of multijet background and theory uncertainties in the SM \ttbar prediction 
are discussed.

As discussed in Section~\ref{sec:background}, in both the resolved and boosted analyses the multijet background in the SRs is estimated
by extrapolating the \mttreco shape obtained from the regions where the $b$-tagging criterion is loosened compared with that in the SRs.
Uncertainties in the \mttreco shape and the yield of the multijet background are estimated separately as follows.

The different $b$-tagging criteria between the signal and control regions could produce a bias in the predicted \mttreco distributions.
In the resolved analysis this effect is estimated by comparing the \mttreco distributions in the validation regions \cAz and \cCz
(see Table~\ref{tab:bucketsABCDPurity})
and the difference observed is assigned as a systematic uncertainty in the multijet background shape. 
The assumption that the potential bias is caused by the 
$b$-tagging instead of top-quark tagging is verified by repeating the same procedure using the validation regions \cAf and \cCf,
which gives a result comparable to the one from the validation regions \cAz and \cCz.
For the boosted analysis, 
the variations of the correction factor applied to the $\pt^{\text{sum}}$ distribution (see Section~\ref{sec:background}) are 
considered as an uncertainty in the multijet background shape. These include the statistical uncertainty of the multijet simulation samples and a small residual difference 
observed in the \mttreco distributions after the shape correction. 
A possible bias arising from using the inclusive \lrt range $\left[0.35, 1.0\right]$ for the multijet template extraction from $\text{TR}i(\text{Inclusive},0b)$
is also taken into account as a source of systematic uncertainty.
The multijet \mttreco distribution obtained from $\text{TR}i(\text{Inclusive},0b)$ is compared with those obtained from the individual \lrt regions ($\text{TR}i(j,0b); j=$\,Medium and Tight) 
and the maximum difference in shape is considered.

The impact on the multijet yield due to correlation between the top- and $b$-quark tagging variables in the resolved analysis is 
evaluated by using the (\tz, \tz) or (\tm, \tm) categories instead of the (\tw, \tm) category. 
As a result, an uncertainty of 20\% is added to the normalization of the multijet background, resulting in a 3\% uncertainty in the total background yield. 
In the boosted analysis, the uncertainty in the multijet background normalization is estimated by 
taking the maximum deviation of the expected yields in the four CRs from the average. 
This leads to a 3\% uncertainty in the overall background yield.

There are several sources of theoretical uncertainties affecting the modeling of SM \ttbar background processes in all regions including signal, control and validation regions.
The cross-section uncertainty given in Section~\ref{sec:data_sim} accounts for the choice of PDF and strong coupling constant
calculated using the PDF4LHC prescription~\cite{Botje:2011sn}
with the MSTW2008 68\% CL NNLO~\cite{Martin:2009iq,Martin:2009bu}, CT10 NNLO~\cite{Lai:2010vv, Gao:2013xoa}
and NNPDF2.3 5f FFN~\cite{Ball:2012cx} PDF sets, as well as the 
 renormalization and factorization scale uncertainties.
In addition to this pure normalization uncertainty, the following modeling uncertainties
affecting both the acceptance and shape of the \ttbar kinematic distributions are considered.
The impact from the modeling of extra QCD radiation is evaluated
using  \POWPYTHIA\ samples in which the renormalization and
factorization scales and the $h_{\text{damp}}$ parameter are varied within the ranges
consistent with the measurements of \ttbar production in association with 
jets~\cite{TOPQ-2011-21,TOPQ-2012-03,,ATL-PHYS-PUB-2015-002}.
Additionally, the uncertainty in the \ttbar event kinematics due to higher-order QCD effects 
is considered by adding an uncertainty covering the difference between NLO and NNLO QCD calculations
of \ttbar production.
The recent QCD calculations in Ref.~\cite{Czakon:2016dgf} are used to derive the difference, which is applied 
as a function of top-quark \pt and the transverse momentum of the \ttbar system at the particle level taking into account the final-state radiation, to estimate this uncertainty.
The variation of the event yield at the reconstruction level is less than 4\% at \mttreco below 500~\GeV,
but approaches 11\% at \mttreco of 1.2~\TeV\ in the resolved analysis and 20\% above 3~\TeV\ in the boosted analysis.
The electroweak corrections to top-quark kinematics in \ttbar events have an associated uncertainty of about 10\%,
which varies as a function of \mttreco~\cite{Kuhn:2013zoa}. 
The uncertainty associated with the choice of event generator is evaluated by taking the difference between 
the predictions from 
the \ttbar samples generated with \POWHEGBOX\ and \AMCatNLO\ both 
interfaced to {\textsc{Herwig++}}~2.7.1~\cite{Bahr:2008pv}.
The uncertainty in the parton shower modeling is evaluated by comparing the \ttbar events simulated 
with the default  \POWPYTHIA\ with those with the same version of \POWHEGBOX\ but interfaced to \HERWIGV{7}~\cite{Bahr:2008pv,Bellm:2015jjp}.
The uncertainty arising from the choice of PDF set is estimated by taking into account the variations from 
the PDF4LHC15 PDF set, which includes 30 separate uncertainty eigenvectors~\cite{Butterworth:2015oua}, and the difference
between the nominal PDF4LHC15 and CT10 PDF sets.
For the boosted analysis, an additional uncertainty is considered in the \mttreco shape due to
the extrapolation procedure using an exponential function at high \mttreco above 1.5~\TeV\ (Section~\ref{sec:background}). 
This includes the statistical uncertainty in the exponential fit and the stability of the fit results estimated by varying the fit range.
The overall impact on the SM \ttbar event yields from these uncertainties is estimated to be 
29\% in the resolved analysis and 24\% in the boosted analysis.


\section{Statistical analysis}
\label{sec:stat_analysis}
A binned maximum-likelihood fit to the \mttreco distributions is performed to estimate the signal and background yields, 
separately in the resolved and boosted analyses.
The likelihood is defined as a product of the Poisson probabilities to observe $n_{i}$ events when $\lambda_{i}$ events are 
expected in bin $i$. The $\lambda_{i}$ is expressed as $\lambda_{i} = \mu s_{i} ({\bm \theta }) + b_{i}({\bm \theta})$ 
where $\mu$ is the signal strength, defined as a signal cross section in units of the theoretical prediction, to be determined by the fit, and $s_{i}( {\bm \theta })$ and $b_{i}( {\bm \theta })$ are the 
expected numbers of signal and background events, respectively. 
The fit includes two background components; \ttbar and multijet processes, which are estimated by the simulated samples and the data-driven 
methods, respectively, as described in Section~\ref{sec:background}.
The systematic uncertainties are taken into account as nuisance parameters, ${\bm \theta}$, constrained by Gaussian or log-normal penalty terms 
in the likelihood.
Nuisance parameters are also determined by the fit, varying the normalization and shape of the \mttreco distribution for each component of 
the signal and background.

In the resolved analysis, the likelihood fit is performed simultaneously in the three multijet-enriched regions A--C and the main signal region D.
In each region, the \mttreco distribution is divided into 19 bins spanning the range 0 to 2~\TeV. 
The shape of the multijet background is determined by bin-by-bin unconstrained normalization factors. 
Assuming that the \mttreco shape does not depend on the $b$-tagging requirement, the bin-by-bin multijet normalization factors for regions A and C
as well as for regions B and D are treated 
as fully correlated.
In order to consider the normalization component not depending on the top-tagging requirement but  depending on the $b$-tagging requirement, a common free-floating normalization factor is additionally applied to regions C and D. Thus, the correlation between the (\tw, \tm) and (\tw, \tw) categories is introduced in the background parameterization.

The SRs in the boosted analysis cover the \mttreco range between 1 and 6~\TeV, which is divided into 19 bins. 
The fit is performed simultaneously in the eight SRs defined in Section~\ref{subsec:categorization}.
The \mttreco shape and normalization of the multijet background are constrained by the variations due to systematic uncertainties estimated in Section~\ref{sec:systematics} by using them as nuisance parameters in the fit.

A test statistic based on the profile likelihood ratio~\cite{asymptotics} is used to
extract information about $\mu$ from a likelihood fit to data under the signal-plus-background hypothesis,
separately for each model considered.
The distributions of the test statistic under the signal-plus-background and the background-only hypotheses are obtained from pseudo experiments.
The probability that the observed data is compatible with the SM prediction is estimated by computing the local $p_0$-value, 
defined as the probability to observe an excess at least as large as the 
one observed in data, under the background-only hypothesis.
The global $p_0$-value is computed by considering the look-elsewhere effect~\cite{Lyons:1900zz, Gross:2010qma} associated with the multiple testing to scan the signal mass points.
If no significant excess is observed over the background, expected and observed upper 
limits on the signal strength are set at 95\% confidence level (CL) using the CL$_\text{s}$ prescription~\cite{CLs_2002}.
The results of the resolved and boosted analyses are compared in the \mtt region covered by both analyses and the one providing
the better expected limit is selected.
The upper limits on $\mu$ are converted into limits on the cross-section times branching fraction of new particles decaying into \ttbar.


\section{Results}
\label{sec:result}
\newcommand*{\zprimeOnepLimitLowmassExp}{0.57}
\newcommand*{\zprimeOnepLimitHighmassExp}{2.8}
\newcommand*{\zprimeOnepLimitLowmassObs}{0.58}
\newcommand*{\zprimeOnepLimitHighmassObs}{3.1}
\newcommand*{\zprimeThreepLimitLowmassExp}{0.51}
\newcommand*{\zprimeThreepLimitHighmassExp}{3.6}
\newcommand*{\zprimeThreepLimitLowmassObs}{0.53}
\newcommand*{\zprimeThreepLimitHighmassObs}{3.6}
\newcommand*{\zprimeVLimitLowmassExp}{2.0}
\newcommand*{\zprimeVLimitHighmassExp}{2.1}
\newcommand*{\zprimeVLimitLowmassObs}{2.0}
\newcommand*{\zprimeVLimitHighmassObs}{2.2}
\newcommand*{\zprimeVLimitLowmassExpResolved}{0.75}
\newcommand*{\zprimeVLimitHighmassExpResolved}{1.07}
\newcommand*{\zprimeVLimitLowmassObsResolved}{0.74}
\newcommand*{\zprimeVLimitHighmassObsResolved}{0.97}
\newcommand*{\zprimeAVLimitLowmassExp}{1.99}
\newcommand*{\zprimeAVLimitHighmassExp}{2.04}
\newcommand*{\zprimeAVLimitLowmassObs}{2.0}
\newcommand*{\zprimeAVLimitHighmassObs}{2.2}
\newcommand*{\zprimeAVLimitLowmassExpResolved}{--}
\newcommand*{\zprimeAVLimitHighmassExpResolved}{--}
\newcommand*{\zprimeAVLimitLowmassObsResolved}{0.80}
\newcommand*{\zprimeAVLimitHighmassObsResolved}{0.92}
\newcommand*{\kkGLimitLowmassExp}{xxxx}
\newcommand*{\kkGLimitHighmassExp}{xxxx}
\newcommand*{\kkGLimitLowmassObs}{xxxx}
\newcommand*{\kkGLimitHighmassObs}{xxxx}
\newcommand*{\kkgThirtypLimitExp}{3.3}
\newcommand*{\kkgThirtypLimitObs}{3.4}
\newcommand*{\kkgTenpLimitExp}{3.5}
\newcommand*{\kkgTenpLimitObs}{3.4}
\newcommand*{\kkgTwentypLimitExp}{3.4}
\newcommand*{\kkgTwentypLimitObs}{3.4}
\newcommand*{\kkgFortypLimitExp}{3.2}
\newcommand*{\kkgFortypLimitObs}{3.4}

The observed \mttreco distributions in the regions A--D for the resolved analysis and 
in the signal regions SR1 and SR2 for the boosted analysis after the fit (Post-Fit) with the background-only hypothesis 
are shown in Figures~\ref{fig:tbMttbarPostfitUnblinded}, \ref{fig:boosted_signalR1} and \ref{fig:boosted_signalR2}, respectively.
The expected signal and background yields as well as the observed number of data events are summarized in 
Tables~\ref{tab:tbYields} and \ref{tab:boostedYields} for the resolved and boosted analyses, respectively.
The systematic uncertainties with the largest post-fit impact on the signal strength parameter $\mu$ in the resolved and boosted analyses are presented in Table~\ref{tab:postFitImpactOnMu}.
The observed data agree well with the estimated SM background and no significant excess is observed.
Assuming a narrow-width resonance modeled by the \zprime signal, the minimum local $p_0$-value is observed 
in the boosted analysis to be 0.02 ($2.1\sigma$) at $m=1.75$~\TeV. 
The observed excess corresponds to a global significance of less than $1\sigma$.
While the excess is mostly driven by $\text{SR}1(\text{Tight},2b)$ region, it is worth noting that the other regions  contribute significantly to the overall sensitivity, e.g. adding the $\text{SR}2$ regions can improve the sensitivity by up to 20\% (for a 3~\TeV\ signal) and adding the $1b$ regions to the $2b$ ones adds about 10\% more sensitivity.
The data and expected background spectra are also compared using 
BumpHunter~\cite{Choudalakis:2011qn}, which performs a hypothesis test to look for local excesses or deficits in data 
relative to the background, taking the look-elsewhere effect into account as well. No 
significant deviation from the background is found.

\begin{figure}[p]
\centering
\subfloat[]{\includegraphics[width=.49\textwidth]{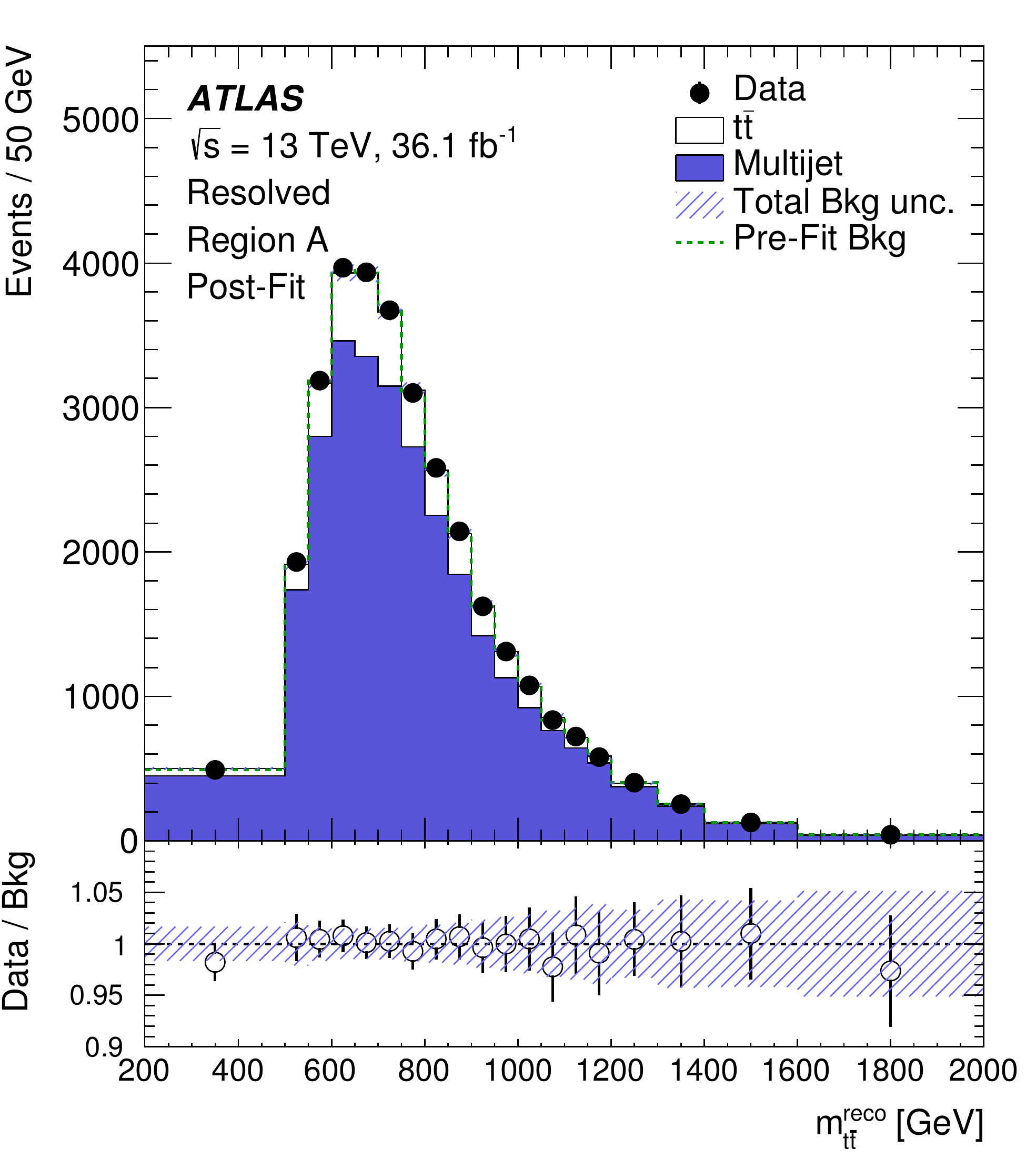}}
\subfloat[]{\includegraphics[width=.49\textwidth]{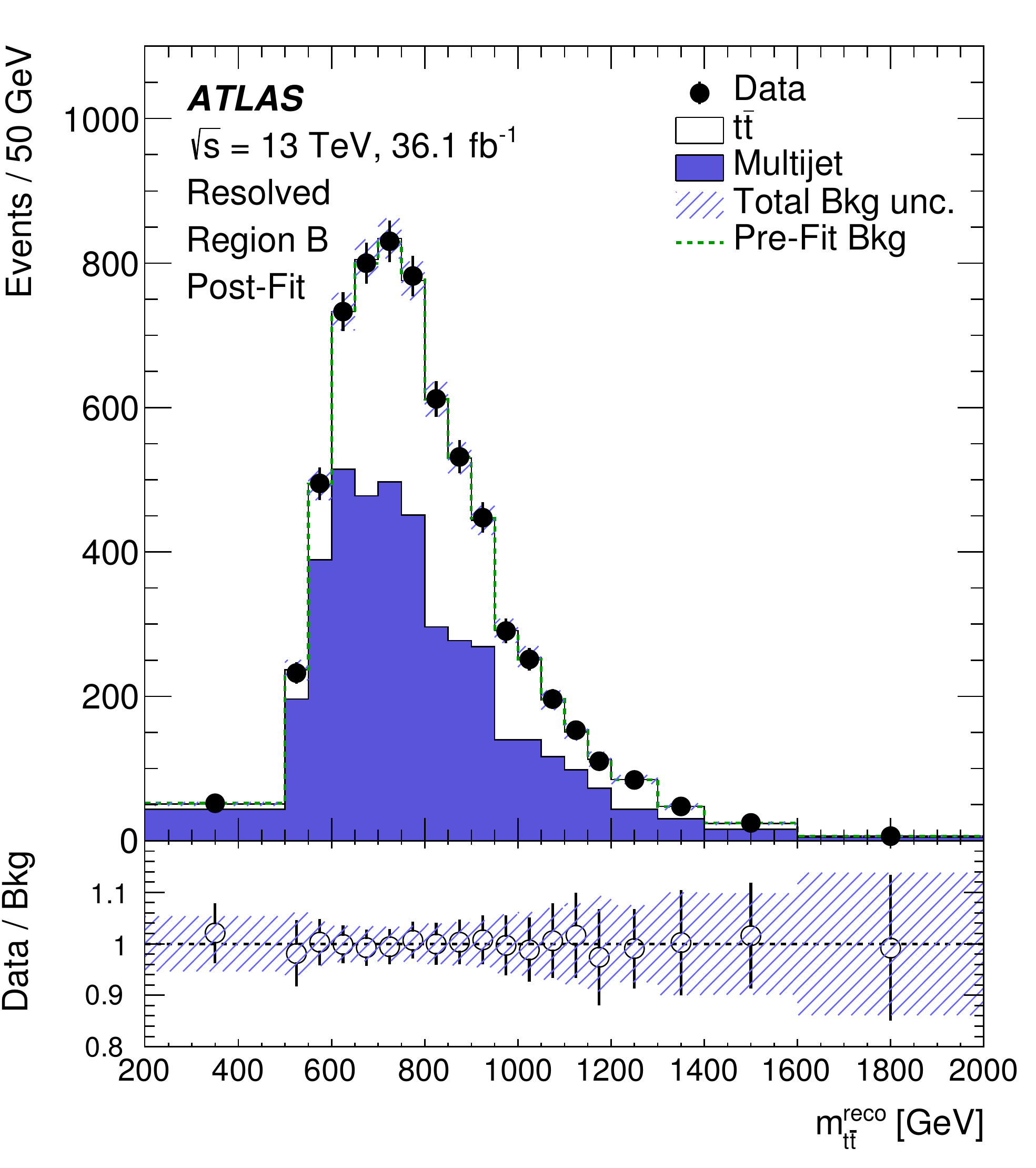}}\\
\subfloat[]{\includegraphics[width=.49\textwidth]{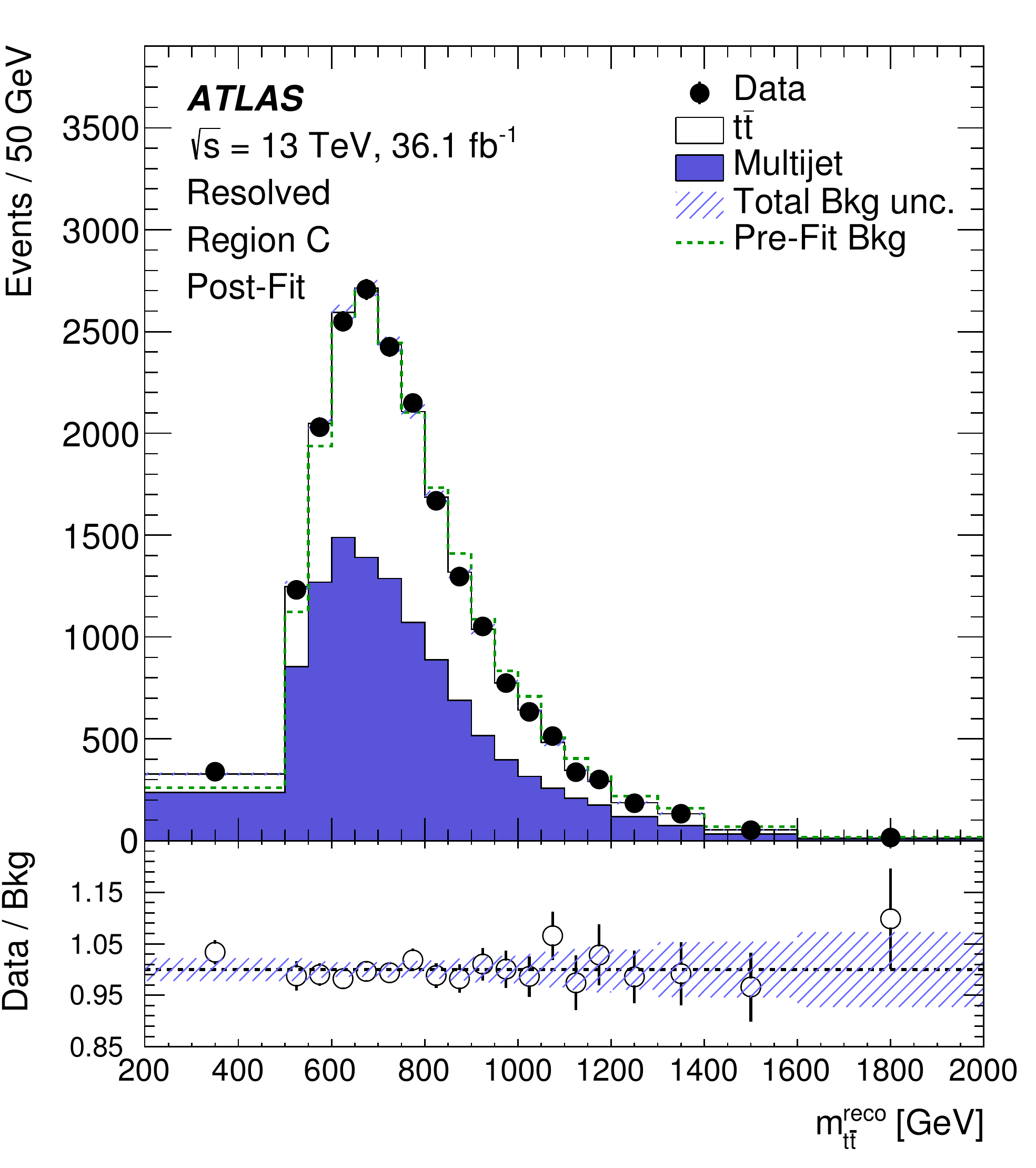}}
\subfloat[]{\includegraphics[width=.49\textwidth]{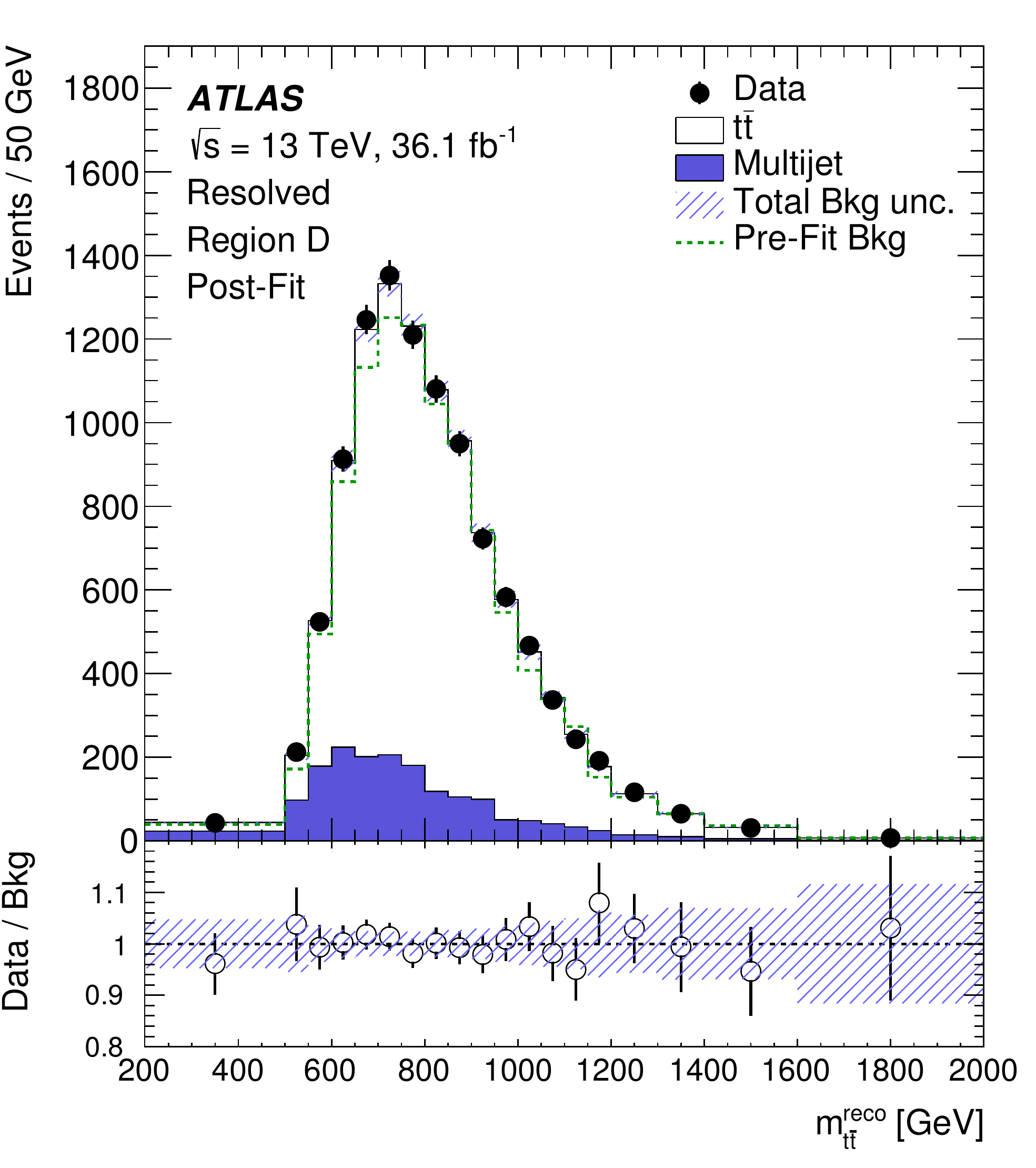}}
\caption{Observed \mttreco distributions in the multijet-enriched regions (a) A, (b) B, (c) C and  (d) the main signal region D after the fit (``Post-Fit'') under the background-only hypothesis for the resolved analysis. The shaded areas around the histograms indicate the total  uncertainties in the background. 
The lower panel of the distribution shows the ratio of data to the fitted background prediction.
The distributions before the fit are shown by the dashed lines and the background components are shown as stacked histograms. The multijet contribution also contains all other small non-\ttbar backgrounds.} 
\label{fig:tbMttbarPostfitUnblinded}
\end{figure}

\begin{figure}[p]
\centering
\subfloat[\label{fig:boosted_medium_1bR1}]{
        \includegraphics[width=0.49\textwidth]{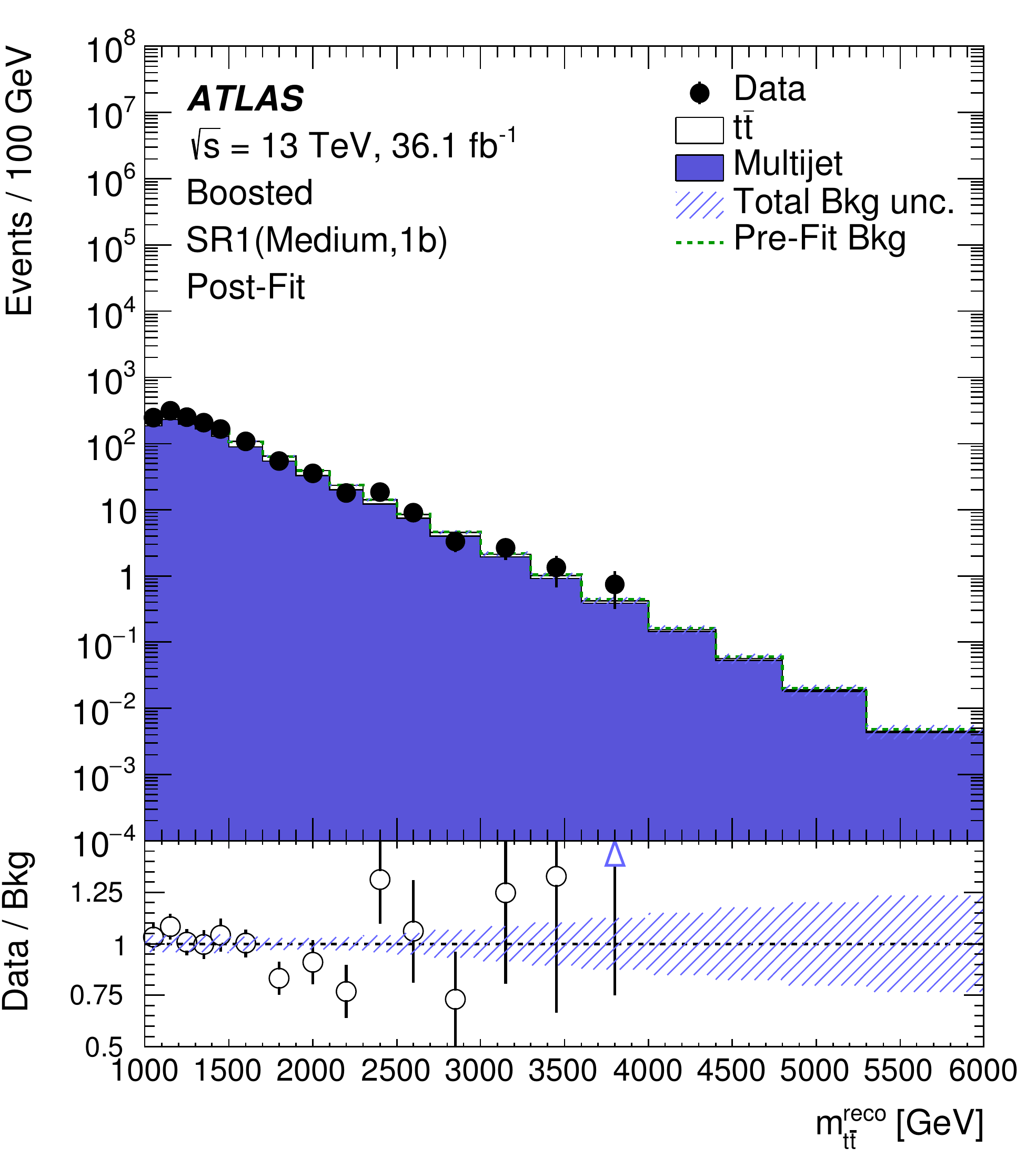}
}
\subfloat[\label{fig:boosted_medium_2bR1}]{
        \includegraphics[width=0.49\textwidth]{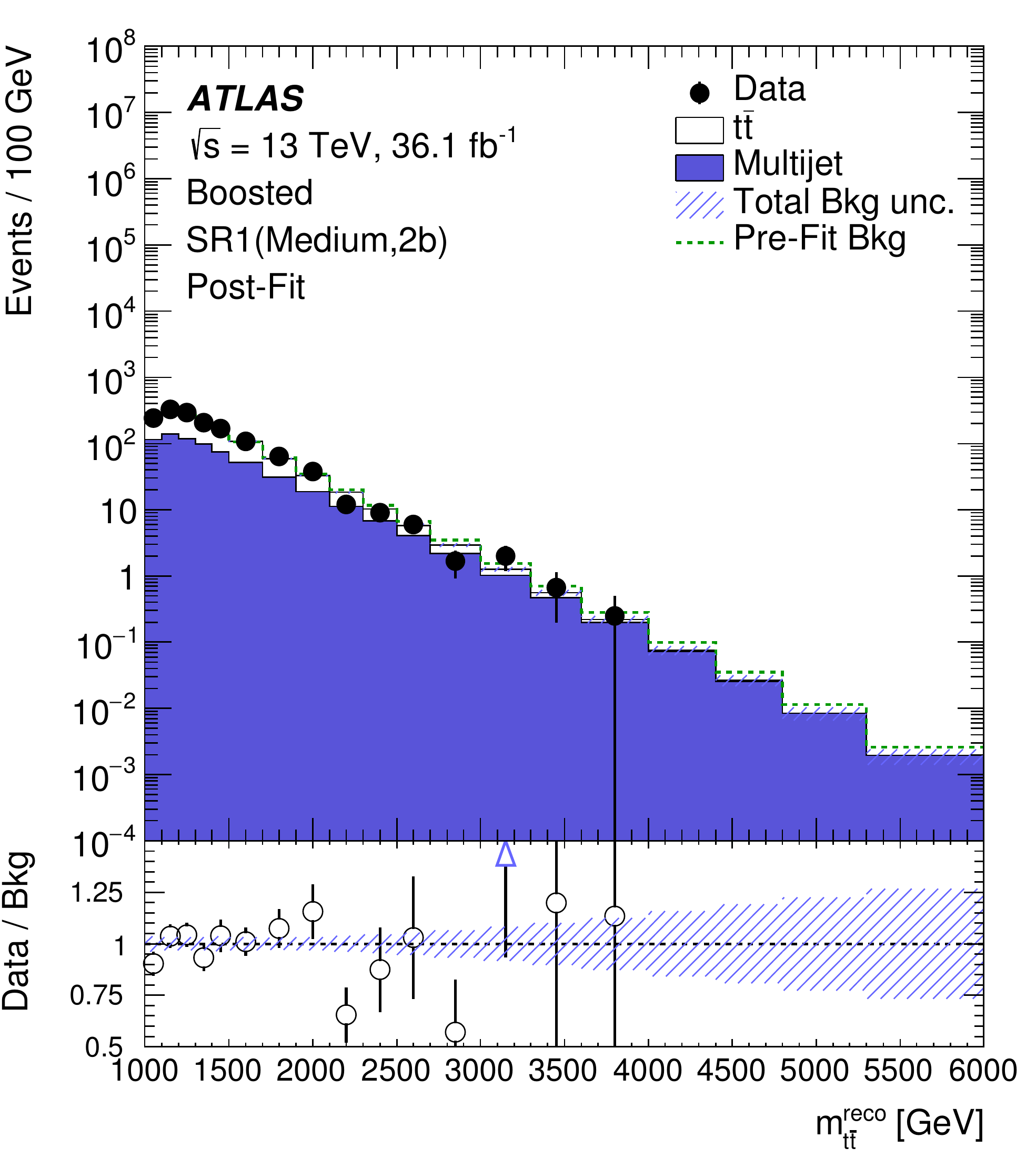}
}\\
\subfloat[\label{fig:boosted_tight_1bR1}]{
        \includegraphics[width=0.49\textwidth]{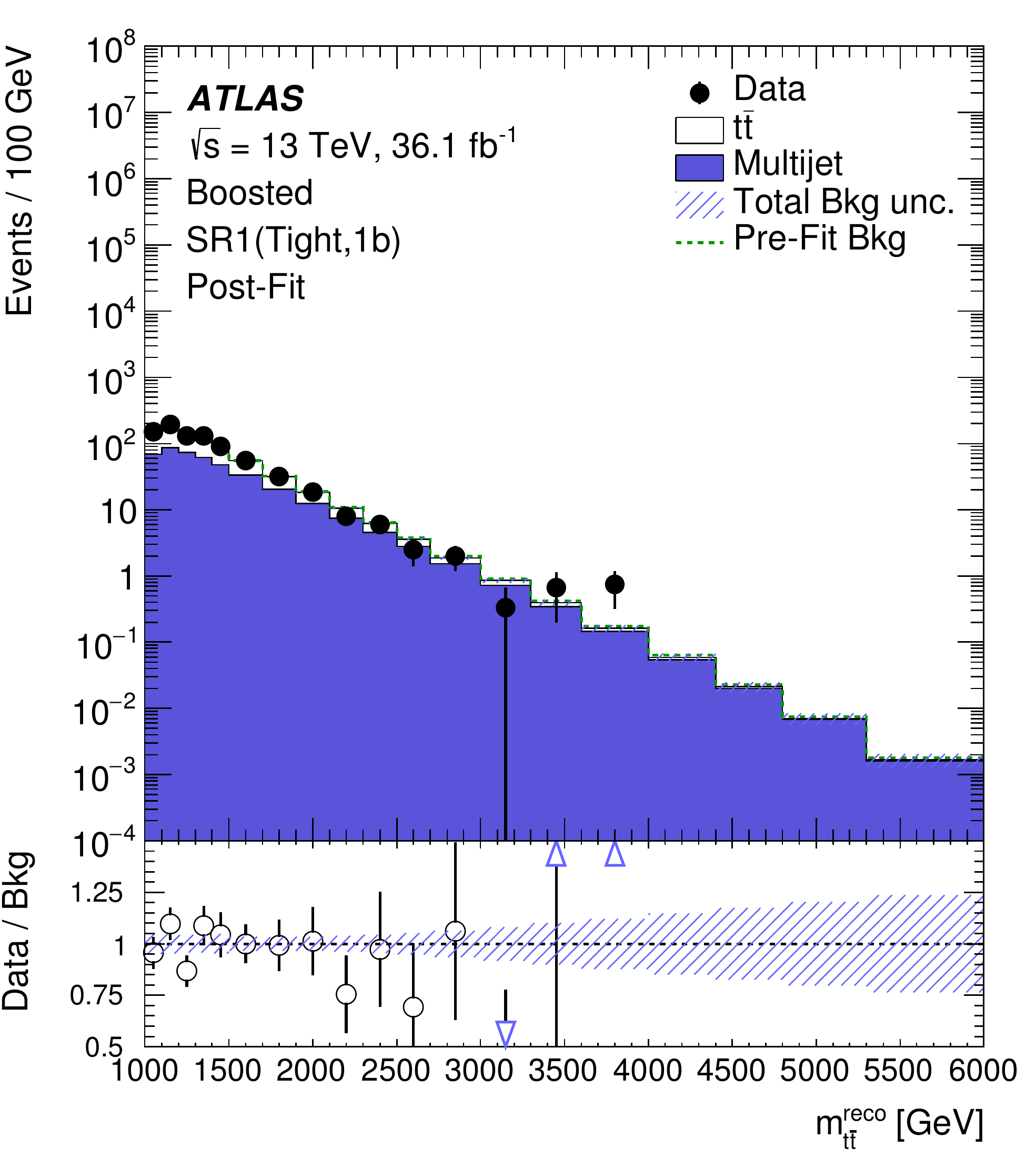}
}
\subfloat[\label{fig:boosted_tight_2bR1}]{
        \includegraphics[width=0.49\textwidth]{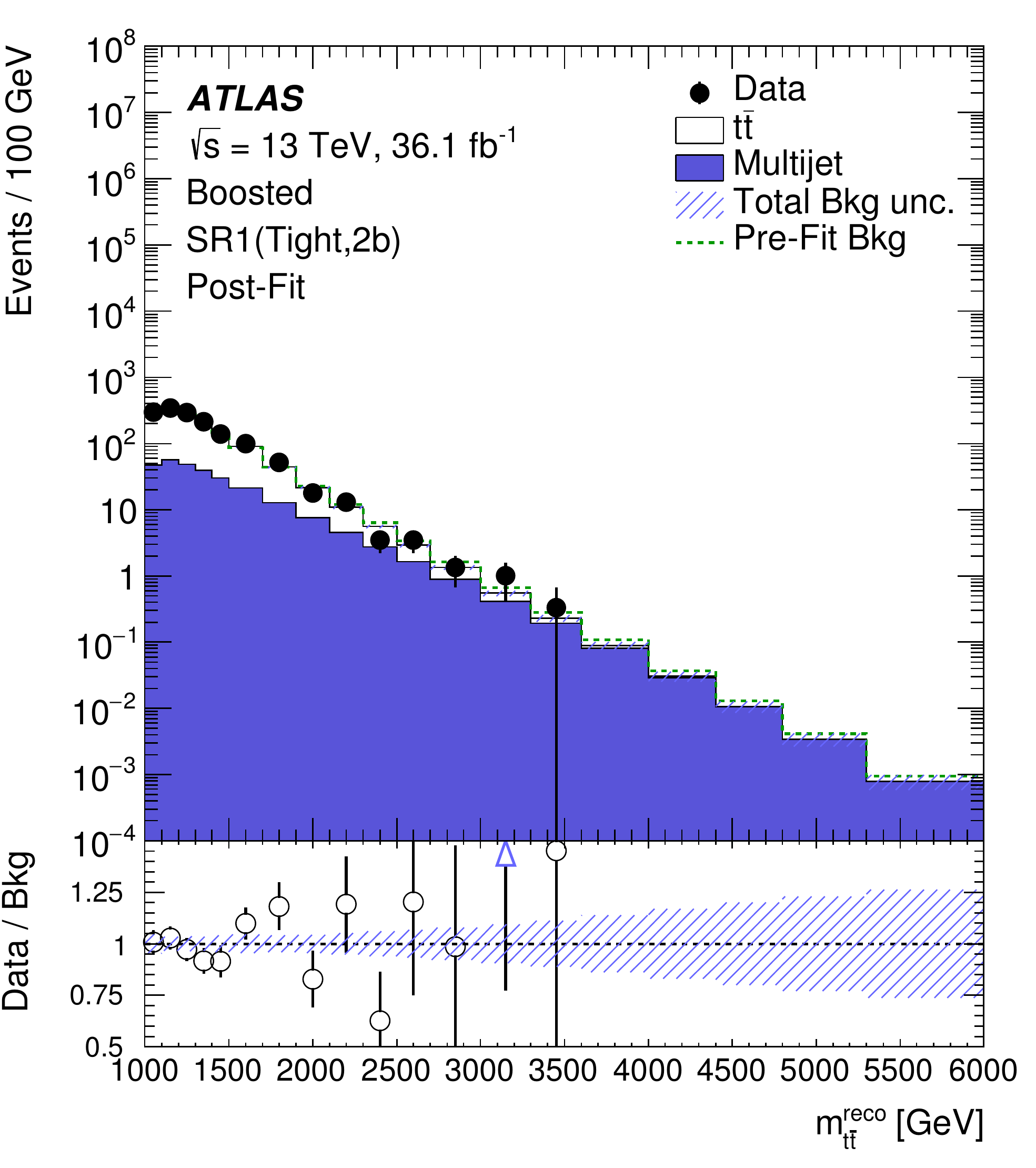}
}
\caption{Observed \mttreco distributions in (a) Medium R1 1b (b) Medium R1 2b (c) Tight R1 1b and (d) Tight R1 2b after the fit (``Post-Fit'') under the background-only hypothesis for the boosted analysis. 
The shaded areas around the histograms indicate the total uncertainties in the background. The lower panel of the distribution shows the ratio of data to the fitted background prediction. The open triangles indicate that the ratio values are outside the plotted range.
The distributions before the fit are shown by the dashed lines and the background components are shown as stacked histograms.
The multijet contribution also contains all other small non-\ttbar backgrounds.
}
\label{fig:boosted_signalR1}
\end{figure}

\begin{figure}[p]
\centering
\subfloat[\label{fig:boosted_medium_1bR2}]{
        \includegraphics[width=0.49\textwidth]{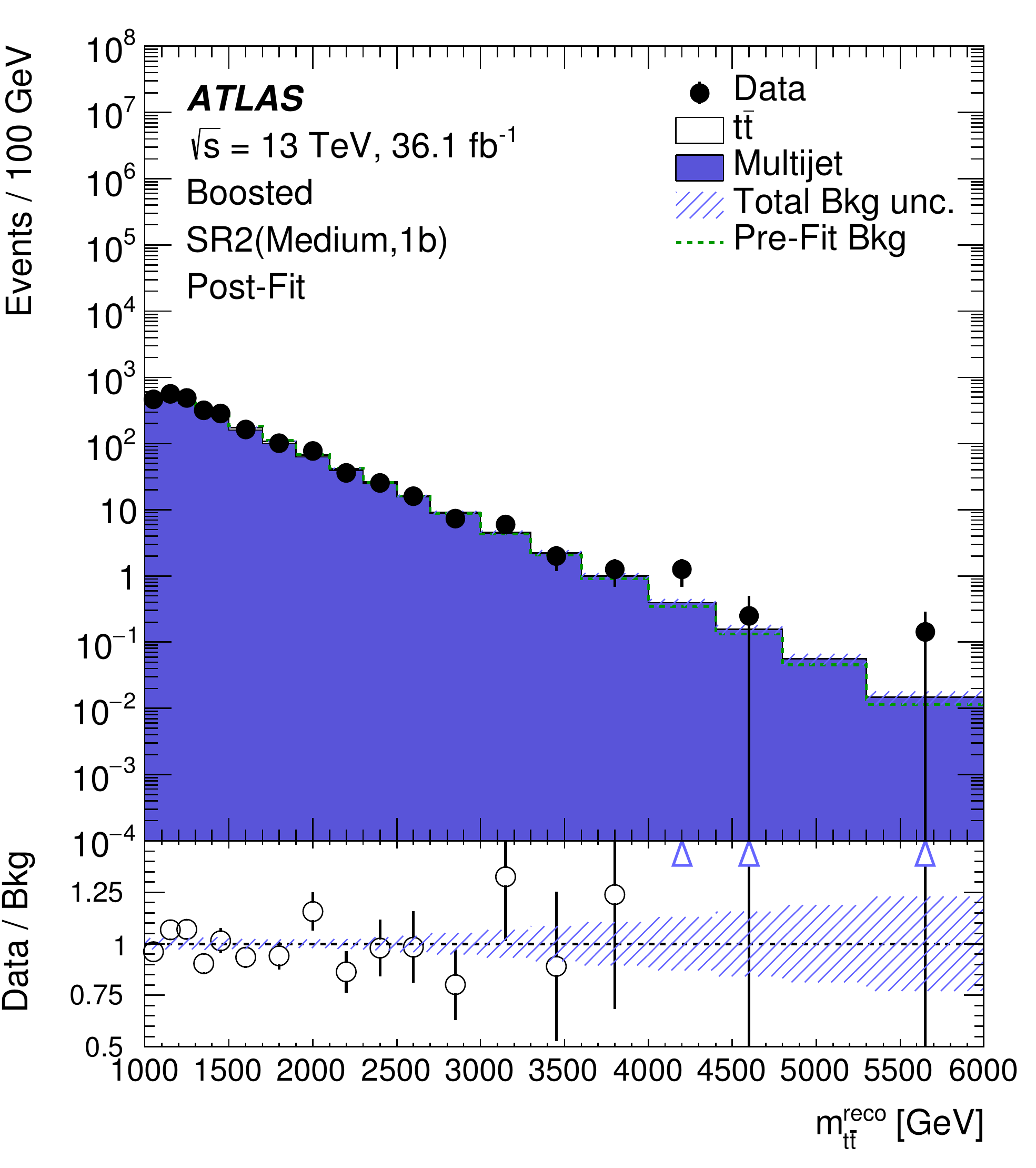}
}
\subfloat[\label{fig:boosted_medium_2bR2}]{
        \includegraphics[width=0.49\textwidth]{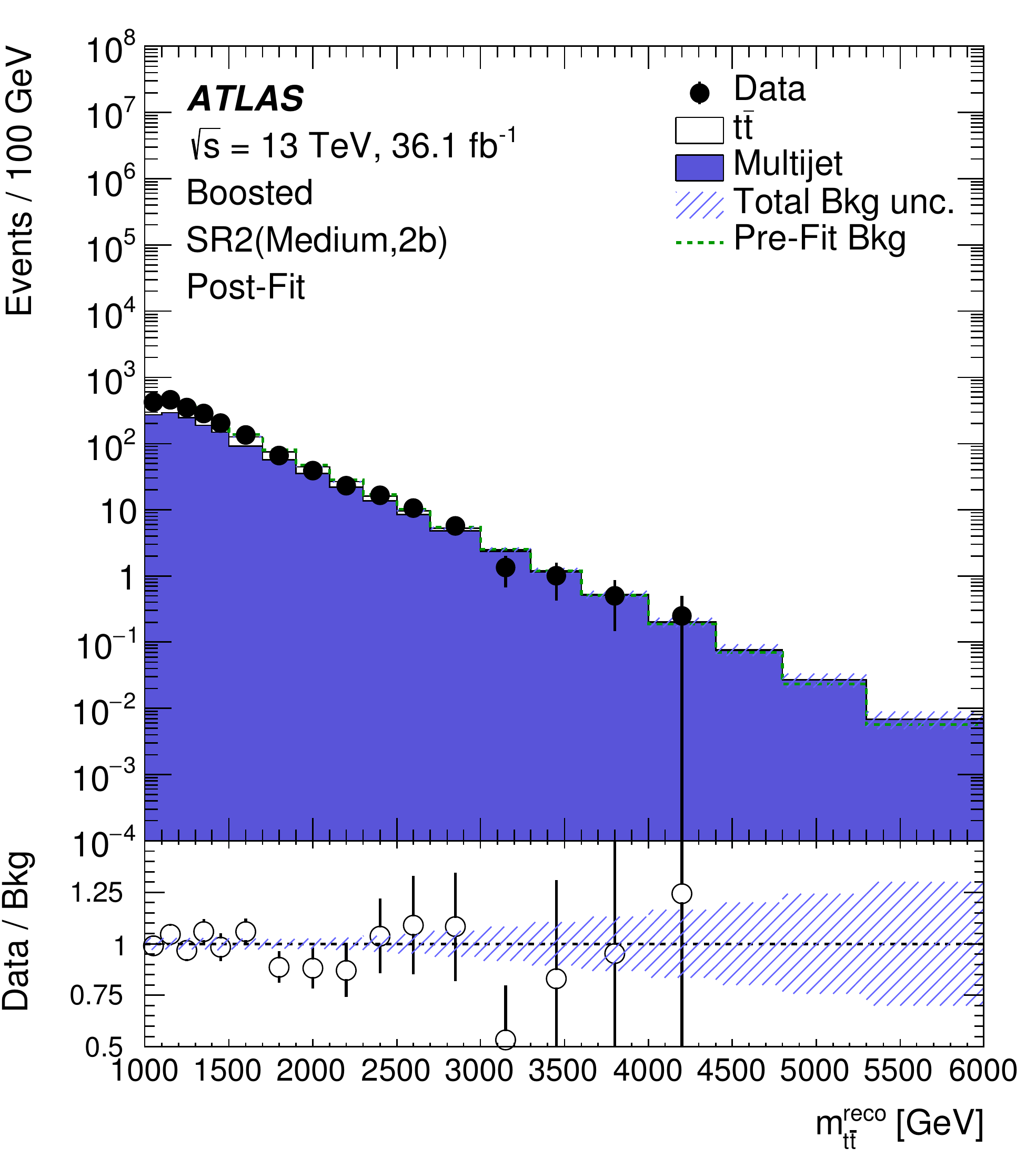}
}\\
\subfloat[\label{fig:boosted_tight_1bR2}]{
        \includegraphics[width=0.49\textwidth]{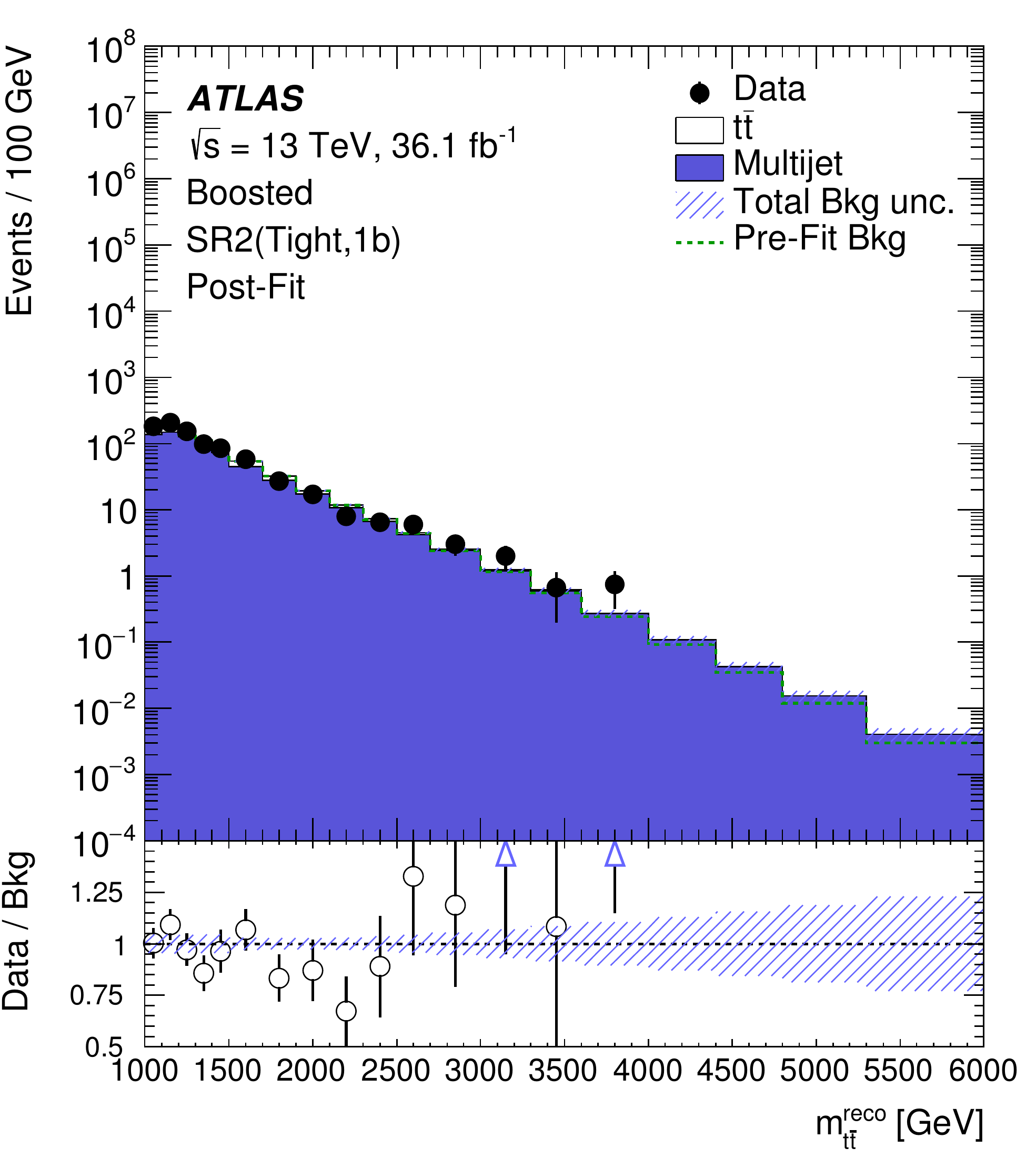}
}
\subfloat[\label{fig:boosted_tight_2bR2}]{
        \includegraphics[width=0.49\textwidth]{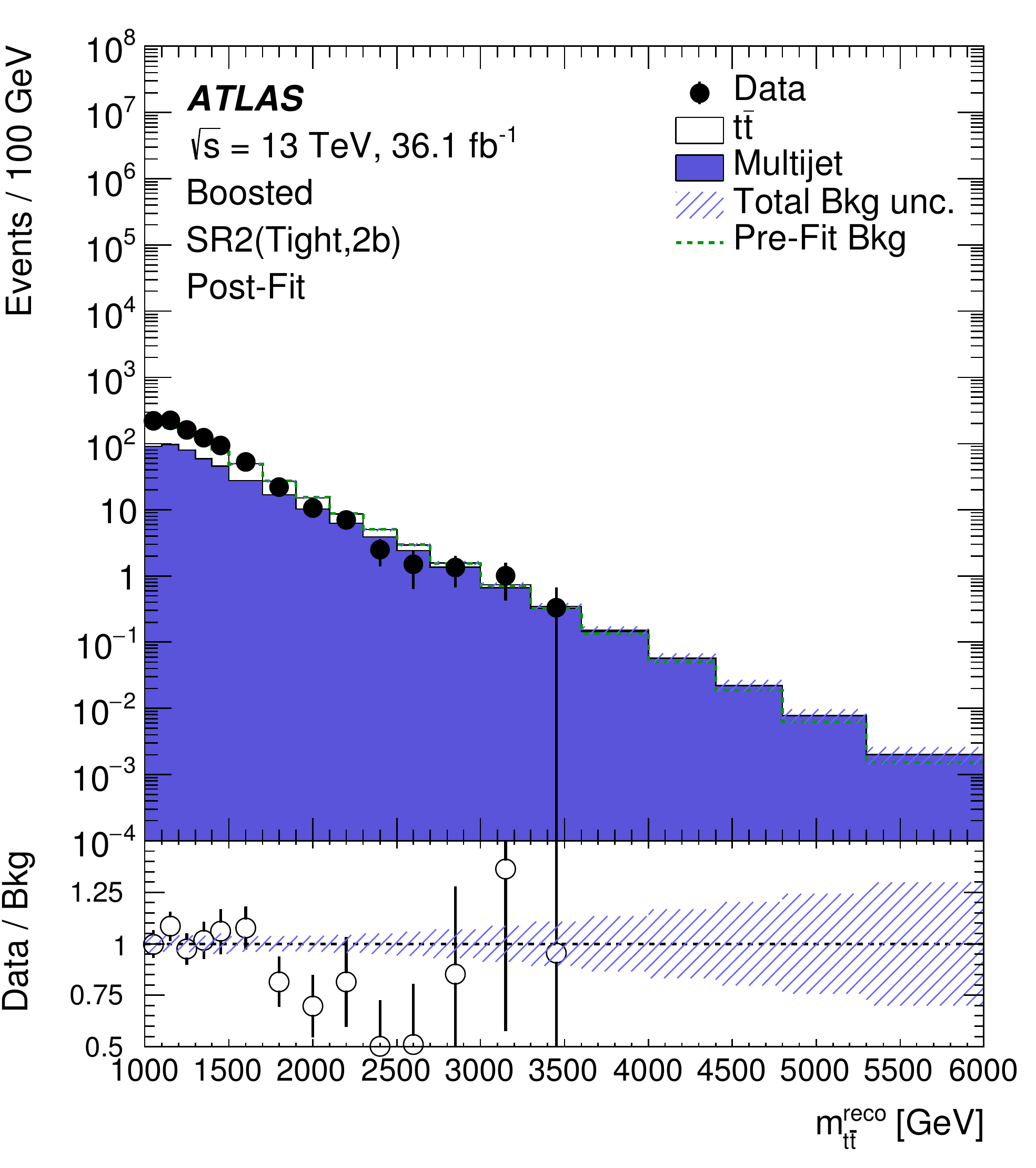}
}
\caption{Observed \mttreco distributions in (a) Medium R2 1b (b) Medium R2 2b (c) Tight R2 1b and (d) Tight R2 2b
after the fit (``Post-Fit'') under the background-only hypothesis for the boosted analysis. The shaded areas around the histograms indicate the total uncertainties in the background. The lower panel of the distribution shows the ratio of data to the fitted background prediction. The open triangles indicate that the ratio values are outside the plotted range.
The distributions before the fit are shown by the dashed lines and the background components are shown as stacked histograms. 
The multijet contribution also contains all other small non-\ttbar backgrounds.
}
\label{fig:boosted_signalR2}
\end{figure}

\begin{table}[t]
\centering
\caption{Expected and observed yields in the main signal region D and multijet-enriched regions A--C for the resolved analysis. The yields and their uncertainties are evaluated after the fit to data under the background-only hypothesis.
The expected $\zprime$ signal yields with masses of 0.75 and 1~\TeV\ are calculated using the $\mu=1$ hypothesis.
The multijet contribution also contains all other small non-\ttbar backgrounds.
}
\label{tab:tbYields}
\typeout{ATTeXLiveVersion is [\ATTeXLiveVersion]}
\ifthenelse{\ATTeXLiveVersion < 2012}{%
\sisetup{round-mode=figures, round-precision=2,
group-integer-digits=true, group-four-digits=true}
}{%
\sisetup{round-mode=figures, round-precision=2,
group-digits=integer, group-minimum-digits=5}
}
\begin{tabular}{l
S[table-format=6, table-number-alignment=right,
round-mode=figures, round-precision=3]@{$\,\pm\,$}
S[table-format=3.1, table-number-alignment=left,
round-mode=figures, round-precision=2]@{$\,$}
S[table-format=5, table-number-alignment=right,
round-mode=figures, round-precision=3]@{$\,\pm\,$}
S[table-format=3.1, table-number-alignment=left,
round-mode=figures, round-precision=2]@{$\,$}
S[table-format=6, table-number-alignment=right,
round-mode=figures, round-precision=3]@{$\,\pm\,$}
S[table-format=3.1, table-number-alignment=left,
round-mode=figures, round-precision=2]@{$\,$}
S[table-format=6, table-number-alignment=right,
round-mode=figures, round-precision=3]@{$\,\pm\,$}
S[table-format=3.1, table-number-alignment=left,
round-mode=figures, round-precision=2]@{$\,$}}
\toprule
Type                           & \multicolumn{2}{c}{Region A} & \multicolumn{2}{c}{Region B} & \multicolumn{2}{c}{Region C}     & \multicolumn{2}{c}{Region D}   \\
\midrule
$t \bar t$                     &  4299                   & 277  & 2740 & 188 &   9816                   & 458 &   8986                   & 248   \\
Multijet (template)            & \numRFSI{\pm}{4}{31418} & 771  & 4440 & 364 & \numRFSI{\pm}{4}{12842}  & 531 &   1816                   & 251   \\ 
\midrule
Total background               & \numRFSI{\pm}{4}{35717} & 772  & 7180 & 367 & \numRFSI{\pm}{4}{22658}  & 347 &  \numRFSI{\pm}{4}{10801} & 186   \\
\midrule
Data                           & \multicolumn{1}{S[table-format=6, table-number-alignment=left,round-mode=figures, round-precision=5]@{\hspace{0.5em}}}{35722} & &                          \multicolumn{1}{S[table-format=4, table-number-alignment=left,round-mode=figures, round-precision=4]@{\hspace{0.5em}}}{7 186} &  &                         \multicolumn{1}{S[table-format=6, table-number-alignment=left,round-mode=figures, round-precision=5]@{\hspace{0.5em}}}{22 665} & &                          \multicolumn{1}{S[table-format=6, table-number-alignment=left,round-mode=figures, round-precision=5]@{\hspace{0.5em}}}{10 821} &        \\
\midrule
$\zprime(0.75~\text{\TeV})$    &   470 &  68  &  367 &  91 &  1196  & 141 &   1203 & 182   \\ 
$\zprime(1~\text{\TeV})$     &   460 &  65  &  296 &  37 &  1018  & 127 &   1014 & 148   \\
\bottomrule
\end{tabular}
\end{table}

\begin{table}[t]
\centering
\caption{Expected and observed yields in the signal regions for the boosted analysis. The yields and their uncertainties are evaluated after the background-only fit to the data. 
The expected $\zprime$ signal yields with masses of 1.5 and 3~\TeV\ are calculated using the $\mu=1$ hypothesis.
The multijet contribution also contains all other small non-\ttbar backgrounds.
}
\label{tab:boostedYields}
\begin{tabular}{l
S[table-format=5, table-number-alignment=right,
round-mode=figures, round-precision=3]@{$\,\pm\,$}
S[table-format=2.1, table-number-alignment=left,
round-mode=figures, round-precision=2]@{$\,$}
S[table-format=5, table-number-alignment=right,
round-mode=figures, round-precision=3]@{$\,\pm\,$}
S[table-format=2.1, table-number-alignment=left,
round-mode=figures, round-precision=2]@{$\,$}
S[table-format=5, table-number-alignment=right,
round-mode=figures, round-precision=3]@{$\,\pm\,$}
S[table-format=2.1, table-number-alignment=left,
round-mode=figures, round-precision=2]@{$\,$}
S[table-format=5, table-number-alignment=right,
round-mode=figures, round-precision=3]@{$\,\pm\,$}
S[table-format=2.1, table-number-alignment=left,
round-mode=figures, round-precision=2]@{$\,$}}
\toprule
Type                        & \multicolumn{2}{c}{$\text{SR}1(\text{Medium},1b)$}        & \multicolumn{2}{c}{$\text{SR}1(\text{Medium},2b)$}      & \multicolumn{2}{c}{$\text{SR}1(\text{Tight},1b)$}     & \multicolumn{2}{c}{$\text{SR}1(\text{Tight},2b)$} \\
\midrule
$t \bar t$                  &  320 & 50      &  930 &  50 & 440 & 70  & 1350 & 70\\
Multijet (template)         & 1360 & 60      &  810 &  50 & 510 & 70  &  330 & 50\\
\midrule                                                                         
Total background            & 1680 & 40      & 1740 &  40 & 950 & 30  & 1680 & 50\\
\midrule
Data                           & \multicolumn{1}{S[table-format=4, table-number-alignment=right,round-mode=figures, round-precision=4]@{\hspace{1em}}}{1689} & &
                                 \multicolumn{1}{S[table-format=4, table-number-alignment=right,round-mode=figures, round-precision=4]@{\hspace{1em}}}{1730} & &
                                 \multicolumn{1}{S[table-format=4, table-number-alignment=right,round-mode=figures, round-precision=4]@{\hspace{1em}}}{ 952} & &
                                 \multicolumn{1}{S[table-format=4, table-number-alignment=right,round-mode=figures, round-precision=4]@{\hspace{1em}}}{1676} &        \\
\midrule
$\zprime(1.5~\text{\TeV})$  & 100 & 20     & 280 & 20    & 150 & 20 & 460 & 30 \\
$\zprime(3~\text{\TeV})$    &   4 &  1     &   8 &  1    &   4 &  1 &   8 &  1 \\
\bottomrule
\toprule
Type                        & \multicolumn{2}{c}{$\text{SR}2(\text{Medium},1b)$}   & \multicolumn{2}{c}{$\text{SR}2(\text{Medium},2b)$}   & \multicolumn{2}{c}{$\text{SR}2(\text{Tight},1b)$}   & \multicolumn{2}{c}{$\text{SR}2(\text{Tight},2b)$} \\
\midrule
$t \bar t$                  & 250  & 40    &  690 &  60   & 190  & 30      &  510 & 40 \\
Multijet (template)         & 2760 & 60    & 1640 &  70   & 820  & 50      &  510 & 50 \\
\midrule                                                                              
Total background            & 3010 & 50    & 2330 &  50   & 1010 & 30      & 1020 & 30 \\
\midrule
Data                           & \multicolumn{1}{S[table-format=4, table-number-alignment=right,round-mode=figures, round-precision=4]@{\hspace{1em}}}{3006} & &
                                 \multicolumn{1}{S[table-format=4, table-number-alignment=right,round-mode=figures, round-precision=4]@{\hspace{1em}}}{2322} & &
                                 \multicolumn{1}{S[table-format=4, table-number-alignment=right,round-mode=figures, round-precision=4]@{\hspace{1em}}}{ 989} & &
                                 \multicolumn{1}{S[table-format=4, table-number-alignment=right,round-mode=figures, round-precision=4]@{\hspace{1em}}}{1021} &        \\
\midrule
$\zprime(1.5~\text{\TeV})$ & 80 & 10     & 210 & 20     & 70 & 10    & 190 & 20 \\
$\zprime(3~\text{\TeV})$   &  4 &  1     &   6 &  1     &  3 &  1    &   5 &  1 \\
\bottomrule
\end{tabular}
\end{table}

\begin{table}
\centering
\caption{The relative impact of the post-fit uncertainties on the signal strength parameter $\mu$ using the \zprime benchmark model with $m=0.75$ (3)~\TeV\ in the resolved (boosted) analysis.
The eight systematic uncertainties with the highest impact on the signal strength parameter in the resolved and boosted analyses, respectively, are shown. 
The uncertainty on the extrapolation using an exponential function at high \mttreco above 1.5~\TeV\ applies to the boosted analysis only.
To estimate the impact from a given source of systematic uncertainty, the fit is performed with the nuisance parameter for the test fixed to the $\pm 1\sigma$ value after the nominal fit and the other nuisance parameters floated. The differences between the best-fit $\mu$ values in the tests and the nominal fit are divided by total post-fit uncertainty in $\mu$ are shown in this table. The total systematic uncertainty is different from the sum in quadrature of the different components due to correlations between nuisance parameters built by the fit. The statistical uncertainty in the data is evaluated by fixing all the nuisance parameters in the fit to the best-fit values except for the free-floating normalization factors.}
\label{tab:postFitImpactOnMu}
 
\begin{tabular}{lcc}
\toprule
& Resolved (\zprime $m=0.75\,\TeV$) & Boosted (\zprime $m=3~\TeV$) \\
\midrule
Source of uncertainty             & Relative impact on $\mu$    & Relative impact  on $\mu$\\
\midrule
Luminosity                        & {$<0.01$}      & $+0.03$/$-0.03$\\
$b$-tagging efficiency                             & $+0.05$/$-0.04$         & $+0.07$/$-0.07$\\
Small- and large-$R$ JES and JER                  & $+0.20$/$-0.24$         & $+0.21$/$-0.09$\\
$t \overline t$ modeling          & $+0.34$/$-0.33$         & $+0.10$/$-0.09$\\
Multijet estimation               & $+0.25$/$-0.27$         & $+0.16$/$-0.13$\\
Extrapolation                     & {--}                & $+0.34$/$-0.33$\\
PDF                               & $+0.07$/$-0.08$         & $+0.10$/$-0.10$\\
Pileup reweighting                & $+0.07$/$-0.05$         & {$<0.01$}\\
Simulation statistical uncertainty & $\pm 0.41$         & {--}\\
\midrule
Total systematic uncertainty      & $\pm 0.92$       & $\pm 0.67$\\
\midrule
Data statistical uncertainty      & $\pm 0.39$       & $\pm 0.74$\\
\bottomrule
\end{tabular} 

\end{table}

In the absence of a significant excess above the background prediction, 95\% CL upper limits on the cross-section times branching fraction of new particles decaying into \ttbar are calculated at each mass value for the different benchmark signal models considered.
The expected and observed upper limits on the cross-section times branching fraction of $\zprime\to\ttbar$ are presented in 
Figure~\ref{fig:massLimitUnblinded}.
Due to the strength of the expected limits, results from the resolved analysis are shown at $m_{\zprime}$ below 1.2~\TeV, whereas the results of the boosted analysis are shown above that value.
The NLO theory cross-section predictions for the \zprime with $\Gamma=1\%$ and 3\%, as well as those at LO with $\Gamma=1.2\%$ are overlaid.
The observed (expected) 95\% CL exclusion range is set for the \zprime masses between 
\zprimeOnepLimitLowmassObs\ and \zprimeOnepLimitHighmassObs~\TeV\ (\zprimeOnepLimitLowmassExp\ and \zprimeOnepLimitHighmassExp~\TeV) and \zprimeThreepLimitLowmassObs\ and \zprimeThreepLimitHighmassObs~\TeV\ (\zprimeThreepLimitLowmassExp\ and \zprimeThreepLimitHighmassExp~\TeV) 
for $\Gamma=1\%$ and 3\%, respectively.
Limits are also set on the cross-section times branching fraction of the vector and 
axial-vector mediators \zprimemed in the simplified DM model, as shown in Figure \ref{fig:massLimitDM}.
The vector (axial-vector) mediator \zprimemed is excluded in the mass ranges of $\zprimeVLimitLowmassObsResolved\,\TeV < m_{\zprimevec} < \zprimeVLimitHighmassObsResolved\,\TeV$ and $\zprimeVLimitLowmassObs\,\TeV < m_{\zprimevec} < \zprimeVLimitHighmassObs\,\TeV$ ($\zprimeAVLimitLowmassObsResolved\,\TeV < m_{\zprimeaxvec} < \zprimeAVLimitHighmassObsResolved\,\TeV$ and $\zprimeAVLimitLowmassObs\,\TeV < m_{\zprimeaxvec} < \zprimeAVLimitHighmassObs\,\TeV$) at 95\% CL by the data with the corresponding expected mass ranges of $\zprimeVLimitLowmassExpResolved\,\TeV < m_{\zprimevec} < \zprimeVLimitHighmassExpResolved\,\TeV$ and $\zprimeVLimitLowmassExp\,\TeV < m_{\zprimevec} < \zprimeVLimitHighmassExp\,\TeV$ ($\zprimeAVLimitLowmassExp\,\TeV < m_{\zprimeaxvec} < \zprimeAVLimitHighmassExp\,\TeV$).
The upper limit on the cross-section times branching fraction of the \kkG in the bulk RS model is shown in 
Figure~\ref{fig:massLimitGravitonUnblinded}. 
The cross-section times branching fraction for \kkG production with the model
parameters described in Section~\ref{sec:signal} is too low to be excluded with the sensitivity of this measurement,
hence the limit is presented only up to 3~\TeV.
Figure~\ref{fig:massLimitGluonUnblinded} shows the upper limit on the cross-section times branching fraction of the \kkg with $\Gamma=30\%$ in the RS model with a single warped extra dimension. 
The observed and expected lower limits on the \kkg mass are \kkgThirtypLimitObs\ and \kkgThirtypLimitExp~\TeV, respectively. 
The exclusion limit is also extracted for the \kkg as a function of the width at representative mass values. 
Figures~\ref{fig:massLimitKKGluonWidth500}, \ref{fig:massLimitKKGluonWidth1000}, \ref{fig:massLimitKKGluonWidth1500},~\ref{fig:massLimitKKGluonWidth2000} and~\ref{fig:massLimitKKGluonWidth5000} show the results for  $m_{\kkg} = 0.5$, 1.0, 1.5, 2.0 and 5.0~\TeV, respectively.
For $m_{\kkg}>0.5$~\TeV, the limits on the cross-section times branching fraction deteriorate with increasing \kkg width as the signal peak of the reconstructed \mttreco distribution becomes broad.
The limit at $m_{\kkg}=0.5$~\TeV\ does not depend on the signal width since the events with reconstructed $\mttreco<0.5$~\TeV\ are covered by one bin, as shown in Figure~\ref{fig:tbMttbarPostfitUnblinded}.

The extracted lower limits on the masses for various signal hypotheses where the sensitivity of the analysis allows for it are summarized in Table~\ref{tab:excluded_masses}.

\begin{figure}
\centering
\includegraphics[width=.49\textwidth, angle =-0]{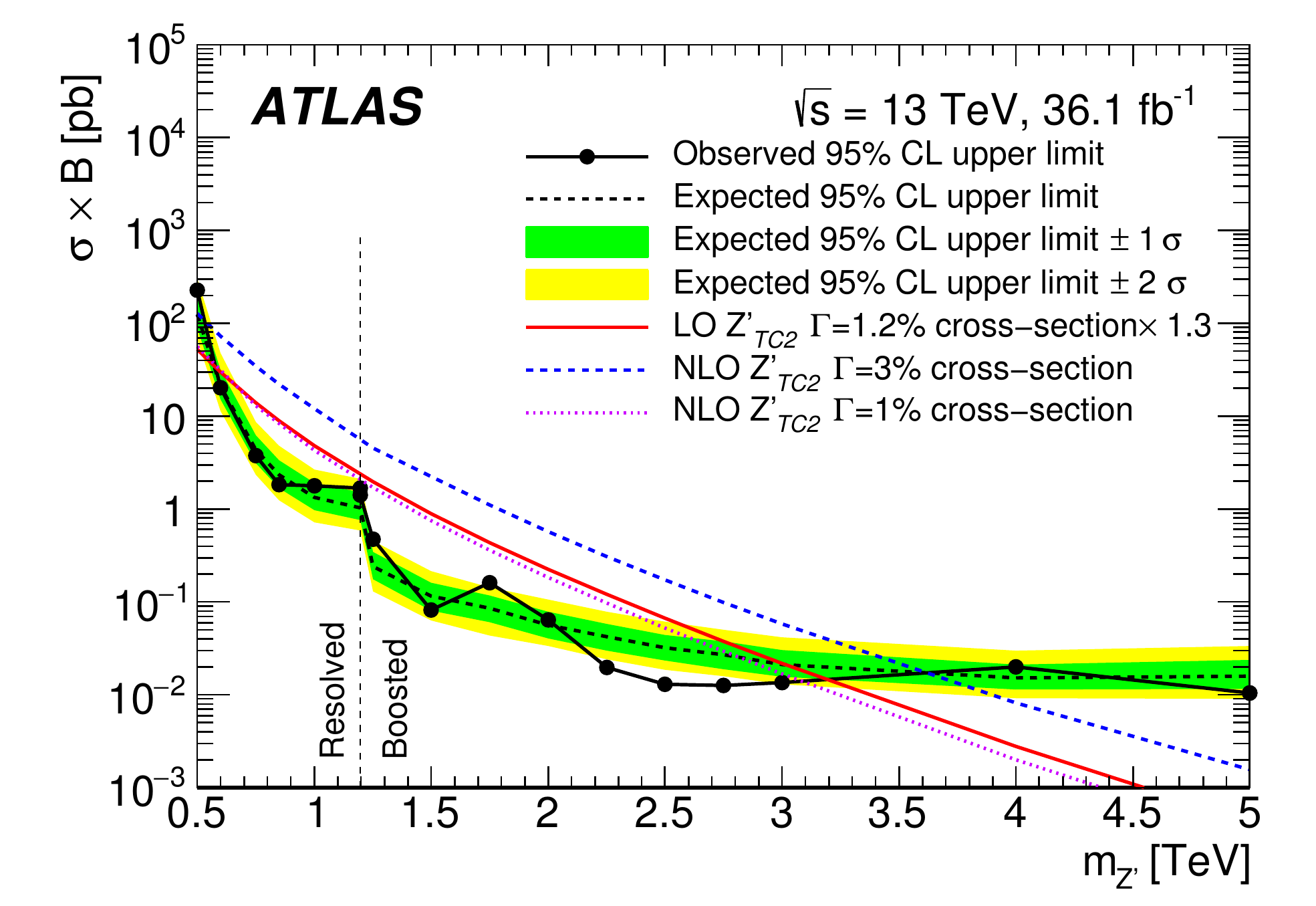}
\caption{Observed and expected upper limits on the cross-section times branching fraction of \zprime decaying into \ttbar as a function of the \zprime mass. The theory predictions of the cross sections for the \zprime with $\Gamma=1\%$ and 3\% are shown by the dotted and dashed lines at NLO and by the solid line with $\Gamma=1.2\%$ at LO, respectively. The results from the resolved and boosted analyses are shown to the left and right of the vertical dashed line, respectively.}
\label{fig:massLimitUnblinded}
\end{figure}

\begin{figure}
\centering
\subfloat[]{
	\includegraphics[width=.49\textwidth, angle =-0]{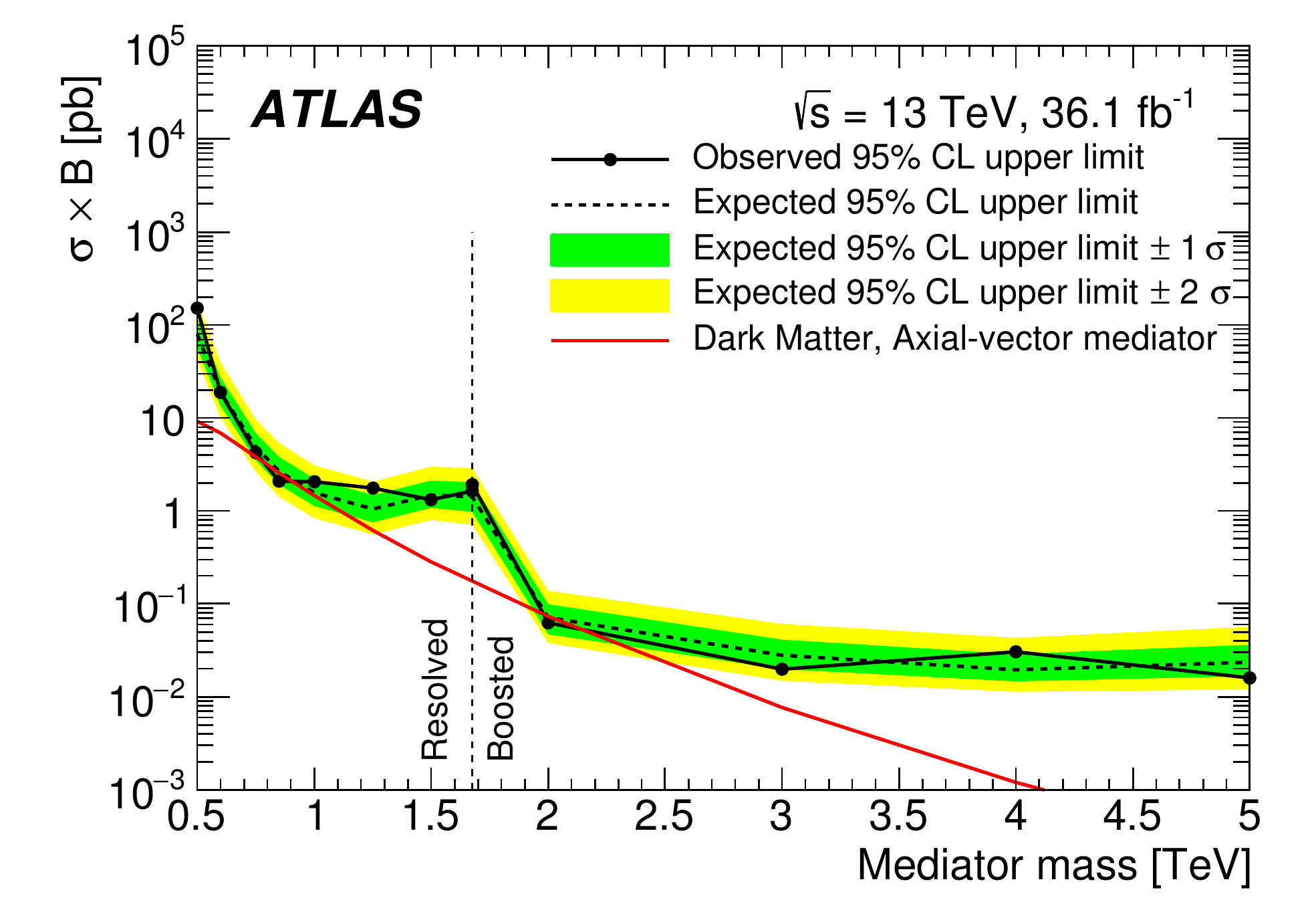}
	}
\subfloat[]{
	\includegraphics[width=.49\textwidth, angle =-0]{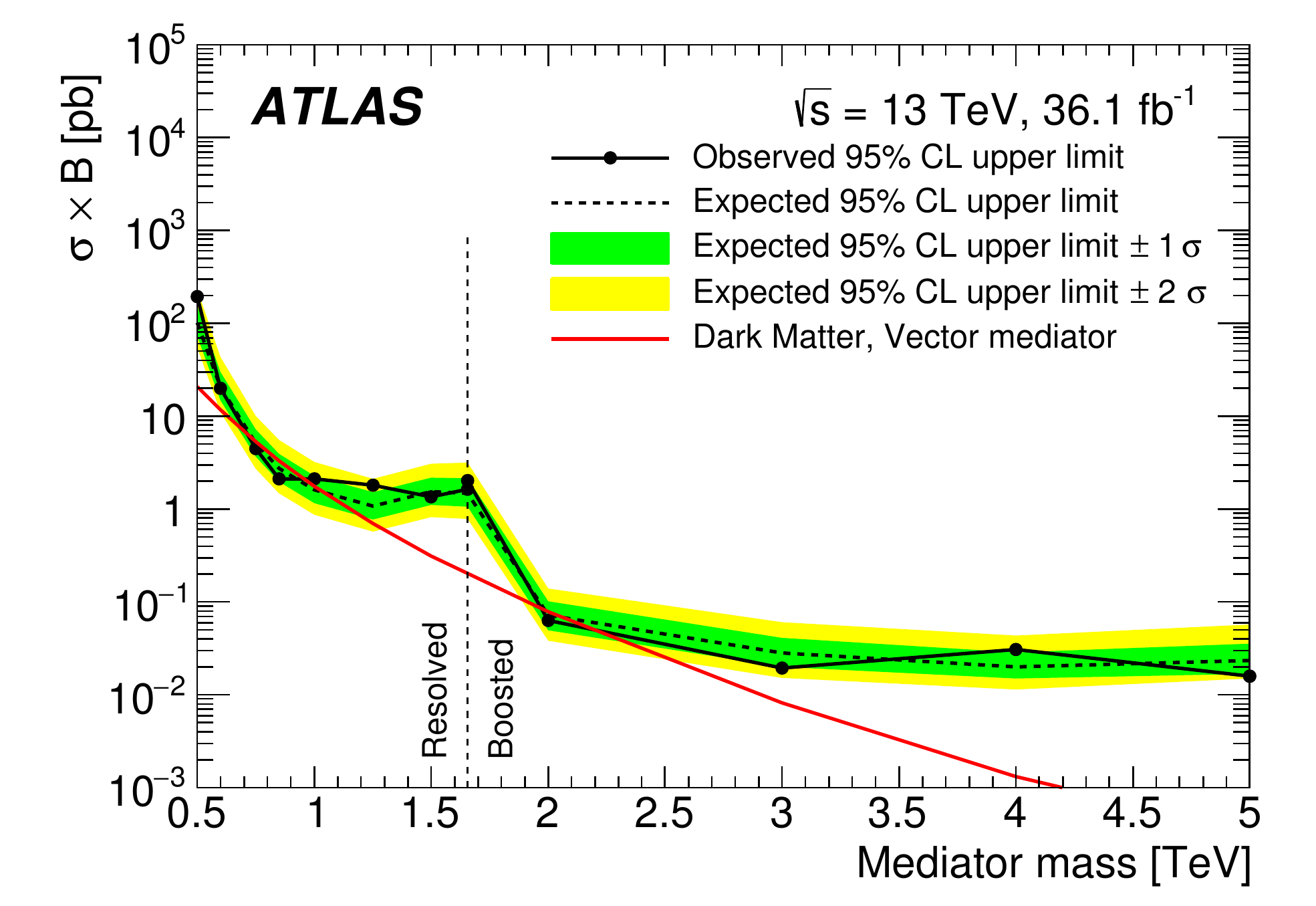}
	}
\caption{Observed and expected 95\% CL upper limits on the cross-section times branching fraction of \zprimemed decaying into \ttbar as a function of the \zprimemed mass. The theoretical predictions of the cross sections for the \zprimemed in the (a) A1 axial-vector mediator and (b) V1 vector mediator scenarios of the benchmark DM models are shown by the solid lines. The resolved and boosted analyses are shown to the left and right of the vertical dashed line, respectively.}
\label{fig:massLimitDM}
\end{figure}

\begin{figure}
\centering
\includegraphics[width=.49\textwidth, angle =-0]{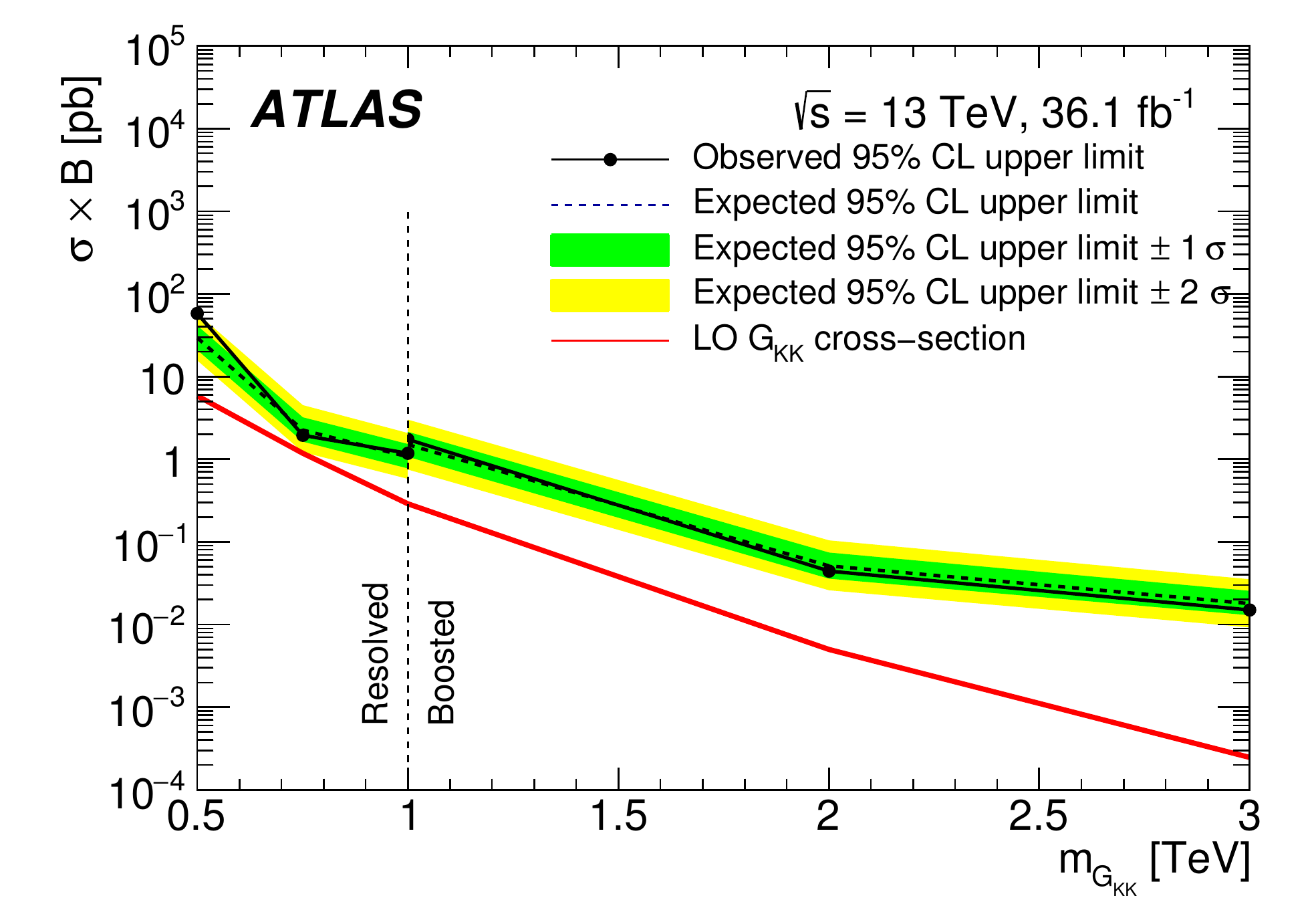}
\caption{Observed and expected 95\% CL upper limits on the cross-section times branching fraction of \kkG decaying into \ttbar as a function of the \kkG mass. The theoretical prediction of the cross section for the \kkG in the bulk RS model with $k/\overline{M}_{\mathrm{Pl}}=1.0$ is shown by the solid line. The resolved and boosted analyses are shown to the left and right of the vertical dashed line, respectively.}
\label{fig:massLimitGravitonUnblinded}
\end{figure}

\begin{figure}
\centering
\includegraphics[width=.49\textwidth, angle =-0]{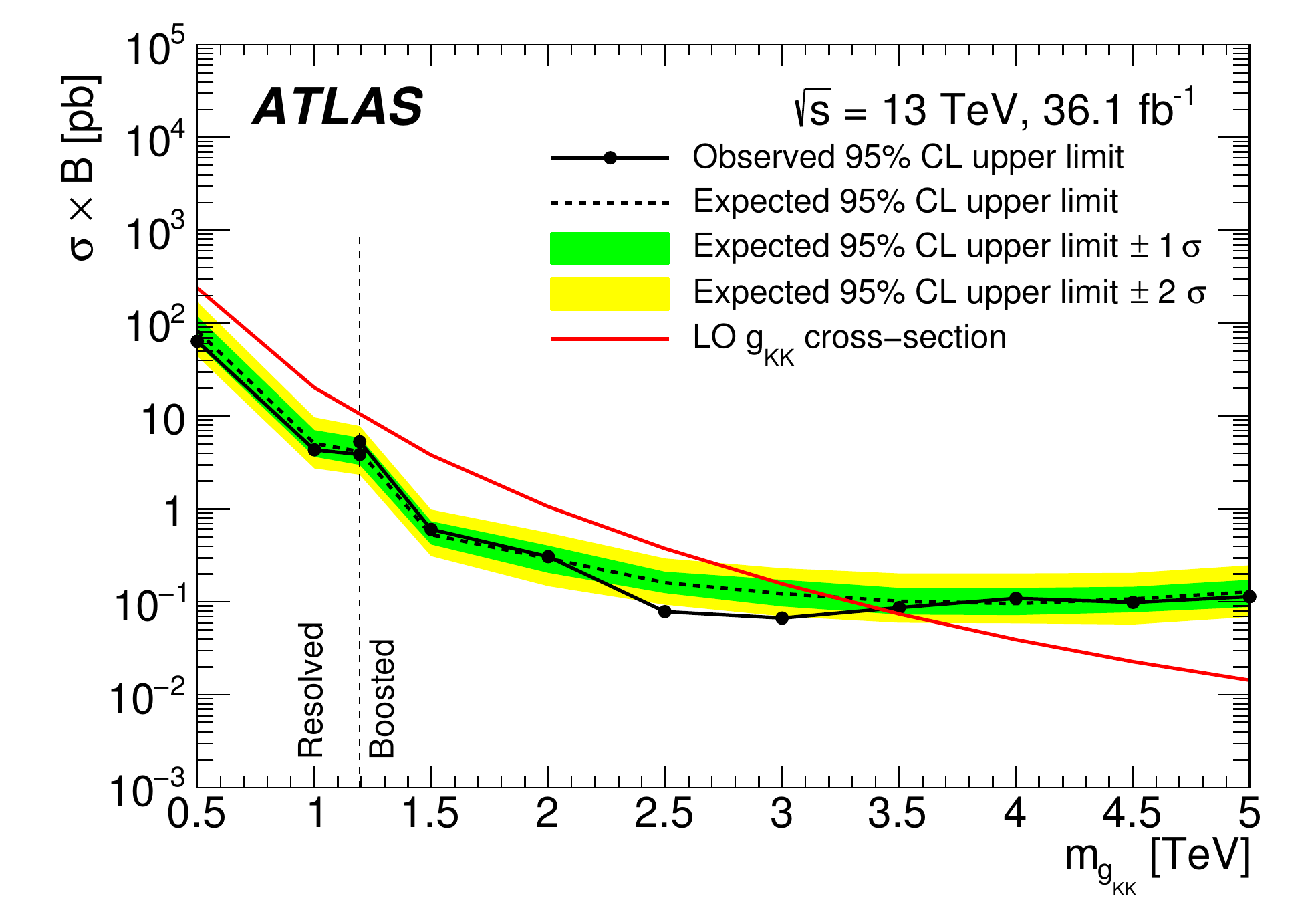}
\caption{Observed and expected 95\% CL upper limits on the cross-section times branching fraction of \kkg decaying into \ttbar as a function of the \kkg mass with $\Gamma=30\%$. The theoretical prediction of the cross section for the \kkg in the RS model with a single warped extra dimension is shown by the solid line. The resolved and boosted analyses are shown to the left and right of the vertical dashed line, respectively.}
\label{fig:massLimitGluonUnblinded}
\end{figure}

\begin{figure}
\centering
\subfloat[\label{fig:massLimitKKGluonWidth500}]{\includegraphics[width=.49\textwidth, angle =-0]{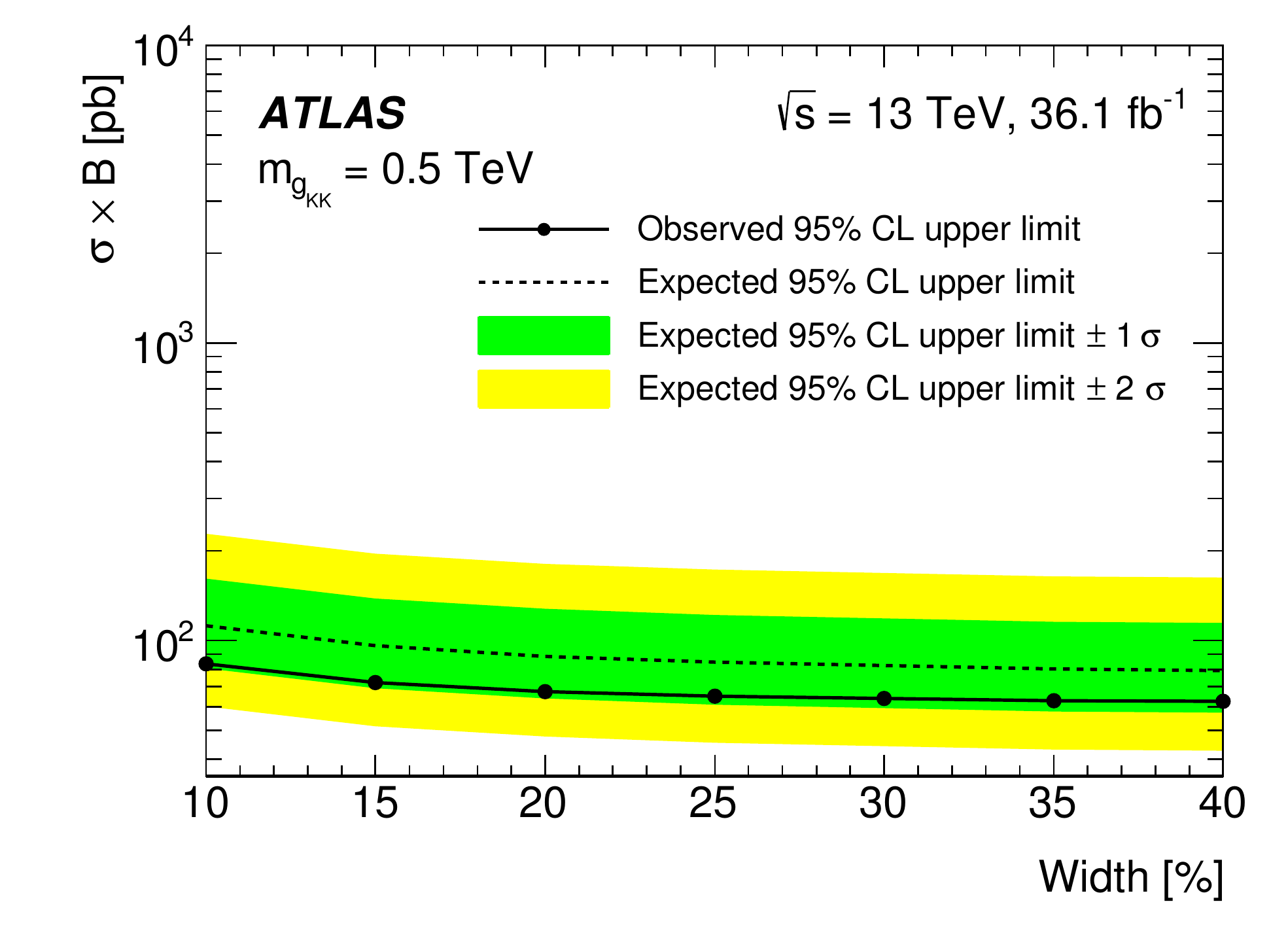}}
\subfloat[\label{fig:massLimitKKGluonWidth1000}]{\includegraphics[width=.49\textwidth, angle =-0]{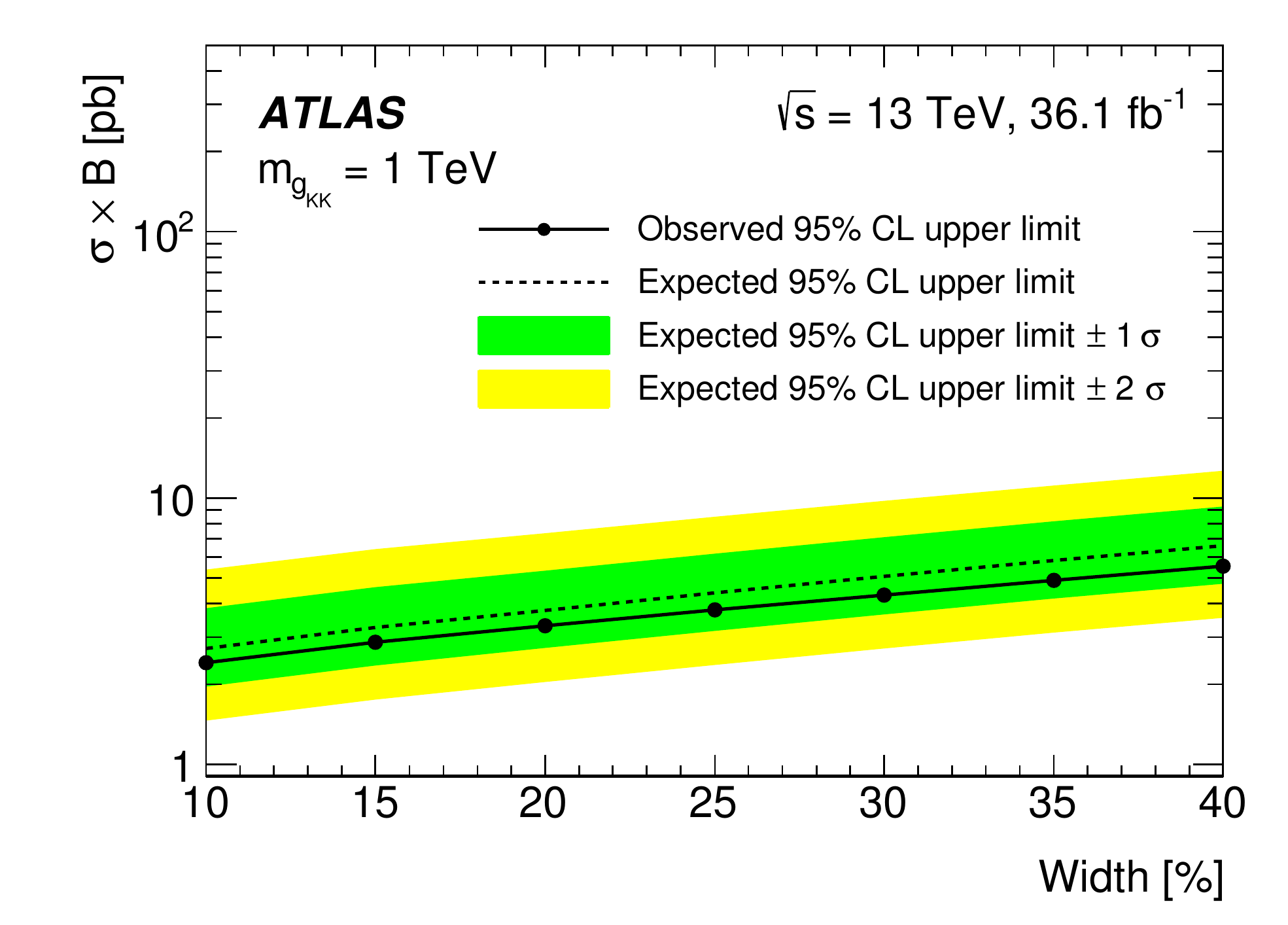}}\\
\subfloat[\label{fig:massLimitKKGluonWidth1500}]{\includegraphics[width=.49\textwidth, angle =-0]{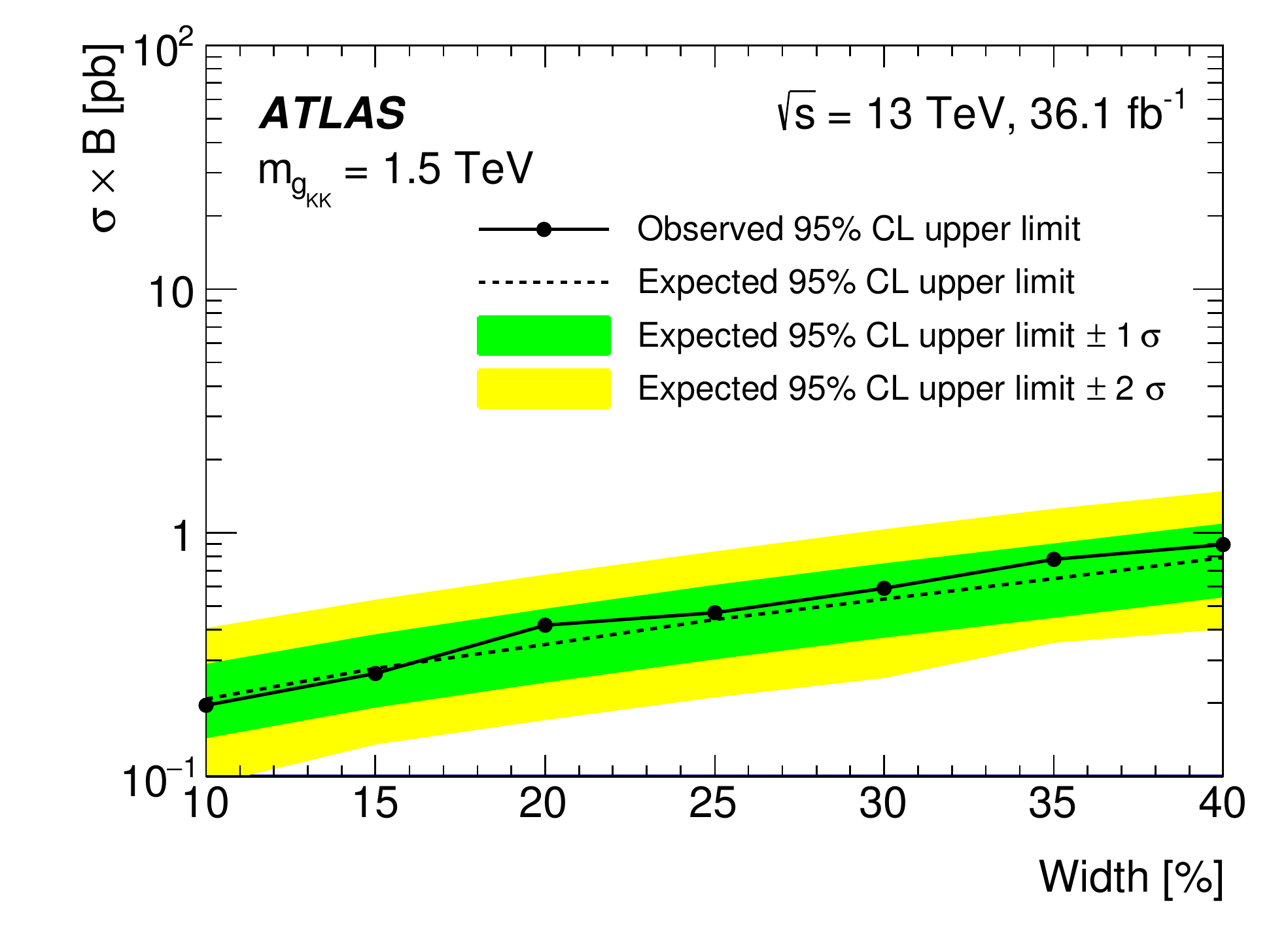}}
\subfloat[\label{fig:massLimitKKGluonWidth2000}]{\includegraphics[width=.49\textwidth, angle =-0]{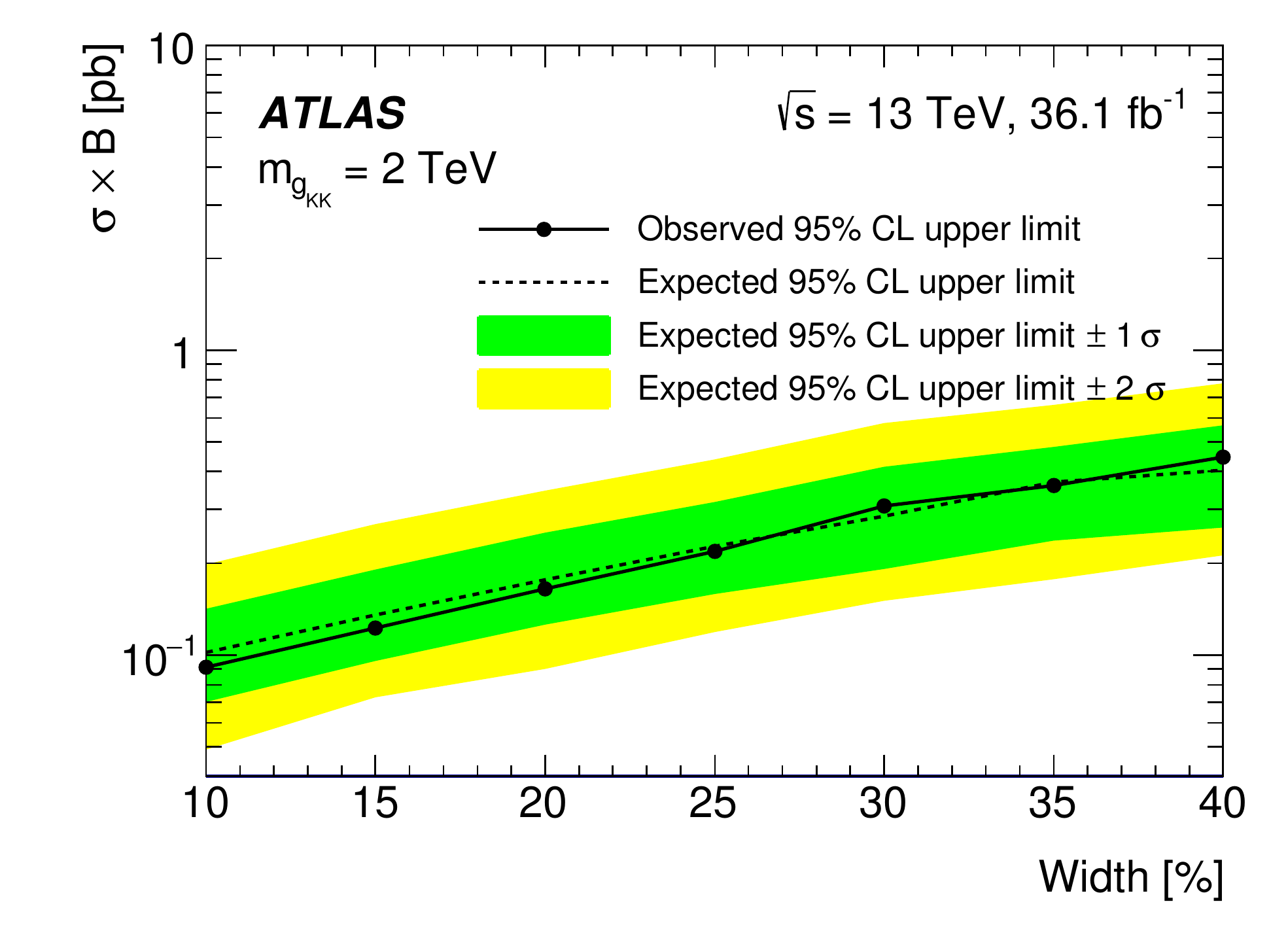}}\\
\subfloat[\label{fig:massLimitKKGluonWidth5000}]{\includegraphics[width=.49\textwidth, angle =-0]{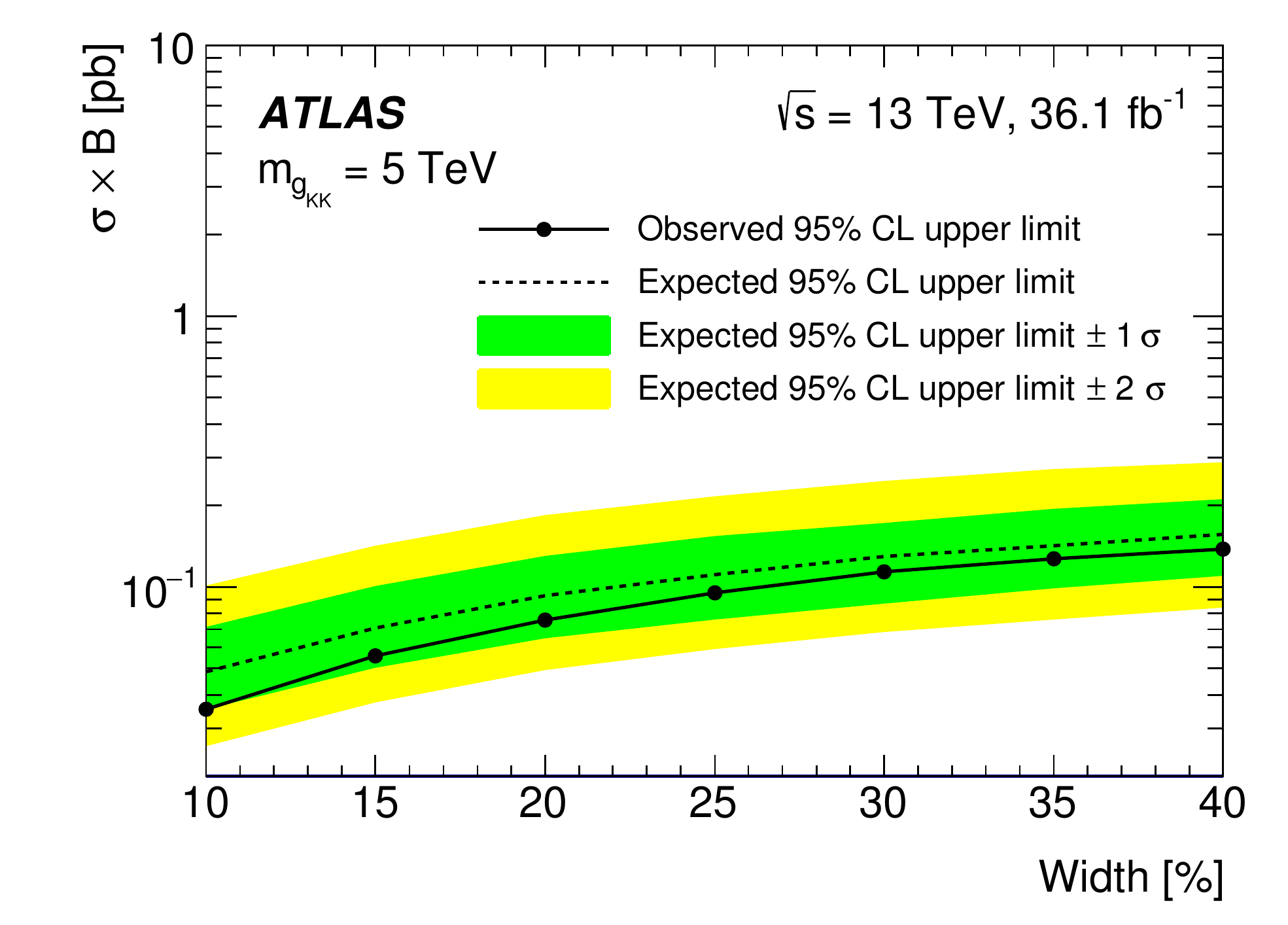}}\\
\caption{Observed and expected 95\% CL upper limits on cross-section times branching fraction of \kkg decaying into \ttbar as a function of the width of \kkg for masses of (a) 0.5~\TeV\ and (b) 1~\TeV\ (using the resolved analysis) and  (c) 1.5~\TeV\, (d) 2.0~\TeV\ and (e) 5.0~\TeV\ (using the boosted analysis). The width refers to the decay width of a resonance divided by the resonance mass.}
\label{fig:massLimitKKGluonWidth}
\end{figure}

\begin{table}[t]
\centering
\caption{Summary of expected and observed  excluded mass ranges at 95\% CL for the benchmark models studied.}
\label{tab:excluded_masses}
\begin{tabular}{llcc}
\toprule
Signal & & Expected excluded mass [\TeV] & Observed excluded mass [\TeV] \\
\midrule
\multirow{2}{*}{\zprime}                & ($\Gamma = 1\%$)  & $\left[ \zprimeOnepLimitLowmassExp, \zprimeOnepLimitHighmassExp \right]$ & $\left[ \zprimeOnepLimitLowmassObs, \zprimeOnepLimitHighmassObs \right]$ \\
                                        & ($\Gamma = 3\%$)  & $\left[ \zprimeThreepLimitLowmassExp, \zprimeThreepLimitHighmassExp \right]$ & $\left[ \zprimeThreepLimitLowmassObs, \zprimeThreepLimitHighmassObs \right]$ \\
\midrule
\multirow{2}{*}{\zprimemed}             & (vector)       & $\left[ \zprimeVLimitLowmassExpResolved, \zprimeVLimitHighmassExpResolved \right] \cup \left[ \zprimeVLimitLowmassExp, \zprimeVLimitHighmassExp \right]$ & $\left[ \zprimeVLimitLowmassObsResolved, \zprimeVLimitHighmassObsResolved \right] \cup \left[ \zprimeVLimitLowmassObs, \zprimeVLimitHighmassObs \right]$ \\
                                        & (axial-vector) & $\left[ \zprimeAVLimitLowmassExp, \zprimeAVLimitHighmassExp \right]$ & $\left[ \zprimeAVLimitLowmassObsResolved, \zprimeAVLimitHighmassObsResolved \right] \cup \left[ \zprimeAVLimitLowmassObs, \zprimeAVLimitHighmassObs \right]$ \\
\midrule
\multirow{4}{*}{\kkg}                   & ($\Gamma = 10\%$) & $< \kkgTenpLimitExp$ & $< \kkgTenpLimitObs$ \\
                                        & ($\Gamma = 20\%$) & $< \kkgTwentypLimitExp$ & $< \kkgTwentypLimitObs$ \\
                                        & ($\Gamma = 30\%$) & $< \kkgThirtypLimitExp$ & $< \kkgThirtypLimitObs$ \\
                                        & ($\Gamma = 40\%$) & $< \kkgFortypLimitExp$ & $< \kkgFortypLimitObs$ \\
\bottomrule
\end{tabular}
\end{table}

\FloatBarrier 
\section{Conclusion}
\label{sec:conclusion}
A search for resonant production of \ttbar decaying into the fully hadronic final state is performed using 
36.1~\ifb\ of $pp$ collision data recorded at $\sqrt{s}=13$~\TeV\ with the ATLAS detector at the LHC\@. 
Depending on the mass of new hypothetical particles, the search exploits two analysis techniques optimized 
for the reconstruction of a top-quark pair and background suppression.
No significant deviation from the Standard Model expectation is observed over the search range 
considered. Upper limits are set on the production cross-section times branching fraction for several 
benchmark signals, such as \zprime boson predicted in the topcolor-assisted-technicolor model, 
vector and axial-vector mediators \zprimemed in the dark-matter simplified model, and 
the Kaluza--Klein excitations of the graviton \kkG and gluon \kkg in the specific models 
based on the Randall--Sundrum scenario of warped extra dimensions.
The \zprime boson is excluded in the mass range of \zprimeOnepLimitLowmassObs\ and
\zprimeOnepLimitHighmassObs~\TeV\ (\zprimeThreepLimitLowmassObs\ and \zprimeThreepLimitHighmassObs~\TeV)
for the decay width of 1\% (3\%).
The vector (axial-vector) mediator \zprimemed is excluded in the mass ranges of $\zprimeVLimitLowmassObsResolved\,\TeV < m_{\zprimevec} < \zprimeVLimitHighmassObsResolved\,\TeV$ and $\zprimeVLimitLowmassObs\,\TeV < m_{\zprimevec} < \zprimeVLimitHighmassObs\,\TeV$ ($\zprimeAVLimitLowmassObsResolved\,\TeV < m_{\zprimeaxvec} < \zprimeAVLimitHighmassObsResolved\,\TeV$ and $\zprimeAVLimitLowmassObs\,\TeV < m_{\zprimeaxvec} < \zprimeAVLimitHighmassObs\,\TeV$).
The lower limit on the \kkg mass is set at \kkgThirtypLimitObs~\TeV\ for the decay width of 30\%.
The cross-section limits for the \zprime boson are comparable at a \zprime mass above $\sim1$~\TeV\ 
to those from the previous ATLAS {\em lepton-plus-jets} analysis performed at 13~\TeV~\cite{EXOT-2015-04}.


\section*{Acknowledgments}

We thank CERN for the very successful operation of the LHC, as well as the
support staff from our institutions without whom ATLAS could not be
operated efficiently.

We acknowledge the support of ANPCyT, Argentina; YerPhI, Armenia; ARC, Australia; BMWFW and FWF, Austria; ANAS, Azerbaijan; SSTC, Belarus; CNPq and FAPESP, Brazil; NSERC, NRC and CFI, Canada; CERN; CONICYT, Chile; CAS, MOST and NSFC, China; COLCIENCIAS, Colombia; MSMT CR, MPO CR and VSC CR, Czech Republic; DNRF and DNSRC, Denmark; IN2P3-CNRS, CEA-DRF/IRFU, France; SRNSFG, Georgia; BMBF, HGF, and MPG, Germany; GSRT, Greece; RGC, Hong Kong SAR, China; ISF and Benoziyo Center, Israel; INFN, Italy; MEXT and JSPS, Japan; CNRST, Morocco; NWO, Netherlands; RCN, Norway; MNiSW and NCN, Poland; FCT, Portugal; MNE/IFA, Romania; MES of Russia and NRC KI, Russian Federation; JINR; MESTD, Serbia; MSSR, Slovakia; ARRS and MIZ\v{S}, Slovenia; DST/NRF, South Africa; MINECO, Spain; SRC and Wallenberg Foundation, Sweden; SERI, SNSF and Cantons of Bern and Geneva, Switzerland; MOST, Taiwan; TAEK, Turkey; STFC, United Kingdom; DOE and NSF, United States of America. In addition, individual groups and members have received support from BCKDF, CANARIE, CRC and Compute Canada, Canada; COST, ERC, ERDF, Horizon 2020, and Marie Sk{\l}odowska-Curie Actions, European Union; Investissements d' Avenir Labex and Idex, ANR, France; DFG and AvH Foundation, Germany; Herakleitos, Thales and Aristeia programmes co-financed by EU-ESF and the Greek NSRF, Greece; BSF-NSF and GIF, Israel; CERCA Programme Generalitat de Catalunya, Spain; The Royal Society and Leverhulme Trust, United Kingdom. 

The crucial computing support from all WLCG partners is acknowledged gratefully, in particular from CERN, the ATLAS Tier-1 facilities at TRIUMF (Canada), NDGF (Denmark, Norway, Sweden), CC-IN2P3 (France), KIT/GridKA (Germany), INFN-CNAF (Italy), NL-T1 (Netherlands), PIC (Spain), ASGC (Taiwan), RAL (UK) and BNL (USA), the Tier-2 facilities worldwide and large non-WLCG resource providers. Major contributors of computing resources are listed in Ref.~\cite{ATL-GEN-PUB-2016-002}.

\clearpage
\printbibliography

\clearpage
\input{atlas_authlist}

\end{document}

%% file: atlas_authlist.tex
 
\begin{flushleft}
{\Large The ATLAS Collaboration}

\bigskip

M.~Aaboud$^\textrm{\scriptsize 34d}$,    
G.~Aad$^\textrm{\scriptsize 99}$,    
B.~Abbott$^\textrm{\scriptsize 125}$,    
D.C.~Abbott$^\textrm{\scriptsize 100}$,    
O.~Abdinov$^\textrm{\scriptsize 13,*}$,    
B.~Abeloos$^\textrm{\scriptsize 129}$,    
D.K.~Abhayasinghe$^\textrm{\scriptsize 91}$,    
S.H.~Abidi$^\textrm{\scriptsize 164}$,    
O.S.~AbouZeid$^\textrm{\scriptsize 39}$,    
N.L.~Abraham$^\textrm{\scriptsize 153}$
% ...
\bigskip
\\

% ...
$^{13}$Institute of Physics, Azerbaijan Academy of Sciences, Baku; Azerbaijan.\\
% ...
$^{34}$$^{(a)}$Facult\'e des Sciences Ain Chock, R\'eseau Universitaire de Physique des Hautes Energies - Universit\'e Hassan II, Casablanca;$^{(b)}$Centre National de l'Energie des Sciences Techniques Nucleaires (CNESTEN), Rabat;$^{(c)}$Facult\'e des Sciences Semlalia, Universit\'e Cadi Ayyad, LPHEA-Marrakech;$^{(d)}$Facult\'e des Sciences, Universit\'e Mohamed Premier and LPTPM, Oujda;$^{(e)}$Facult\'e des sciences, Universit\'e Mohammed V, Rabat; Morocco.\\
% ...
$^{39}$Niels Bohr Institute, University of Copenhagen, Copenhagen; Denmark.\\
% ...
$^{91}$Department of Physics, Royal Holloway University of London, Egham; United Kingdom.\\
% ...
$^{99}$CPPM, Aix-Marseille Universit\'e, CNRS/IN2P3, Marseille; France.\\
$^{100}$Department of Physics, University of Massachusetts, Amherst MA; United States of America.\\
% ...
$^{125}$Homer L. Dodge Department of Physics and Astronomy, University of Oklahoma, Norman OK; United States of America.\\
% ...
$^{129}$LAL, Universit\'e Paris-Sud, CNRS/IN2P3, Universit\'e Paris-Saclay, Orsay; France.\\
% ...
$^{153}$Department of Physics and Astronomy, University of Sussex, Brighton; United Kingdom.\\
% ...
$^{164}$Department of Physics, University of Toronto, Toronto ON; Canada.\\
% ...
$^{a}$ Also at Borough of Manhattan Community College, City University of New York, NY; United States of America.\\
$^{b}$ Also at California State University, East Bay; United States of America.\\
$^{c}$ Also at Centre for High Performance Computing, CSIR Campus, Rosebank, Cape Town; South Africa.\\
$^{d}$ Also at CERN, Geneva; Switzerland.\\
$^{e}$ Also at CPPM, Aix-Marseille Universit\'e, CNRS/IN2P3, Marseille; France.\\
% ...
$^{*}$ Deceased

\end{flushleft}
